\documentclass[a4paper,11pt]{article}
\pdfoutput=1 % if your are submitting a pdflatex (i.e. if you have
\usepackage{jcappub} % for details on the use of the package, please
\usepackage[T1]{fontenc} % if needed

\usepackage{bbm}
\usepackage{orcidlink}
\DeclareMathOperator{\SU}{\mathrm{SU}}
\DeclareMathOperator{\su}{\mathrm{su}}
\DeclareMathOperator{\U}{\mathrm{U}}
\DeclareMathOperator{\Sp}{\mathrm{Sp}}
\DeclareMathOperator{\USp}{\mathrm{USp}}

\DeclareMathOperator{\Li}{\mathrm{Li}}

\DeclareMathOperator{\Mat}{\mathrm{Mat}}
\DeclareMathOperator{\SO}{\mathrm{SO}}
\DeclareMathOperator{\so}{\mathrm{so}}
\DeclareMathOperator{\Z}{\mathbb{Z}}
\DeclareMathOperator{\C}{\mathbb{C}}
\DeclareMathOperator{\F}{\mathbb{F}}
\DeclareMathOperator{\Tr}{\mathrm{Tr}}

\DeclareMathOperator{\ide}{\mathbbm{1}}
\DeclareMathOperator{\zeromat}{0}
\DeclareMathOperator{\sgn}{sgn}

\newcommand{\transp}[1]{{#1}^{\tiny{T}}}
\newcommand{\dd}{{\rm{d}}}
\newcommand{\Tc}{T_{\mbox{\tiny{c}}}}
\newcommand{\TCMB}{T_{\mbox{\tiny{CMB}}}}
\newcommand{\Treh}{T_{\mbox{\tiny{reh}}}}
\newcommand{\spac}{a_{\mbox{\tiny{lat}}}}
\newcommand{\betac}{\beta_{\mbox{\tiny{c}}}}

\newcommand{\gimp}{g_{\mbox{\tiny{imp}}}}
\newcommand{\Lambdaimp}{\Lambda_{\mbox{\tiny{imp}}}}
\newcommand{\betaimp}{\beta_{\mbox{\tiny{imp}}}}
\newcommand{\Ntau}{N_\tau}
\newcommand{\Ns}{N_s}
\newcommand{\cs}{c_{\mbox{\tiny{s}}}}
\newcommand{\csSB}{c_{\mbox{\tiny{s,\,SB}}}}
\newcommand{\Lh}{L_{\mbox{\tiny{h}}}}
\newcommand{\CV}{C_V}
\newcommand{\CmaxV}{C^{\mbox{\tiny{max}}}_V}
\newcommand{\Tdark}{T_{\mbox{\tiny{dark}}}}
\newcommand{\TSM}{T_{\mbox{\tiny{SM}}}}
\newcommand{\sigmacd}{\sigma_{\mbox{\tiny{cd}}}}
\newcommand{\pc}{p_{\mbox{\tiny{c}}}}
\newcommand{\pd}{p_{\mbox{\tiny{d}}}}
\newcommand{\entropydensityc}{s_{\mbox{\tiny{c}}}}
\newcommand{\entropydensityd}{s_{\mbox{\tiny{d}}}}
\newcommand{\redchisq}{\chi^2_{\mbox{\tiny{red}}}}
\newcommand{\tpt}{t_{\mbox{\tiny{pt}}}}
\newcommand{\Tfo}{T_{\mbox{\tiny{f-o}}}}
\newcommand{\OmegaDM}{\Omega_{\mbox{\tiny{dm}}}}
\newcommand{\OmegaGW}{\Omega_{\mbox{\tiny{gw}}}}
\newcommand{\GeV}{\mathrm{GeV}}
\newcommand{\rc}{r_{\mbox{\tiny{c}}}}
\newcommand{\mPlanck}{m_{\mbox{\tiny{Planck}}}}
\newcommand{\gstareff}{g^{\mbox{\tiny{eff}}}_{\star}}
\newcommand{\vw}{v}
\newcommand{\vf}{v_{\mbox{\tiny{f}}}}
\newcommand{\gammaf}{\gamma_{\mbox{\tiny{f}}}}
\newcommand{\Fgwzero}{F_{\mathrm{gw}, 0}}
\newcommand{\fpzero}{f_{\mathrm{p}, 0}}
\newcommand{\Hn}{H_{\mbox{\tiny{n}}}}
\newcommand{\Tn}{T_{\mbox{\tiny{n}}}}
\newcommand{\eq}[1]{\begin{align}\label{#1}}
\newcommand{\en}{\end{align}}
\newcommand{\eqar}[1]{\begin{align}\label{#1}}
\newcommand{\enar}{\end{align}}
\newcommand{\CC}{C\nolinebreak\hspace{-.05em}\raisebox{.4ex}{\tiny\bf +}\nolinebreak\hspace{-.10em}\raisebox{.4ex}{\tiny\bf +}}

\newcommand{\revision}[1]{#1}

\title{\boldmath Thermal evolution of dark matter and gravitational-wave production in the early universe from a symplectic glueball model}

\author[a,b]{Mattia~Bruno,\orcidlink{0000-0002-5127-4461}}
\author[c]{Niccol\`o~Forzano,\orcidlink{0000-0003-0985-8858}}
\author[d,e,f]{Marco~Panero,\orcidlink{0000-0001-9477-3749}}
\author[c,g]{and Antonio~Smecca\orcidlink{0000-0002-8887-5826}}

\affiliation[a]{Department of Physics ``Giuseppe Occhialini'', Milano-Bicocca University,\\ Piazza della Scienza 3, I-20126 Milan, Italy}
\affiliation[b]{INFN, Milano-Bicocca unit,\\ Piazza della Scienza 3, I-20126 Milan, Italy}
\affiliation[c]{Department of Physics, Faculty of Science and Engineering, Swansea University,\\ Singleton Park, SA2 8PP Swansea, Wales, United Kingdom}
\affiliation[d]{Department of Physics, University of Turin,\\ Via Pietro Giuria 1, I-10125 Turin, Italy}
\affiliation[e]{INFN, Turin unit,\\ Via Pietro Giuria 1, I-10125 Turin, Italy}
\affiliation[f]{Department of Physics \& Helsinki Institute of Physics, University of Helsinki,\\ PL 64, FIN-00014 Helsinki, Finland}
\affiliation[g]{INFN, Roma Tre unit,\\ Via della Vasca Navale 84, I-00146 Rome, Italy}

\emailAdd{mattia.bruno@unimib.it}
\emailAdd{niccolo.forzano@swansea.ac.uk}
\emailAdd{marco.panero@unito.it}
\emailAdd{antonio.smecca@roma3.infn.it}

\abstract{The hypothesis that dark matter could be a bound state of a strongly coupled non-Abelian gauge theory is theoretically appealing and has a variety of interesting phenomenological implications. In particular, an interpretation of dark matter as the lightest glueball state in the spectrum of a dark Yang--Mills theory, possibly coupled to the visible sector only through gravitational interactions, has been discussed quite extensively in the literature, but most of previous work has been focused on dark $\SU(N)$ gauge theories. In this article, we consider an alternative model, based on a symplectic gauge group, which has a first-order confinement/deconfinement phase transition at a finite critical temperature. We first determine the equation of state of this theory, focusing on temperatures close to the transition, and evaluating the associated latent heat. Then we discuss the evolution of this dark-matter model in the early universe, commenting on the mechanisms by which it could indirectly interact with the visible sector, on the spectrum of gravitational waves it could produce, and on the relic abundances it would lead to. Our discussion includes an extensive review of relevant literature, a number of comments on similarities and differences between our model and dark $\SU(N)$ gauge theories, as well as some possible future extensions of the present study.
}

\begin{document}
\maketitle
\flushbottom

\section{Introduction}
\label{sec:introduction}

Identifying the nature and properties of dark matter remains a crucial open problem in fundamental physics, with major implications for astrophysics and for the evolution of the universe at the cosmological level~\cite{Cirelli:2024ssz}. By now, the hypothesis of the existence of dark matter is supported by observational evidence from many effects, on widely separated scales, ranging from the motion, structure, and evolution of many galaxies, to weak-gravitational-lensing effects in galaxy clusters, to the cosmic microwave background, to other effects relevant for cosmology~\cite{Begeman:1991iy, Burkert:1995yz, Bartelmann:1999yn, Sofue:2000jx, Springel:2006vs, Planck:2018vyg}. In particular, Big-Bang nucleosynthesis and observations of the cosmic microwave background indicate that roughly a quarter of the present energy density of the universe is accounted for by dark matter; in addition, the latter also plays a crucial r\^ole in explaining the inhomogeneities associated with the existence of structures like galaxies and clusters: if a matter species decoupled from photons much before baryons did during the evolution of the early universe, it would have allowed density perturbations to grow sufficiently, before the baryon recombination time. Even though some of these observations may be explained by modified theories of gravity~\cite{Milgrom:1983ca}, others cannot~\cite{Aguirre:2001fj, Clowe:2006eq, Dodelson:2011qv, McGaugh:2014nsa}, leaving a particle interpretation of dark matter as the most likely option~\cite{Bertone:2004pz, Hooper:2009zm, Bertone:2010zza}.

The main features that a particle dark matter should have include, in particular, being subject to gravitational interaction (as the evidence for dark matter comes from observations of effects related to gravity) while being ``dark'', namely electromagnetically neutral; in fact, the negative results from experiments aimed at direct detection of dark matter suggest that dark matter should be neutral under all of the interactions of the Standard Model of particle physics -- both the electroweak and the strong one. Finally, the lifetime of a dark-matter candidate is expected to be at least of the order of the age of the universe. In addition to these properties, there are hints that dark matter may be (at least partially) self-interacting, even though the allowed scattering cross section is constrained by observational bounds related to structure formation~\cite{Spergel:1999mh, Dave:2000ar, Vogelsberger:2012ku, Rocha:2012jg, Peter:2012jh, Tulin:2017ara, Banerjee:2019bjp, Yamanaka:2019aeq, Han:2023olf}.

Among many models that have been proposed (for an overview, see, for example, ref.~\cite{Feng:2010gw}), the possibility that dark matter is made of composite states of a novel, strongly coupled theory is especially appealing, both for theoretical and for phenomenological reasons~\cite{Kribs:2016cew}. Even at a very heuristic level, one should note that already the largest fraction of visible mass in the universe is generated dynamically by a strongly coupled gauge theory, namely quantum chromodynamics (QCD), so the same mechanism may be at work for dark matter, too---and, much like the dynamically generated hadronic scale in quantum chromodynamics, the mass scale of these composite states would not be affected by the naturalness problem. The composite states of a strongly coupled sector are ``dark'' if they are globally neutral under SM, even if their elementary constituents had non-vanishing SM charges. Moreover, their stability against decays could be due to some internal symmetry: if the leading operators that violate the latter have a large dimension, then the induced decay rate can be sufficiently small to make the lifetime of dark matter comparable with or longer than the age of the universe; note that this is analogous to the accidental baryon-number $\U(1)$ symmetry of the renormalizable SM Lagrangian, which, being broken only by dimension-$6$ operators, explains the very high lower bound on the proton lifetime~\cite{SNO:2018ydj}. In addition, this mechanism could also explain the (so far) experimentally unobserved interaction between dark and ordinary matter in terms of the suppression of the corresponding interaction terms by inverse powers of the dynamically generated dark confinement scale. A dark confining gauge theory supporting bound states with sufficiently large self-interaction cross section, but negligible annihilation or dissipation, may be consistent with observational constraints related to structure formation. Finally, in the literature it has been suggested that an interpretation of dark matter in terms of a theory akin to QCD could even explain the fact that the contributions to the total energy density of the universe from visible matter and from dark matter are of comparable magnitude~\cite{Bai:2013xga}.

A particularly simple realization of this type of dark-matter models is given by a confining non-Abelian gauge theory, without matter fields---a model similar to the purely gluonic sector of QCD. While the \emph{fundamental} degrees of freedom of this model correspond to massless, ``gluon-like'' particles not coupled to the Standard Model (SM) fields, the confinement phenomenon and the dynamical generation of a mass gap implies that the \emph{physical} states in the spectrum of the theory are massive ``glueball'' states, which are neutral under the fundamental ``color-like'' charge of the theory and under the SM. In particular, the lightest such state, which has spin, parity and charge quantum numbers $J^{PC}=0^{++}$, can be a dark matter candidate~\cite{Carlson:1992fn, Faraggi:2000pv}, as it is predicted to be completely neutral under the SM charges, to be stable on cosmological timescales by virtue of the confining nature of the new gauge interaction, and to interact with the SM particles only gravitationally, thereby providing a very secluded type of dark matter. Among the phenomenologically appealing features of this type of models, it is worth remarking that they naturally lead to a type of dark matter that could simultaneously be warm and self-interacting, and that would not be ruled out by the absence of direct experimental detection at colliders. In addition, the model depends only on a very limited number of free parameters and, therefore, has strong predictive power: once the gauge group is specified, all dimensionful physical quantities depend on a single energy scale. As we will discuss in detail below, this means, for example, that the masses of the entire spectrum of the theory are proportional to the lightest one, through ratios that are fixed and exactly computable from the microscopic Lagrangian of the theory -- as long as a non-perturbative calculation tool is available. In turn, the running of the renormalized coupling of the theory (in a given scheme) is specified by a $\Lambda$ parameter, which is also proportional to the mass of the lightest physical state in the spectrum; similarly, the force between a pair of probe sources in the fundamental representation of the gauge group at asymptotically large distances is proportional to the square of the mass of the lightest ``glueball'' of the theory.

Further interesting implications of a glueball dark-matter model can be derived, once one considers its behavior at a finite temperature $T$: in particular, at sufficiently high temperatures the color-singlet states cease to exist and give way to a plasma of gluon-like excitations interacting with each other through thermally screened long-distance forces. Again, the value of the critical deconfinement temperature $\Tc$ separating these two phases is proportional to the mass gap of the theory, and, if the confining and the plasma phase are separated by a first-order transition, the associated latent heat per unit volume is proportional to the fourth power of the mass gap. Remarkably, the liberation of a large number of light degrees of freedom at high temperatures would have an impact on the total density of free energy and other thermodynamic quantities in the early universe, affecting its cosmological cooling rate. 

The simplest formulation of a glueball-like model for dark matter is one in which not only the ``dark glueballs'' but also their constituent ``dark gluons'' are neutral under the SM. However, it is interesting to note that, if instead the constituents \emph{do} carry SM charges, when they are deconfined at the very high temperatures of the early universe they could interact sufficiently strongly with the SM particles: in principle, this could explain the fact that the present densities of baryonic matter and dark matter are of the same order of magnitude, and may also have interesting implications for the baryon asymmetry of the universe~\cite{Sakharov:1967dj, Nussinov:1985xr, Kaplan:2009ag, Zurek:2013wia}.

In the present work, we study the viability as a dark-matter model of a Yang-Mills theory based on local invariance under the $\Sp(2)$ compact symplectic group, which is the compact real form of the symplectic group over the field of complex numbers $\Sp(4,\C)$, and the intersection of the latter with the unitary group $\U(4)$.\footnote{Note that the compact symplectic groups like the one considered in this work are sometimes denoted as $\Sp(2N)$, or, alternatively, as $\USp(2N)$, in order to emphasize that, in the defining representation, they are given by matrices of size $2N \times 2N$. In this work, we use, instead, the convention in which they are denoted as $\Sp(N)$, as was done, for example, in ref.~\cite{Holland:2003kg}.} In particular, we focus on the behavior of this model at temperatures of the order of and above the deconfinement transition, and determine its equation of state non-perturbatively, by means of numerical calculations in the lattice regularization~\cite{Wilson:1974sk}.

An interesting feature of the model that we consider, is that the dark sector may have undergone a strong first-order thermal phase transition in the early universe, possibly producing gravitational waves~\cite{Kosowsky:1992rz, Schwaller:2015tja, Caprini:2015zlo, Hindmarsh:2020hop, Athron:2023xlk, Croon:2024mde}. Given that in the past decade the latter have finally entered the arena of experimental science~\cite{LIGOScientific:2016aoc}, it is now particularly timely and important to understand the features of the gravitational-wave spectrum that new-physics models could produce. To this purpose, it should be remarked that, in addition to those produced by non-equilibrium processes at a discontinuous transition, any plasma in thermal equilibrium produces gravitational waves, with an emission rate proportional to the shear viscosity~\cite{Ghiglieri:2015nfa}; while the latter can become relatively large at weak coupling, this contribution to the gravitational-wave spectrum remains subdominant with respect to those generated at a first-order phase transition. Gravitational waves produced in a dark sector contribute to the total energy density (and to the effective number of massless degrees of freedom), so the experimental detection of gravitational waves with a characteristic spectrum could provide useful hints about new physics. For example, a first-order phase transition at temperatures of the order of hundreds $\GeV$, i.e., the scale characteristic of electroweak physics, could generate gravitational waves with peak frequencies of the order of $10^{-3}$~Hz, which are potentially within the reach of space-based gravitational-wave observatories~\cite{Corbin:2005ny, Kawamura:2011zz, TianQin:2015yph, Hu:2017mde, LISA:2017pwj, Barausse:2020rsu, Colpi:2024xhw}. Given that the SM predicts only a smooth crossover at the electroweak scale~\cite{Kajantie:1996mn, Kajantie:1996qd, Karsch:1996yh, Gurtler:1997hr, Csikor:1998eu, Aoki:1999fi, DOnofrio:2014rug, DOnofrio:2015gop}, the experimental detection of gravitational waves in this frequency range could be interpreted as evidence of physics beyond the Standard Model.

The expectation of a first-order transition in our model is suggested by known results in theories based on $\SU(N)$ gauge groups: while in QCD with physical quark masses there is no exact global symmetry distinguishing the hadronic from the quark-gluon plasma phase, and the confinement/deconfinement transition is actually a crossover~\cite{Aoki:2006we, Bazavov:2011nk}, for sufficiently large quark masses the change of phase is a first-order transition~\cite{Brown:1990ev} as in the purely gluonic $\SU(3)$ theory~\cite{Boyd:1996bx, Borsanyi:2012ve} (for a recent discussion, see also ref.~\cite{Shao:2024dxt}). In fact, when fermions are absent (or sufficiently heavy to be effectively quenched), the change of state between the deconfined and the confining phase of an $\SU(N)$ gauge theory can be defined unambiguously: traditionally, this has been expressed in terms of the restoration of center symmetry~\cite{Polyakov:1975rs, Svetitsky:1982gs} (while a more modern take on the subject, based on generalized global symmetries, is discussed in ref.~\cite{Iqbal:2024pee} and in the references therein). The transition is known to become more and more strongly discontinuous when the number of color charges $N$ is increased~\cite{Lucini:2003zr, Panero:2009tv}; this was conjectured to be due to the larger and larger imbalance between the number of degrees of freedom in the confining phase, where the thermodynamics can be modeled in terms of a gas of color-singlet glueballs, and in the deconfined phase, in which the multiplicity of states in the plasma is expected to scale like the dimension of the Lie algebra of the group~\cite{Pepe:2004rc}. Beside the possibility of a gravitational-wave signature, it has been pointed out that glueball-like dark matter may lead to a very rich pattern of phenomenological implications, possibly including the existence of massive dark stars that could be detected through gravitational lensing~\cite{Soni:2016gzf}.

Models featuring dark matter candidates in the form of glueball states have been studied in several recent articles; a partial list of references includes refs.~\cite{Juknevich:2009ji, Juknevich:2009gg, Feng:2011ik, Boddy:2014yra, GarciaGarcia:2015fol, Pappadopulo:2016pkp, Dienes:2016vei, Forestell:2016qhc, Soni:2016gzf, Soni:2016yes, Soni:2017nlm, Forestell:2017wov, Acharya:2017szw, Mendes:2018ifl, Archer-Smith:2019gzq, LatticeStrongDynamics:2020jwi, Halverson:2020xpg, Huang:2020crf, Bigazzi:2020avc, Kang:2021epo, Garcia-Bellido:2021zgu, Garani:2021zrr, Reichert:2021cvs, Reichert:2022naa, Morgante:2022zvc, Carenza:2022pjd, Curtin:2022tou, Curtin:2022oec, Kolesova:2023yfp, Carenza:2023eua, Zu:2023olm, Yamada:2023thl, Pasechnik:2023hwv, Batz:2023zef, Biondini:2024cpf, McKeen:2024trt, Carenza:2024avj}. This family of models can be thought of as examples of the ``hidden valley'' paradigm~\cite{Strassler:2006im, Strassler:2006ri, Han:2007ae} (which admits realizations in various theoretical scenarios~\cite{Blumenhagen:2005mu, Chacko:2005pe, Burdman:2006tz, Zurek:2008qg, Arkani-Hamed:2008kxc, Nelson:2008hj}) and offer phenomenologically attractive features; in particular, while they evade current bounds from colliders~\cite{Pospelov:2008zw}, they could be coupled to SM particles through portal interactions~\cite{Holdom:1985ag, Patt:2006fw, March-Russell:2008lng, Djouadi:2011aa}.

The spectrum and related properties of glueballs in $\SU(N)$ gauge theories have been studied in the lattice regularization in various works~\cite{Morningstar:1999rf, Lucini:2004my, Meyer:2004gx, Chen:2005mg, Loan:2006gm, Meyer:2008tr, Lucini:2010nv, Caselle:2011fy, Yamanaka:2019yek, Yamanaka:2019aeq, Athenodorou:2020ani, Athenodorou:2021qvs}, but the literature on analogous computations in $\Sp(N)$ gauge theories is more limited. In fact, with the exception of the smallest symplectic group $\Sp(1)$ (which is nothing but $\SU(2)$, and gauge theories based on this group -- with or without dynamical matter fields -- have already been studied very extensively in lattice simulations, both for the qualitative similarities with QCD and for the potential relevance for physics beyond the SM: for a recent review, see ref.~\cite[subsection~5.2]{Aarts:2023vsf}), the first lattice study of symplectic gauge theories was reported in ref.~\cite{Holland:2003kg}; during the past few years, research on this subject has gained significant momentum~\cite{Bennett:2017kga, Bennett:2019jzz, Bennett:2019cxd, Bennett:2020hqd, Bennett:2020qtj, Kulkarni:2022bvh, Bennett:2022yfa, Bennett:2022ftz, Bennett:2022gdz, Vadacchino:2022zcj, Bennett:2023wjw, Bennett:2023gbe, Bennett:2023mhh, Forzano:2023zyr, Bennett:2023qwx, Hsiao:2023nyn, Bennett:2024cqv, Dengler:2024maq, Bennett:2024bhy, Mason:2024dve, Bennett:2024tex, Hsiao:2024tjf}.

While the focus of our present work is on its possible viability as a dark-matter model, we mention that the investigation of gauge theories based on symplectic groups is also interesting for other, more conceptual, reasons. These include: (i) the possibility of investigating the dynamics of (de)confinement and the relevant degrees of freedom in a family of gauge groups having the same center, since the latter is known to play a prominent r\^ole in the confining mechanism~\cite{tHooft:1979rtg, Mack:1980rc, DelDebbio:1996lih, Greensite:2003bk}, and (ii) the correspondences that are expected to relate gauge theories based on symplectic and orthogonal gauge groups to each other~\cite{Mkrtchian:1981bb, Cvitanovic:1982bq, Lovelace:1982hz, Unsal:2007fb, Mkrtchian:2011oda, Caputa:2013vla, Bond:2019npq}.

The structure of this article is as follows. After discussing the formulation of $\Sp(N)$ gauge theories in the lattice regularization and the method to compute the equation of state in section~\ref{sec:formulation_of_the_model}, we present our results for the equilibrium thermodynamic properties of $\Sp(2)$ theory in section~\ref{sec:results}, while in section~\ref{sec:interpretation_as_a_model_for_dark_matter_and_gravitational-wave_production} we discuss the possible interpretation of these results in terms of a model for dark matter, and their implications (including, in particular, the possible generation of primordial gravitational waves at the first-order thermal transition of the theory), before recapitulating our findings in section~\ref{sec:conclusions}. Some general properties of symplectic groups are discussed in appendix~\ref{app:generalities_about_the_compact_symplectic_group}, while appendix~\ref{app:details_of_the_simulation_updates} is devoted to technical details about our Markov-chain Monte~Carlo algorithm. Throughout this paper we assume natural units, whereby the speed of light in vacuum, the reduced Planck's constant and Boltzmann's constant are set to unity.

\section{Formulation of the model}
\label{sec:formulation_of_the_model}

Our non-perturbative study of the equation of state of $\Sp(2)$ Yang-Mills theory is carried out through a regularization of the theory on a Euclidean, isotropic, hypercubic lattice $\Lambda$ of spacing $\spac$. The physical extent of the system in each of the three spatial directions (labeled by indices $1$, $2$ and $3$) is denoted as $L=\spac\Ns$, while the size of the Euclidean-time direction (labeled by the index $0$) is denoted as $\spac\Ntau$; furthermore, let $V=L^3$ denote the spatial volume of the system. Periodic boundary conditions are imposed along the direction of the four main axes, so that the physical temperature $T$ of the system is given by $T=1/(\spac\Ntau)$. The simulations at finite temperature are performed on lattices with $\Ns/\Ntau \gtrsim 4$ to suppress finite-volume corrections, while $\Ntau=\Ns$ for the simulations at $T\simeq 0$. For a theory based on a generic $\Sp(N)$ gauge group, the fundamental degrees of freedom of the theory regularized on the lattice are a set of $4 \Ns^3 \Ntau$ square matrices in the defining representation of the $\Sp(N)$ group; these matrices have size $2N$ and are defined on the oriented bonds between nearest-neighboring lattice sites; in particular, we denote the matrix on the link from the site $x$ to the site $x+\spac\hat\mu$ as $U_\mu(x)$. The Euclidean action of the lattice theory is taken to be the Wilson action~\cite{Wilson:1974sk}
\begin{align}
\label{Wilson_action}
S = - \frac{\beta}{4} \sum_{x \in \Lambda} \sum_{0 \le \mu < \nu \le 3} \Tr U_{\mu\nu} (x),
\end{align}
where $\beta=4N/g_0^2$, with $g_0$ the bare coupling of the theory, and 
\begin{align}
\label{plaquette}
U_{\mu\nu} (x) = U_\mu (x) U_\nu \left(x+\spac\hat{\mu}\right) U_{\mu}^\dagger \left(x+\spac\hat{\nu}\right) U_{\nu}^\dagger (x)
\end{align}
denotes the ``plaquette'' constructed from the path-ordered product of link variables along the sides of the square of vertices $x$, $x+\spac\hat{\mu}$, $x+\spac\hat{\mu}+\spac\hat{\nu}$ and  $x+\spac\hat{\nu}$. For later convenience, we also introduce the normalized trace of the plaquette averaged over the lattice:
\begin{align}
P = \frac{1}{12N \Ntau \Ns^3} \sum_{x \in \Lambda} \sum_{0 \le \mu < \nu \le 3} \Tr U_{\mu\nu} (x).
\end{align}

Note that, if $U_\mu(x)$ is interpreted as a parallel transporter in the internal (``color'') space of the theory along the lattice bond from $x$ to $x+\spac\hat{\mu}$,
\begin{align}
U_\mu(x)=\exp\left( i \int_x^{x+\spac\hat{\mu}}\dd y A_\mu\left( y \right) \right) ,
\end{align}
then in the limit of vanishing lattice spacing at fixed physical volume the Wilson action defined in eq.~(\ref{Wilson_action}) tends to the Euclidean action of the continuum Yang-Mills theory, up to an irrelevant additive constant, and to corrections that are quadratically suppressed with the lattice spacing:
\begin{align}
\label{continuum_limit_of_Wilson_action}
\lim_{\spac \to 0} S = \frac{1}{2} \int \dd^4 x \Tr \left( F_{\mu\nu} (x)F^{\mu\nu} (x) \right)  + \mathrm{const}.
\end{align}
In fact, this remains true also at the level of the full quantum theory: in the $\spac \to 0$ limit, the lattice theory correctly reproduces the continuum theory.

The partition function of the lattice theory is given by
\begin{align}
\label{lattice_partition_function}
Z = \int \prod_{x \in \Lambda} \prod_{\mu = 0}^{3} \dd U_\mu(x) \exp \left[ -S (U) \right],
\end{align}
where $\dd U_\mu(x)$ is the Haar measure for the $U_\mu(x)$ group element. The expectation value of a generic quantity $\mathcal{O}$ is then defined as
\begin{align}
\label{lattice_expectation value}
\langle \mathcal{O} \rangle = \frac{1}{Z} \int \prod_{x \in \Lambda} \prod_{\mu = 0}^{3} \dd U_\mu(x) \mathcal{O} \exp \left[ -S (U) \right].
\end{align}
The multiple integrals in eq.~(\ref{lattice_expectation value}) are estimated numerically by the Monte~Carlo method; to this purpose, we adapted the {\CC} code that was originally used in refs.~\cite{Panero:2009tv, Mykkanen:2012ri} for the simulation of $\Sp(N)$ theories: it creates ensembles of matrix configurations in a Markov chain based on a combination of local heat-bath~\cite{Creutz:1980zw, Kennedy:1985nu} and over-relaxation updates~\cite{Adler:1981sn, Brown:1987rra}, which, following ref.~\cite{Cabibbo:1982zn}, are applied to a sequence of different blocks of each $\Sp(N)$ matrix, as discussed in the appendix~\ref{app:details_of_the_simulation_updates}. To produce a new configuration, these updates are applied to all link matrices of the lattice. The expectation values $\langle \mathcal{O} \rangle_T$ are then estimated from the average over the configurations that remain after discarding those produced during a thermalization transient at the beginning of the simulations, and the uncertainties on these estimates are evaluated with the jackknife method~\cite{bootstrap_jackknife_book}.

In order to express the results extracted from lattice simulations (which are dimensionless numbers, i.e., are measured in units of the appropriate power of the lattice spacing) as physical quantities, one has to set the scale, namely to establish the correspondence between the values of the bare coupling of the lattice theory $g_0$ (or, equivalently, the $\beta$ parameter) and the lattice spacing. In this work we set the scale non-perturbatively, by determining the critical $\beta$ at which the deconfinement transition occurs for different values of $\Ntau$, interpreting the corresponding values of $1/(\spac\Ntau)$ as the deconfinement temperature $\Tc$, and interpolating the numerical results obtained from this procedure to express the lattice spacing as a continuous function of $\beta$ in the range of values that were simulated.\footnote{Since the physics of purely gluonic non-Abelian gauge theories in four spacetime dimensions (and without a topological $\theta$ term) depends only on one parameter, setting the scale using a different dimensionful quantity would lead to the same $\spac(\beta)$ function, up to small corrections due to discretization effects.} In turn, the determination of the $\beta$ value corresponding to the critical deconfinement temperature is carried out by studying the fourth-order Binder cumulant~\cite{Binder:1981aaa}
\begin{align}
\label{Binder_cumulant}
B_4=1-\frac{\langle \mathcal{L}^{4}\rangle}{3\langle \mathcal{L}^{2}\rangle^2}
\end{align}
constructed from the normalized trace of the Polyakov loop winding around the Euclidean-time direction
\begin{align}
\label{Polyakov_loop}
\mathcal{L}\left(\vec{x}\right) = \frac{1}{4} \Tr \prod_{k=0}^{\Ntau-1} U_0\left(x+k\spac\hat{0}\right)
\end{align}
averaged over the spatial volume
\begin{align}
\label{spatially_averaged_Polyakov_loop}
\mathcal{L} = \frac{1}{\Ns^3} \sum_{\vec{x}} \mathcal{L}\left(\vec{x}\right).
\end{align}
The critical $\beta$ value in the thermodynamic limit is determined through a finite-size scaling analysis, from the crossing of the curves $B_4(\beta)$ obtained at different values of the linear spatial size of the system $L$. An alternative quantity that can be used to identify the $\beta$ value corresponding to deconfinement is the susceptibility associated with the spatially-averaged Polyakov loop:\footnote{Note that the definition of $\chi_{\mathcal{L}}$ is actually based on the absolute value of $\mathcal{L}$ (rather than $\mathcal{L}$ itself) because all simulations are carried out in finite volume and at finite lattice spacing, and thus involve a strictly finite number of degrees of freedom; as a consequence, any non-vanishing expectation value of $\mathcal{L}$ would simply be an artifact of the finiteness of the sample of configurations. This problem is solved by using $|\mathcal{L}|$, instead of $\mathcal{L}$, in the definition of $\chi_{\mathcal{L}}$.}
\begin{align}
\chi_{\mathcal{L}} = \Ns^3 \left( \langle \mathcal{L}^2 \rangle - \langle |\mathcal{L}| \rangle^2 \right).
\end{align}

Once the interpolating function $\spac(\beta)$ is known, the temperature of the system can be varied continuously, simply by tuning $\beta$; the temperature range explored in this work for the $\Sp(2)$ gauge theory is in the interval $0.7 \le T/\Tc \le 2.75$.

The numerical determination of the equation of state on the lattice is not a completely trivial problem; in particular, the free energy $F=-T\ln Z$ cannot be directly accessed in a Monte~Carlo simulation. Our approach to compute the equation of state in the present work is based on the ``integral method''~\cite{Engels:1990vr}: starting from the known thermodynamic relation between the pressure $p$ and the free energy per unit volume $f=F/V$,
\begin{align}
\label{pressure_free-energy_density}
p = -f = \frac{T}{V} \ln{Z},
\end{align}
and noting that the derivative of the logarithm of the partition function defined in eq.~(\ref{lattice_partition_function}) with respect to $\beta$ is proportional to the expectation value of the average plaquette trace $P$, the ratio of the pressure over the fourth power of the temperature can be obtained as
\begin{align}
\label{pressure}
\frac{p}{T^4} = 6 \Ntau^4 \int_{\beta_0}^{\beta} \dd \beta^\prime \left( \langle P \rangle_T - \langle P \rangle_0 \right),
\end{align}
where the subscripts indicate that the plaquette expectation values appearing on the right-hand side are respectively evaluated at the target temperature $T$ and at zero temperature, $\beta$ corresponds to the lattice spacing yielding the temperature $T$, whereas $\beta_0$ gives a temperature at which the pressure is approximately coincident with its value in the vacuum.\footnote{This means that eq.~\eqref{pressure} defines the pressure with respect to its value at $T=0$.} Note that the expectation value of the plaquette at $T=0$ is subtracted to remove the contribution of non-thermal, ultraviolet quantum fluctuations, which is divergent in the $\spac \to 0$ limit. The integral on the right-hand side of eq.~\eqref{pressure} is computed numerically, dividing the integration range into $n$ intervals of equal size $h$ and using the formula~\cite{Caselle:2007yc}:
\begin{align}
\label{integral_discretization}
\int_{x_0}^{x_n} f(x) \dd x = h \biggl\{ &\frac{17}{48}\left[ f(x_0)+f(x_n) \right] + \frac{59}{48}\left[ f(x_1)+f(x_{n-1}) \right]+ \frac{43}{48}\left[ f(x_2)+f(x_{n-2}) \right]\nonumber \\
& + \frac{49}{48}\left[ f(x_3)+f(x_{n-3}) \right] + \sum_{j=4}^{n-4} f(x_j) \biggl\} + O\left(n^{-4}\right),
\end{align}
where $x_j=x_0+j h$.

The trace of the energy-momentum tensor, denoted as $\Delta$, in units of $T^4$ is given by 
\begin{align}
\label{Delta}
\frac{\Delta}{T^4} = T \frac{\partial}{\partial T} \left( \frac{p}{T^4} \right) = -6\Ntau^4\spac \frac{\partial \beta}{\partial \spac} \left( \langle P \rangle_T -  \langle P \rangle_0 \right),
\end{align}
From $p$ and $\Delta$ one can then derive the energy density per unit volume in units of $T^4$
\begin{align}
\label{energy_density}
\frac{\epsilon}{T^4} = \frac{1}{VT^2} \left. {\frac{\partial \ln Z}{\partial T}}\right|_V = \frac{\Delta + 3p}{T^4}
\end{align}
and the entropy density per unit volume in units of $T^3$
\begin{align}
\label{entropy_density}
\frac{s}{T^3} = \frac{\epsilon}{T^4} + \frac{\ln Z}{VT^3} = \frac{\Delta + 4p}{T^4}.
\end{align}
Note that, from the pressure and the energy density, it is also possible to derive an estimate for the square of the speed of sound $\cs$ in terms of the stiffness:
\begin{align}
\label{squared_speed_of_sound}
\cs^2(T) = \frac{\partial p}{\partial \epsilon}.
\end{align}

Another interesting physical quantity to characterize the deconfining transition of the theory is the latent heat $\Lh$, which can be defined in terms of the discontinuity in energy density (or, equivalently, in $\Delta$) across the critical temperature,
\begin{align}
\label{Lh_definition}
\Lh = \lim_{V\to\infty} \frac{6}{\spac^4} \frac{\partial \beta}{\partial \ln \spac} \left( \langle P \rangle_{\Tc^-} - \langle P \rangle_{\Tc^+} \right).
\end{align}
In particular, the latent heat is non-zero for a first-order phase transition. 

It is also interesting to consider the specific heat at fixed volume per unit volume $\CV$, which is related to the size of the fluctuations in the density of Euclidean action and can be defined as
\begin{align}
\label{CV_definition}
\CV(T) = \frac{1}{V} \frac{\partial}{\partial T} \left( T^2 \frac{\partial}{\partial T} \ln Z \right) .
\end{align}
$\CV$ is maximal at the deconfinement temperature and in the thermodynamic limit is related to the latent heat through~\cite{Billoire:1992sz}
\begin{align}
\label{Lh_and_CV}
\Lh^2 = \lim_{V\to\infty} \frac{4\Tc^2}{V} \CV(\Tc).
\end{align}

Finally, before presenting our original results for the thermodynamics of the $\Sp(2)$ theory, we review the classification of different glueball states whose masses can be computed on the lattice; as will be explained in section~\ref{sec:results}, this is relevant for our discussion of the equation of state of the theory in the confining phase, which can be modelled in terms of a gas of massive, and approximately non-interacting, glueballs. On a hypercubic lattice, the symmetry under spatial rotations is broken down to rotations by integer multiples of $\pi/2$ around the three main spatial axes, so that glueball states are actually classified according to the irreducible representations of the orientation-preserving octahedral group (the symmetry group of the cube, with $24$ elements falling into five distinct conjugacy classes), which is isomorphic to the group of permutations of four elements $S_4$. The latter has five conjugacy classes (respectively containing the identity, the six permutations that can be written as a single cycle of length two, the three permutations that can be written as the product of two disjoint cycles of length two, the eight permutations expressed by a cycle of length three, and the six permutations expressed by a cycle of length four) and hence five irreducible representations; Burnside's theorem fixes their dimensions to be $1$, $1$, $2$, $3$, and $3$: they are the trivial representation, the sign representation, the two-dimensional irreducible representation, the standard representation, and the product of the standard and sign representations. When regarded as irreducible representations of the orientation-preserving octahedral group, they are usually denoted as $A_1$, $A_2$, $E$, $T_1$, and $T_2$, respectively. The restriction of irreducible representations of the $\SO(3)$ group to the orientation-preserving octahedral group, i.e., the subduced representations, can be decomposed into tensor sums of the various irreducible representations $A_1$, $A_2$, $E$, $T_1$, and $T_2$ (for a discussion, see, e.g., refs.~\cite{Berg:1982kp, Lucini:2010nv}). In particular, this implies that the $2J+1$ states of a spin-$J$ representation in the continuum get split among different irreducible representations of the orientation-preserving octahedral group; for example:
\begin{align}
& (J=0) \to A_1 , \label{J0_decomposition} \\
& (J=1) \to T_1 , \label{J1_decomposition} \\
& (J=2) \to E \oplus T_2, \label{J2_decomposition} \\
& (J=3) \to A_2 \oplus T_1 \oplus T_2, \label{J3_decomposition} \\
& (J=4) \to A_1 \oplus E \oplus T_1 \oplus T_2 \label{J4_decomposition} ;\end{align}
note that the dimensions on the left- and right-hand sides match.

\section{Results}
\label{sec:results}

Before presenting the thermodynamic quantities of $\Sp(2)$ Yang--Mills theory, we briefly review some results for this model that are known from previous works~\cite{Holland:2003kg, Bennett:2017kga, Bennett:2020qtj}, and that are relevant for our discussion.

First of all, at zero temperature the theory is linearly confining with string tension $\sigma$ (defined as the strength between probe sources in the fundamental representation, at asymptotically large distances) and has a finite mass gap. The spectrum of physical states consists of glueballs, which in the continuum can be classified according to their spin ($J$) and parity ($P$) quantum numbers.\footnote{Note that there is no charge-conjugation quantum number, due to the absence of complex representations.} As discussed in section~\ref{sec:formulation_of_the_model}, the classification of states of different spin on the lattice can be done according to the theory of subduced representations. In table~\ref{tab:glueball_masses} we reproduce the continuum-extrapolated results for the glueball masses, in channels corresponding to different representations of the orientation-preserving octahedral group and different parities, that were reported in ref.~\cite{Bennett:2020qtj} (including also previous results from ref.~\cite{Bennett:2017kga}): all values are expressed in units of the square root of the string tension. Note, in particular, that the lightest glueball is the scalar one, and that the observed degeneracy among the $T_2^{+}$ and $E^{+}$ states indicates that the second-lightest glueball is the $J^+=2^+$ continuum state.

\begin{table}[h]
\centering
\phantom{-------}
\begin{tabular}{|c|c|}
\hline
channel & $\frac{m}{\sqrt{\sigma}}$ \\
\hline
$A_1^{+}$      & $3.577(49)$ \\
$T_2^{+}$      & $5.050(88)$ \\
$E^{+}$        & $5.05(13)$  \\
$A_1^{-}$      & $5.69(16)$  \\
$A_1^{+\star}$ & $6.049(40)$ \\
$E^{-}$        & $6.65(12)$  \\
$T_2^{-}$      & $6.879(88)$ \\
$A_1^{-\star}$ & $7.809(79)$ \\
$A_2^{+}$      & $7.91(16)$  \\
$T_1^{+}$      & $8.67(28)$  \\
$T_1^{-}$      & $9.24(33)$  \\
$A_2^{-}$      & $9.30(38)$  \\
\hline            
\end{tabular}
\phantom{-------}
\caption{Glueball masses in channels corresponding to different irreducible representations of the orientation-preserving octahedral group and different parities (denoted by the superscript sign), as reported in ref.~\cite{Bennett:2020qtj} (including also previous calculations presented in ref.~\cite{Bennett:2017kga}), and sorted in non-decreasing order. All values are extrapolated to the continuum limit, and are expressed in units of the square root of the string tension; stars denote excited states in a given channel.}
\label{tab:glueball_masses}
\end{table}

As a side remark, it is interesting to note that in ref.~\cite{Hong:2017suj}, it was conjectured that the ratio defined as
\begin{align}
\label{conjectured_m2_sigma_ratio}
\eta(D)=\frac{M^2}{\sigma}\frac{\mathcal{C}_2(\mbox{fund})}{\mathcal{C}_2(\mbox{adj})}
\end{align}
(where $M$ is the mass of the lightest physical state in the scalar channel in continuum, while $\mathcal{C}_2(\mbox{fund})$ and $\mathcal{C}_2(\mbox{adj})$ respectively denote the eigenvalue of the quadratic Casimir operator for the fundamental and for the adjoint representation of the gauge group, defined in the appendix~\ref{app:generalities_about_the_compact_symplectic_group}), depends only on the number of spacetime dimensions $D$, and is universal for all purely gluonic Yang--Mills theories, taking the value $\eta(4)=5.41(12)$ in the case of four spacetime dimensions. Also, note that for the theory we are considering the $\mathcal{C}_2(\mbox{fund})/\mathcal{C}_2(\mbox{adj})$ ratio equals $5/12$, which is close to the value it takes for the $\SU(3)$ Yang--Mills theory, i.e., $4/9$.

As for the behavior of the $\Sp(2)$ theory at finite temperature, the study reported in ref.~\cite{Holland:2003kg} already revealed the existence of a first-order deconfining phase transition at a finite critical temperature which, when expressed in units of the square root of the string tension, is $\Tc/\sqrt{\sigma}=0.6875(18)$.

We now present our original results, starting from those to set the scale of the lattice theory---namely, to establish a relation between the parameter $\beta$ and the corresponding value of the lattice spacing $\spac$. While at sufficiently large $\beta$ one could do this by means of perturbation theory (leaving only an integration constant to be determined), the range of parameters of interest for our lattice simulations requires a non-perturbative procedure to set the scale. While there are different methods to set the scale non-perturbatively in lattice calculations~\cite{Sommer:2014mea}, they can be broadly divided in methods based on unphysical reference scales and methods based on physical ones. The latter consist in identifying the value of a dimensionful quantity, that can be accurately estimated in lattice simulations, with its value in the real world: since every dimensionful quantity extracted from lattice simulations is always expressed in units of the appropriate power of the lattice spacing, this allows one to obtain $\spac$, for the value of $\beta$ at which the calculation was performed. The choice of different physical quantities leads to small differences in the estimate of $\spac$, which tend to vanish in the continuum limit, and which can be included in the budget of systematic uncertainties affecting the calculation. For a theory in which no dimensionful observables are experimentally known yet, the method allows, nevertheless, to determine the ratios of all other quantities with respect to the one chosen as the reference scale. In this work we chose to set the scale with the deconfinement temperature; even though other methods (such as the one based on the Wilson flow~\cite{Luscher:2010iy}) allow one to obtain a numerically even more precise scale setting, this has the advantage of relying on a quantity of direct physical interpretation, making the predictive power of the model clearer.

\begin{figure}
    \centering
    \includegraphics[width=0.7\textwidth]{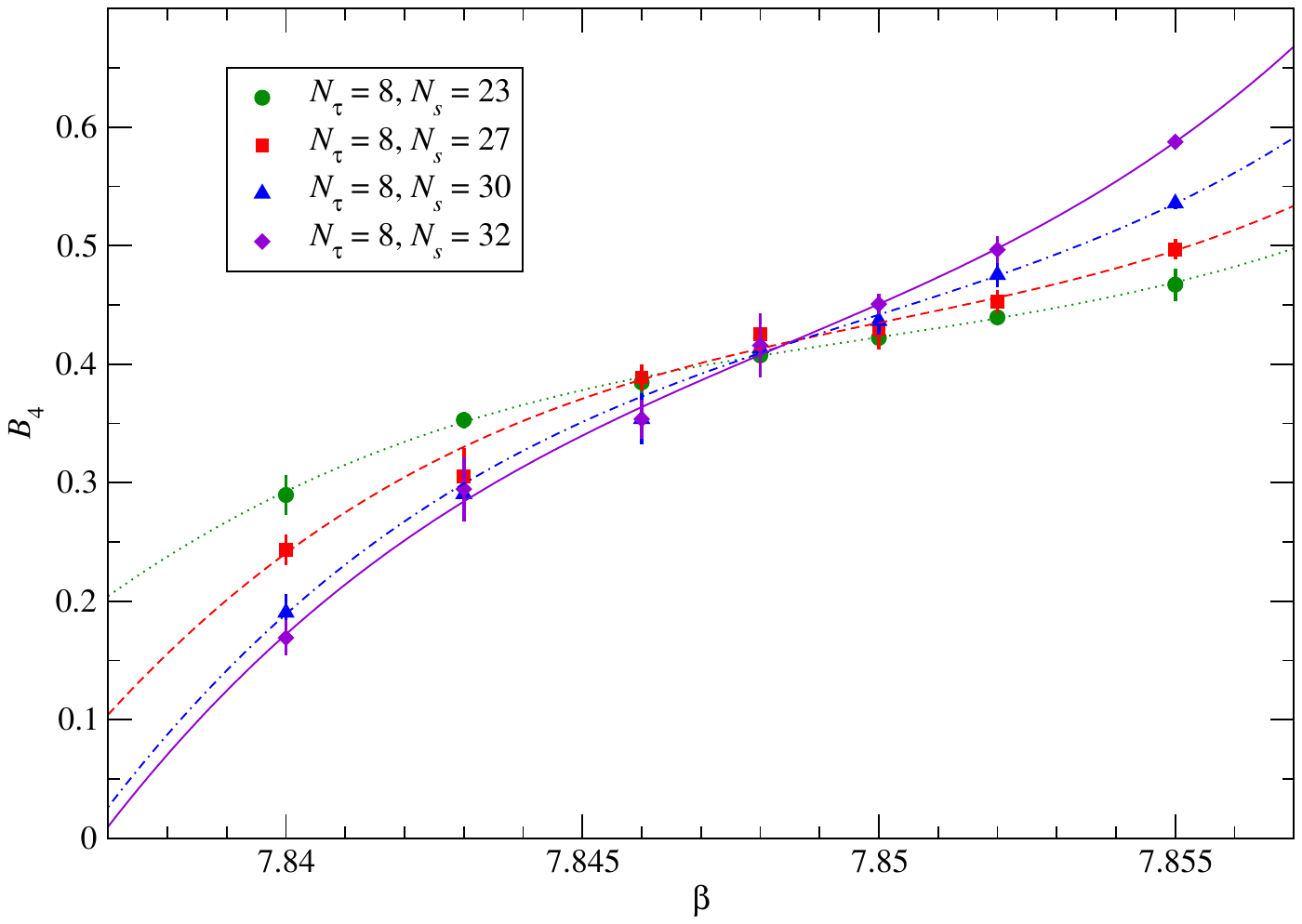}
    \caption{Binder cumulant associated with the Polyakov loop, defined in eq.~\eqref{Binder_cumulant}, extracted from simulations on lattices with $\Ntau=8$ sites in the Euclidean-time direction and for different spatial volumes $L^3$ (displayed by symbols of different shapes and colors), and shown as a function of $\beta=8/g_0^2$. The continuous curves are splines to guide the eye.}
    \label{fig:crossing_of_Binder_cumulants_Nt8}
\end{figure}

Our study of the deconfinement temperature of the $\Sp(2)$ theory is based on a finite-size scaling of Binder cumulants. The plot in figure~\ref{fig:crossing_of_Binder_cumulants_Nt8} shows an example of our results for the behavior of the Binder cumulant associated with the Polyakov loop: the data displayed in the figure were obtained from simulations on a lattice with $\Ntau=8$ sites in the Euclidean-time direction. The data are plotted against $\beta=8/g_0^2$ (where $g_0$ is the bare coupling of the theory), and for different values of the extent of the lattice (in units of the lattice spacing) along the spatial directions, which are denoted by symbols of different shapes and colors. The critical value of $\beta$ corresponding to the deconfinement transition can be estimated from the crossing of the curves interpolating the data at different spatial volumes.

The estimates of the critical values of $\beta$ corresponding to the deconfinement temperature for each $\Ntau$ by means of the Binder cumulant are consistent with those determined from the Polyakov-loop susceptibilities; the results are summarized in table~\ref{tab:scale-setting_values}, which also includes the values originally determined in ref.~\cite{Holland:2003kg} for $\Ntau=3$ and $\Ntau=4$.
\begin{table}[ht]
\centering
\begin{tabular}{|c|cccccccc|}
\hline
$\Ntau$ & $3$ & $4$ & $5$ & $6$ & $7$ & $8$ & $9$ & $10$ \\
\hline
$\betac$ & $7.1228(4)$ & $7.339(1)$ & $7.496(4)$ & $7.624(5)$ & $7.738(2)$ & $7.848(3)$ & $7.932(2)$ & $8.025(3)$ \\
\hline
\end{tabular}
\caption{$\beta$ values corresponding to the deconfinement transition (in the thermodynamic limit), for different values of $\Ntau$. The table also includes the results reported in ref.~\cite{Holland:2003kg} for $\Ntau=3$ and $\Ntau=4$.}
\label{tab:scale-setting_values}
\end{table}

\begin{figure}
    \centering
    \includegraphics[width=0.7\textwidth]{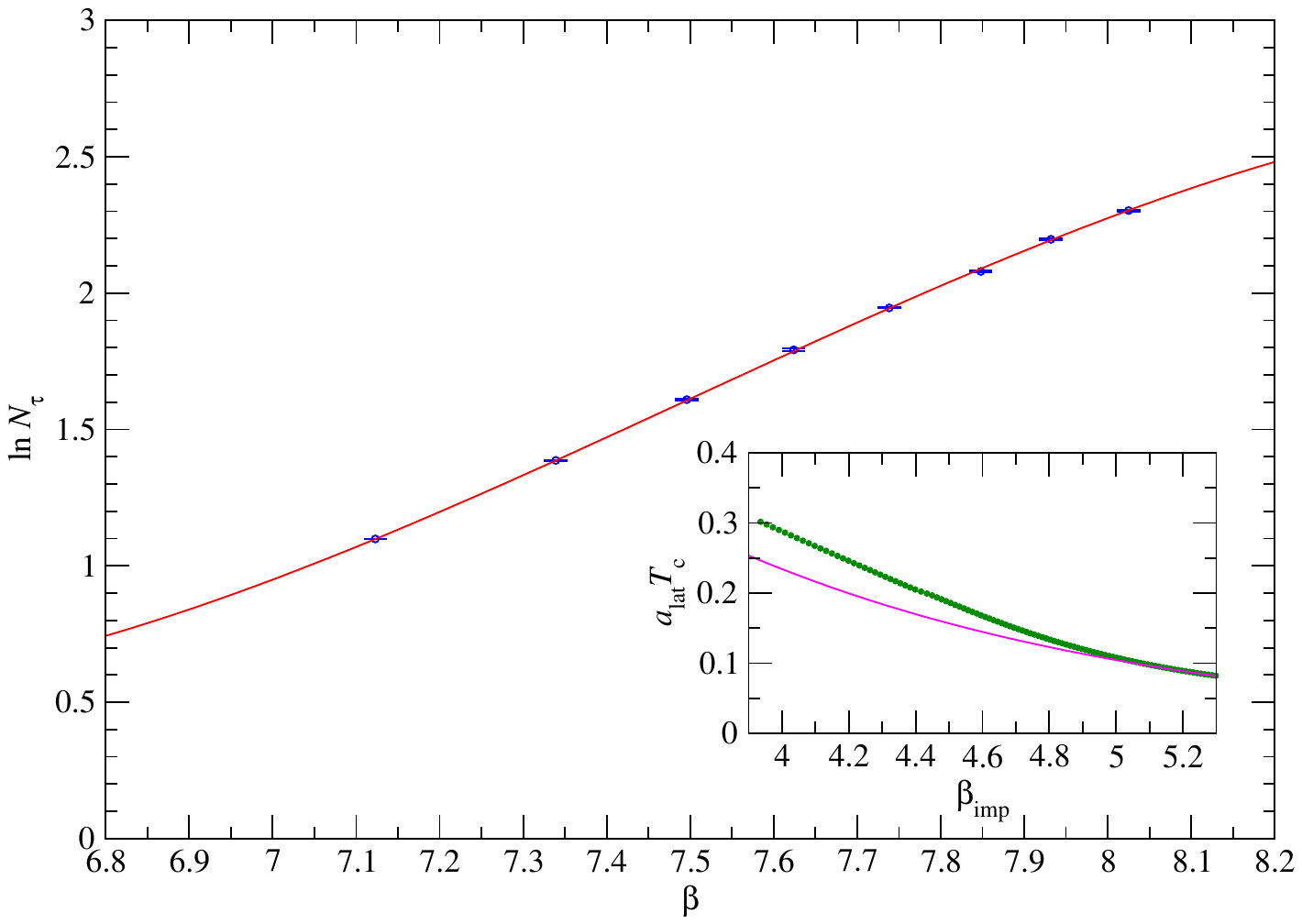}
    \caption{Main plot: fit of the data reported in table~\ref{tab:scale-setting_values} to eq.~\eqref{scale-setting_fit}. Inset plot: comparison of our non-perturbative scale setting (green points) with the mean-field improved perturbative scale setting (magenta line) discussed in the text.}
    \label{fig:scale_setting}
\end{figure}

The data in table~\ref{tab:scale-setting_values} can be fitted to
\begin{align}
\label{scale-setting_fit}
\ln \Ntau (\beta) = \sum_{i=0}^3 k_i (\beta - \beta_\star)^i, \qquad \mbox{with $\beta_\star = 7.7$},
\end{align}
obtaining
\begin{align}
\label{scale-setting_fit_results}
k_0 = 1.892(3), \quad k_1 = 1.371(8), \quad k_2 = -0.21(4), \quad k_3 = -0.35(7).
\end{align}
\revision{This fit, with four degrees of freedom, leads to $\redchisq\simeq 2.2$, corresponding to $p \simeq 0.066$.} The resulting curve, which is our non-perturbative scale setting for the theory, is shown in fig.~\ref{fig:scale_setting}, alongside the data. We \revision{also} studied how the scale setting varies, if the fitting form on the right-hand side of eq.~\eqref{scale-setting_fit} is replaced by a polynomial of different order, and found that the results remain consistent, as the corresponding systematic deviations are less than the statistical uncertainties. \revision{Note that we chose to express the right-hand side of eq.~\eqref{scale-setting_fit} in terms of an expansion around a $\beta_\star$ value close to the center of the $\beta$ interval covered by our lattice simulations: this choice is expected to minimize the numerical artifacts affecting polynomial interpolations.\footnote{\color{blue} A simple and general argument goes as follows. Consider the approximation of a continuous and differentiable function $f$ on the real finite $[x_{\mbox{\tiny{min}}},x_{\mbox{\tiny{max}}}]$ interval through a sequence of polynomials $p_n(x-x_\star)$ of degree $n$. According to Weierstrass' approximation theorem, the maximum error in the approximation, $\sup_{x \in [x_{\mbox{\tiny{min}}}, x_{\mbox{\tiny{max}}}]} |f(x)-p_n(x-x_\star)|$, can be made arbitrarily small (by increasing the polynomial degree $n$), as long as the $[x_{\mbox{\tiny{min}}}, x_{\mbox{\tiny{max}}}]$ interval lies within the radius of convergence of the expansion about $x_\star$. If that is not the case, then the maximum error in the approximation can remain finite---and, in fact, may even become arbitrarily large---regardless of the degree of the polynomial~\cite{Runge:1901uef}. While approximating an unknown function $f$ (which may have poles in the complex plane at unknown locations), choosing $x_\star$ to be the central point of the $[x_{\mbox{\tiny{min}}},x_{\mbox{\tiny{max}}}]$ interval minimizes the area of the smallest circle in the complex plane of $x$ in which the real $[x_{\mbox{\tiny{min}}},x_{\mbox{\tiny{max}}}]$ interval is entirely contained, therefore maximizing the probability that such circle does not contain any complex poles of $f$ and thus it minimizes the chances that the approximation may have potentially unbounded errors  in the considered interval.}}

While the results in table~\ref{tab:scale-setting_values} allow one to set the value of the lattice spacing for a set of eight values of $\beta$ from the relation $\spac(\betac)=1/(\Ntau \Tc)$, the interpolation in eq.~\eqref{scale-setting_fit} provides a continuous parameterization for the lattice spacing as a function of $\beta$. To extend the correspondence between values of the coupling and values of $\spac$ to finer lattice spacings (which, in particular, is necessary in order to study high temperatures, as well as the continuum limit of the lattice theory), we supplement this parameterization with a two-loop perturbative expression in an improved lattice scheme~\cite{Allton:2008ty}, whereby the lattice spacing can be expressed in terms of a mean-field-improved square coupling $\gimp^2=g_0^2/\langle P \rangle$~\cite{Parisi:1980pe} as follows
\begin{align}
\label{perturbative_scale-setting}
\spac = \frac{1}{\Lambdaimp} \exp\left( -\frac{1}{2 \beta_0 \gimp^2} \right) \cdot \left( \frac{\beta_1}{\beta_0^2} + \frac{1}{\beta_0 \gimp^2} \right)^{\frac{\beta_1}{2 \beta_0^2}},
\end{align} 
where $\Lambdaimp$ is an integration constant with the dimensions of an energy, while $\beta_0$ and $\beta_1$ are the scheme-independent one-~\cite{tHooft:1971qjg, tHooft:1971akt, Gross:1973id, Politzer:1973fx} and two-loop~\cite{Caswell:1974gg, Jones:1974mm, Egorian:1978zx} coefficients of the function encoding the dependence of the coupling on the momentum scale
\begin{align}
\label{beta_function}
\mu \frac{d g}{d \mu} = - \beta_0 g^3 - \beta_1 g^5 + O(g^7);
\end{align}
for a purely gluonic non-Abelian gauge theory, they read
\begin{align}
\label{beta0_and_beta1}
\beta_0 = \frac{1}{(4 \pi)^2} \frac{11}{3} \mathcal{C}_2(\mbox{adj}), \qquad \beta_1 = \frac{1}{(4 \pi)^4} \frac{34}{3} \mathcal{C}_2^2(\mbox{adj}),
\end{align}
where, as before, $\mathcal{C}_2(\mbox{adj})$ denotes the eigenvalue of the quadratic Casimir operator for the adjoint representation; for a generic $\Sp(N)$ group, its value is $\mathcal{C}_2(\mbox{adj})=N+1$~\cite{Yamatsu:2015npn}, so that, for the $\Sp(2)$ gauge theory, one has $\beta_0=\frac{11}{(4 \pi)^2}$ and $\beta_1 = \frac{102}{(4 \pi)^4}$. The integration constant $\Lambdaimp$ is fixed by imposing a continuous matching of the perturbative expression~\eqref{perturbative_scale-setting} with the non-perturbative one, derived from eq.~\eqref{scale-setting_fit}.

To compare the mean-field-improved perturbative scale setting with the non-perturbative one, it is convenient to introduce the parameter $\betaimp=4N/\gimp^2=\beta \langle P \rangle$. Note that the $\langle P \rangle$ expectation value is a  monotonically increasing function of $\beta$, ranging from $0$ (at $\beta=0$, i.e., in the strong-coupling limit) to $1$ (for $\beta \to \infty$, namely, in the weak-coupling limit). As a consequence, one always has $\betaimp \le \beta$; in particular, for the $\beta$ values that we considered in our lattice simulations, $\betaimp$ is in the ballpark of $\beta/2$. Therefore, it is interesting to compare the non-perturbative scale setting with the mean-field-improved one for values of $\betaimp$ of the order of four. An example of such comparison is displayed in the inset plot of fig.~\ref{fig:scale_setting}, where the lattice spacing (in units of the inverse of the critical temperature) is plotted against $\betaimp$: as expected, these two parameterizations of the lattice spacing as a function of the coupling tend to be in better and better agreement when $\betaimp$ is increased, while they deviate from each other at larger values of the coupling. \revision{It is interesting to discuss the impact of the systematic uncertainty induced by combining a scale setting obtained from the interpolation of simulation results (at intermediate $\beta$ values), and the fixed-order perturbative expression in eq.~\eqref{perturbative_scale-setting} (in the $\beta>8.025$ region). In particular, one may suspect that the matching between a genuinely non-perturbative (and thus, in a sense, ``all-order'') scale setting and a second-order perturbative formula, imposing that they take the same value at the point where they are matched, could be unjustified, as it corresponds to neglecting the impact of higher-order perturbative contributions. It should be noted, however, that eq.~\eqref{perturbative_scale-setting} is an expansion in terms of the \emph{improved} lattice coupling $\gimp^2=g_0^2/\langle P \rangle$, which can be interpreted as the ``effective coupling'' experienced by a background field in the lattice theory (in a mean-field approximation)~\cite{Parisi:1980pe}: since the definition of $\gimp^2$ involves the plaquette expectation value, which we determined non-perturbatively from lattice simulations, the expression in eq.~\eqref{perturbative_scale-setting} is, in fact, expected to encode some non-perturbative information and to satisfy the key properties expected for a ``good'' coupling for a lattice gauge theory---including, in particular, satisfactory convergence properties in perturbation theory~\cite{Lepage:1996jw}. For $\SU(N)$ gauge theories, it was shown in ref.~\cite{Allton:2008ty} that a fixed-order scale setting in terms of $\gimp$ is in remarkably good agreement with a fully non-perturbative scale setting defined in terms of the Schr\"odinger functional~\cite{Symanzik:1981wd, Luscher:1992an} over a surprisingly wide range of couplings, so that the matching between the two different parametrizations for the lattice scale as a function of $\beta$ is actually smooth. In principle, this should be tested for the $\Sp(2)$ gauge theory, too, by means of a dedicated lattice calculation. We expect that the computational costs for such study could be comparable with those of analogous calculations that were carried out for the $\SU(3)$ theory (for a recent, high-precision study, see ref.~\cite{Francesconi:2020fgi}) and for the $\SU(4)$ theory~\cite{Lucini:2008vi}: the dimension of the Lie algebra of the $\Sp(2)$ theory ($10$) is intermediate between those of these two $\SU(N)$ theories ($8$ and $15$), the size of the gauge-group matrices for the $\Sp(2)$ lattice gauge theory ($4$) is the same as for the $\SU(4)$ theory, while the rank of the Lie algebra of the symplectic theory ($2$) is the same as that of $\SU(3)$ (even though the ``fundamental domain'' involved in the formulation of the Schr\"odinger-functional is, of course, different, since it depends on the weights of the Lie algebra). For our present purposes, however, we limit ourselves to a very crude, order-of-magnitude estimate of the systematic uncertainty on $\Lambdaimp$ from the matching procedure. In particular, we note that, at the matching point (corresponding to $\beta=8.025$, or to $\betaimp \simeq 5.075$) the difference in slope of the tangent to our fully non-perturbative scale setting and to the improved two-loop expression in eq.~\eqref{perturbative_scale-setting} is tiny, $0.017$, suggesting that the effect of residual higher-order corrections (which are expected to contribute to the slope of the curve) is small; at $\beta$ values larger than $8.025$, such corrections are expected to be even smaller (as they originate from higher powers of the coupling). The results for the equation of state that we present in this section are limited to temperatures relatively close to $\Tc$: in particular, we report continuum results for the pressure, energy density, and entropy density at temperatures up to $T=2\Tc$, extrapolating from simulation results up to $\Ntau=7$. This means that the finest lattice spacing that we considered in our calculations of the equation of state is $1/(2\cdot 7 \Tc)$, that is about $70\%$ of the finest lattice spacing that we determined non-perturbatively (which is the one corresponding to the critical temperature for $\Ntau=10$, namely $1/(10 \Tc)$), and corresponds to $\betaimp \simeq 5.48$, i.e., is only slightly out of the range of our non-perturbative scale setting. Assuming (as a conservative upper bound) that the deviation between the improved perturbative scale setting and a fully non-perturbative scale setting can be described as a linear function with slope $0.017$ (and considering that at $\beta=8.025$ we have $\spac \Tc = 1/10$) we arrive at the crude estimate that the largest relative uncertainty induced by our scale-setting procedure is of the order of $7\%$.

As a final comment on the scale-setting procedure, one can also note that the $\Sp(2)$ lattice gauge theory at zero temperature has no bulk phase transition~\cite{Holland:2003kg}, and this is generally expected to make the scale-setting procedure and the continuum limit both conceptually and practically more straightforward than, for example, in $\SU(N)$ gauge theories with more than three color charges~\cite{Lucini:2002ku}. In any case, the range of couplings that we investigated in the present work is far from the strong-coupling regime (corresponding to $\beta \to 0$), in which lattice artifacts are expected to be significant.}

\begin{figure}
    \centering
    \includegraphics[width=0.7\textwidth]{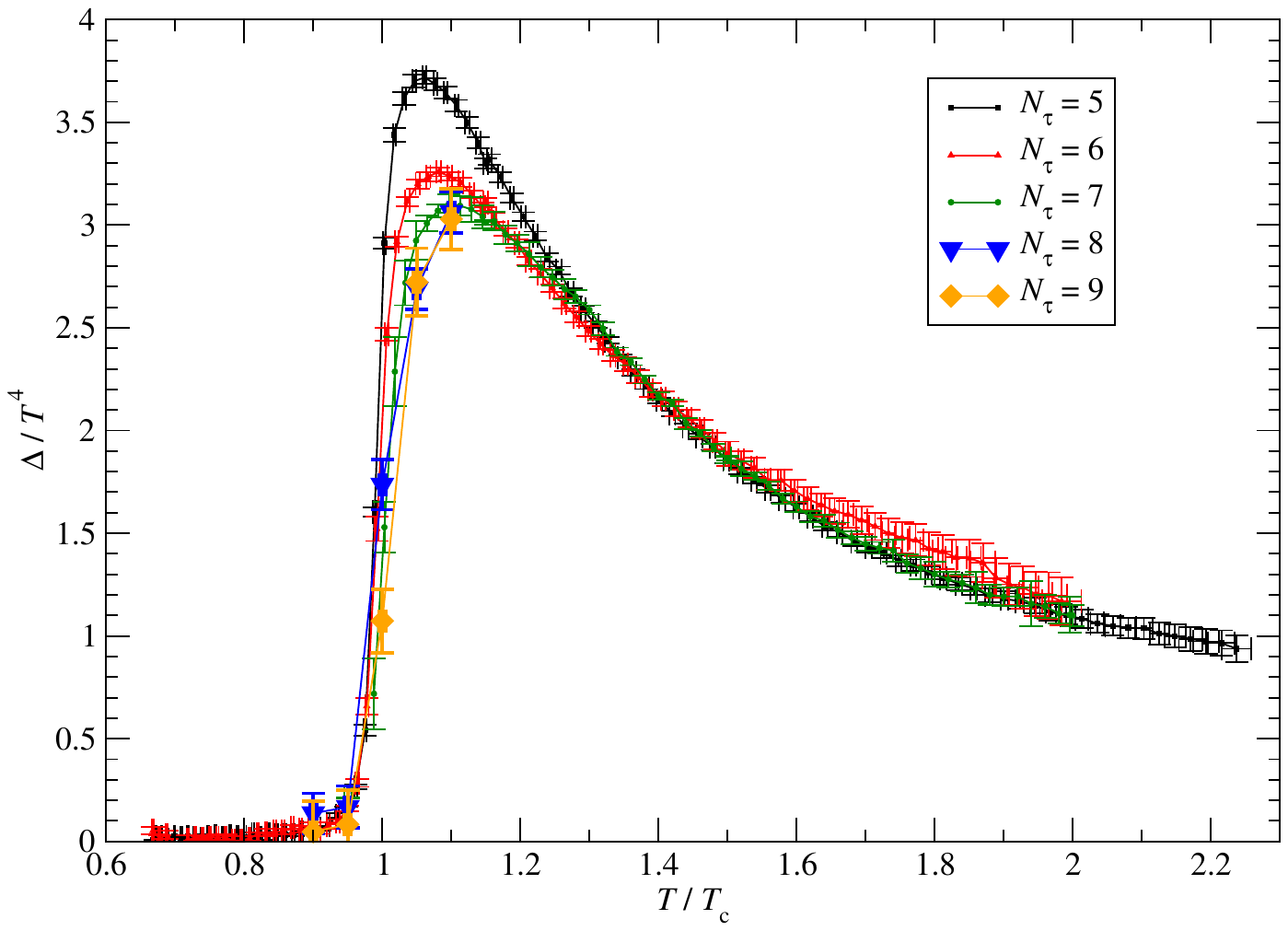}
    \caption{Trace of the energy-momentum tensor $\Delta$ in the $\Sp(2)$ theory, expressed in units of the fourth power of the temperature, as obtained from our simulations at finite values of the lattice spacing, equal to one fifth (black squares), one sixth (red triangles \revision{pointing up}), and one seventh (green circles) of the lattice extent in the Euclidean-time direction. \revision{In addition, the figure also includes the results obtained from simulations on lattices with values of the lattice spacing equal to one eighth (blue triangles pointing down) and one ninth (orange diamonds) of the lattice extent in the Euclidean-time direction, for a more limited set of temperatures ($T/\Tc=0.9$, $0.95$, $1$, $1.05$, and $1.1$).}}
    \label{fig:Delta_at_finite_Ntau}
\end{figure}

Having set the scale, it is then possible to study the thermodynamic quantities in a continuous range of temperatures across the deconfinement phase transition, and to carry out their extrapolation to the continuum limit. As an example, figure~\ref{fig:Delta_at_finite_Ntau} shows our results for the dimensionless ratio of the trace of the energy-momentum tensor $\Delta$ to the fourth power of the temperature, as obtained from eq.~\eqref{Delta}: the plot shows the results of our \revision{main set of} simulations at three different values of the lattice spacing, corresponding to simulations on lattices with $\Ntau=5$, $6$, and $7$ sites along the Euclidean-time direction. \revision{In addition, in order to check that the results at these values of the lattice spacing are already reasonably close to the continuum limit, the plot also includes results obtained from simulations at five temperatures close to the deconfinement transition ($T/\Tc=0.9$, $0.95$, $1$, $1.05$, and $1.1$) from simulations on lattices with $\Ntau=8$ and $9$ sites along the Euclidean-time direction, which, for the sake of clarity, are denoted by larger symbols in the figure. The results obtained from these finer lattices appear to be consistent with those obtained from the three largest ensembles (namely, those with $\Ntau=5$, $6$, and $7$). It is worth noting that the computational costs of the simulations grow quite quickly, when $\Ntau$ is increased. The reason for this can be appreciated by looking at eq.~\eqref{Delta}: the trace of the energy-momentum tensor is proportional to the difference of the plaquette expectation values at finite $T$ and at $T=0$ and to the fourth power of $\Ntau$. This means that the $(\langle P \rangle_T - \langle P \rangle_0)$ difference scales (approximately) as $\Ntau^{-4}$, and if the signal-to-noise ratio in the determination of $\Delta/T^4$ has to be held fixed, then the uncertainties on $\langle P \rangle_T$ and $\langle P \rangle_0$ should also be scaled as $\Ntau^{-4}$. In turn, this would require a number of independent lattice configurations growing as $\Ntau^8$. These are not the only technical challenges that should be faced when $\Ntau$ grows. It is known that lattice simulations on very fine lattices incur severe ergodicity problems, due to the phenomenon of critical slowing down~\cite{Wolff:1989wq}; these include, in particular, the inability to efficiently probe sectors of different topological charge~\cite{Schaefer:2010hu}. Even though these technical reasons limited the precision that we could obtain in our simulations of the $\Sp(2)$ theory on the finest lattices, we note that the general consistency of the lattice results for the $\Ntau$ values that we considered in this work has also been observed elsewhere for $\SU(N)$ gauge theories~\cite{Datta:2010sq, Borsanyi:2012ve}. This gives confidence on the validity of our extrapolations to the continuum limit, within the uncertainties reported here (which are larger than those of previous $\SU(3)$ calculations).} Note that the trace of the energy-momentum tensor is, in some sense, the most basic equilibrium-thermodynamics quantity that is directly extracted from differences of plaquettes computed in our lattice simulations, while the pressure is obtained by numerical integration of the former, according to eq.~\eqref{pressure}, and the densities of energy and entropy are obtained as linear combinations of $\Delta$ and $p$, through eq.~\eqref{energy_density} and eq.~\eqref{entropy_density}, respectively. 

The data shown in figure~\ref{fig:Delta_at_finite_Ntau} reveal a behavior consistent with what is observed in other non-Abelian Yang--Mills theories~\cite{Boyd:1996bx, Panero:2009tv, Borsanyi:2012ve, Bruno:2014rxa, Giudice:2017dor}, including, in particular, rather mild discretization effects, which are most clearly visible close to the peak of $\Delta/T^4$.

It is then possible to derive the results for $p/T^4$, for $\epsilon/T^4$, and for $s/T^3$, and to proceed to their extrapolation to the continuum limit $\spac \to 0$. We carry out the continuum extrapolation of the equilibrium-thermodynamics quantities by first interpolating the values of $\Delta/T^4$ at each $\Ntau$ using splines (a choice that is expected to yield optimal bounds on the interpolation uncertainties~\cite{Hall:1976oeb}), and then extrapolating the values thus obtained at a given temperature as a function of $1/\Ntau^{2}$. Repeating this procedure at a sufficiently large number of temperatures, one can get an arbitrarily precise approximation of a continuous curve, from which the pressure is then obtained by numerical integration, according to eq.~\eqref{Delta}; finally, the energy density and the entropy density are computed from the extrapolated values of $\Delta/T^4$ and $p/T^4$ using eq.~\eqref{energy_density} and eq.~\eqref{entropy_density}.

\revision{While the precision of our continuum-extrapolated results is worse than, for example, the precision that has been achieved in the past few years for the equation of state in the pure-glue sector of QCD, it may be instructive to consider the quality of the extrapolation at some reference temperatures. As an example, in figure~\ref{fig:extrapolation_at_T_over_Tc_1.5} we show the extrapolation to the continuum limit of the large-volume extrapolation of our results for $\Delta/T^4$ at $T/\Tc=1.5$ (denoted by the black circles), at the three values of the lattice spacing for which we could obtain sufficiently precise simulation results, corresponding to $\Ntau=5$, $\Ntau=6$, and $\Ntau=7$. As the leading finite-cutoff corrections in the lattice discretization with Wilson's action are proportional to $\spac^2$, the data are fitted to a constant plus a term linear in $1/\Ntau^2$. The fit (shown by the dashed black line) yields $1.867(12)$ for the intercept (denoted by the red square), while the result for the slope, $-0.2(4)$, is compatible with zero. The reduced $\chi^2$ of the fit is $0.21$, with one degree of freedom, and the non-diagonal elements of the correlation matrix of the fit parameters are $-0.947$.
\begin{figure}
    \centering
    \includegraphics[width=0.7\textwidth]{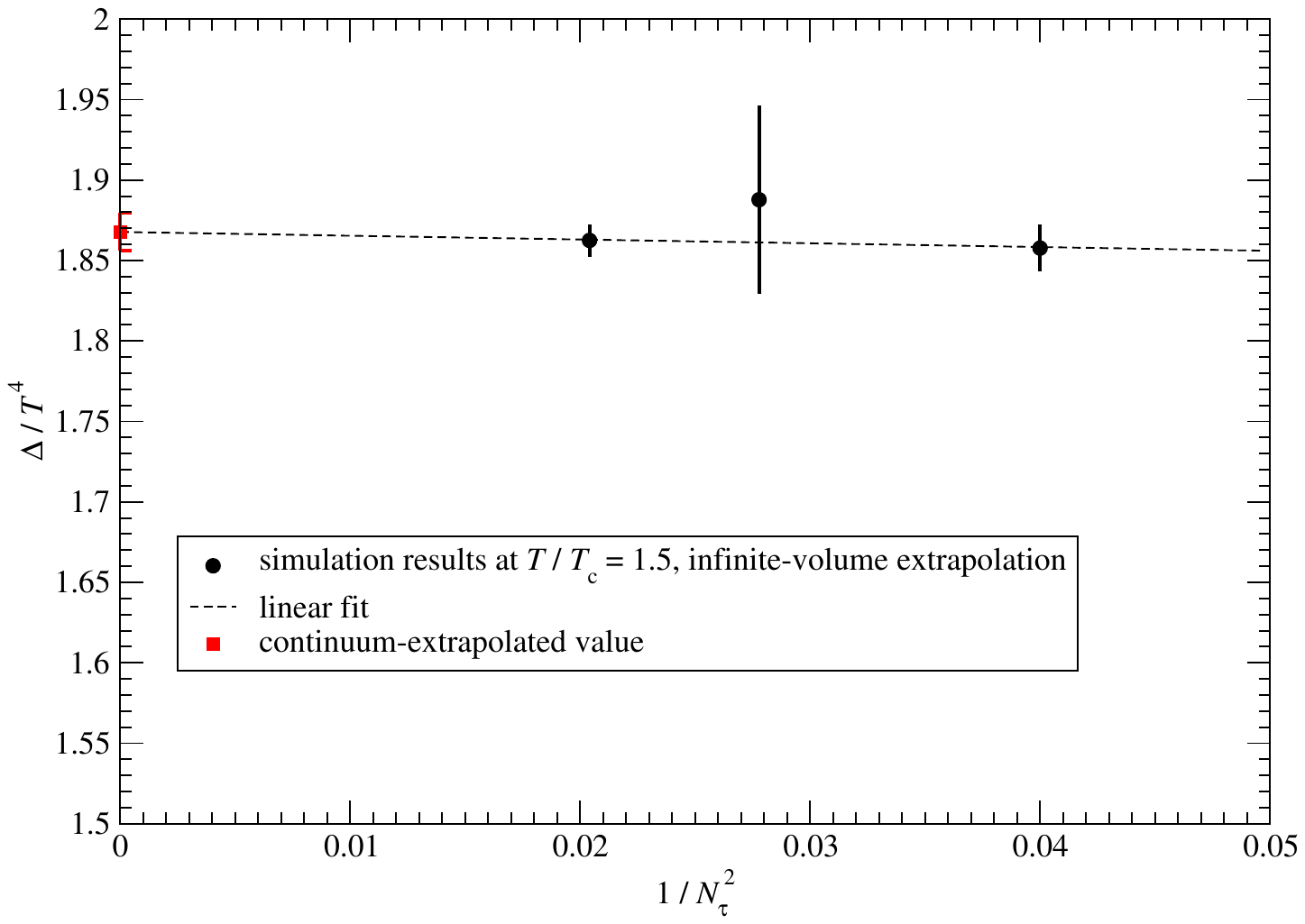}
    \caption{\revision{Continuum extrapolation for $\Delta/T^4$ at $T/\Tc=1.5$: the data obtained from the infinite-volume extrapolation of the results of simulations at $T/\Tc=1.5$ on lattices with $\Ntau=5$, $\Ntau=6$, and $\Ntau=7$ points along the Euclidean-time direction (black circles) are fitted as a function of $1/\Ntau^2$, including a constant and linear term. The fit result is denoted by the dashed black line. The red square denotes the extrapolation to the continuum limit.}}
    \label{fig:extrapolation_at_T_over_Tc_1.5}
\end{figure}

Similar fit results are obtained at the other temperatures that we studied in this work. Considering, for example, some selected temperatures in different parts of the $T/\Tc$ range that we considered, namely $T/\Tc=0.95$ (in the confining phase), $T/\Tc=1.15$ (in the region of the peak of $\Delta/T^4$), and $T/\Tc=2$ (the highest temperature at which we carried out the continuum extrapolation), we respectively obtain 
$\redchisq=0.19$, $\redchisq=0.18$, and $\redchisq=3.81$. The slope of the interpolating straight line as a function of $1/\Ntau^2$ is compatible with zero, with the exception of the $T/\Tc=1.15$ case (for which it is found to be $14.5(6)$).
}

\begin{figure}
    \centering
    \includegraphics[width=0.7\textwidth]{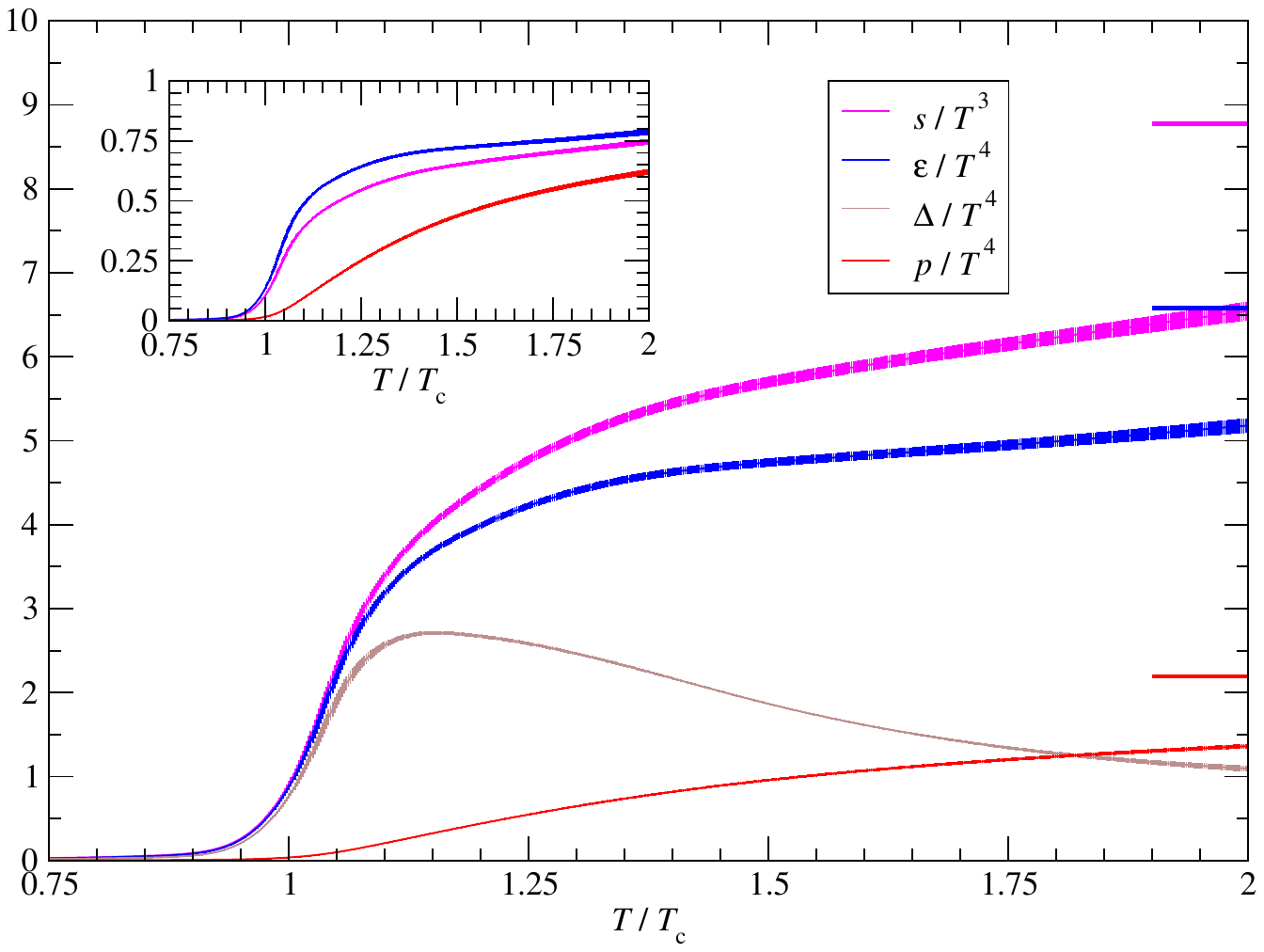}
    \caption{Results for thermodynamic-equilibrium quantities extrapolated to the continuum limit, and plotted as a function of the $T/\Tc$ ratio: the \revision{main plot} shows the entropy density in units of $T^3$ (magenta symbols), the energy density (blue symbols), the trace of the energy-momentum tensor (brown symbols), and the pressure (red symbols) in units of $T^4$. The horizontal bars on the right-hand side of the figure show, from top to bottom, the values corresponding to the Stefan--Boltzmann limit for $s/T^3$, for $\epsilon/T^4$, and for $p/T^4$. \revision{In the inset plot, the same quantities (except for $\Delta/T^4$, whose value in the free limit is zero) are normalized by their values in the Stefan--Boltzmann limit, with the same colors as in the main plot.}}
    \label{fig:extrapolated_equilibrium_quantities}
\end{figure}

The results of the extrapolation to the continuum limit are shown in figure~\ref{fig:extrapolated_equilibrium_quantities}, \revision{while in tables~\ref{tab:Delta_error_budget} and~\ref{tab:p_error_budget} we report a breakdown of the error budget for $\Delta/T^4$ and for $p/T^4$, respectively, at the selected temperatures we mentioned above (while we omit the corresponding discussion for $\epsilon/T^4$ and for $s/T^3$, as the latter quantities were obtained from linear combinations of the former), which deserves some comments. First of all, the various contributions to the total uncertainties are somewhat entangled with each other: for example, given the results for $\Delta/T^4$ obtained at a given $\Ntau$ from three different lattice volumes, with their own (typically: similar) uncertainties, it is somewhat ambiguous to state what is the ``statistical contribution'' to the final error budget; for this reasons, the quoted figures should be considered as order-of-magnitude estimates of the various contributions to the total error budget, whose main purpose is to highlight which are the bottlenecks affecting the precision of the final extrapolated results. A second observation is that usually the extrapolation to the thermodynamic limit is quite smooth: the simulations carried out for this work were performed on lattices with space-time cross-sections characterized by aspect ratios $LT \simeq 4$, which typically is enough to guarantee a sufficiently strong suppression of finite-volume corrections. This is a consequence of the fact that, as table~\ref{tab:glueball_masses} shows, in the confining phase the purely gluonic $\Sp(2)$ theory has a large mass gap: finite-volume corrections are exponentially suppressed with a characteristic length scale given by the Compton wavelength associated with the lightest physical state, which, in natural units, is approximately five times smaller than the inverse of the critical temperature, and thus approximately $20$ times smaller than the linear extent of the lattice in each of the three main spatial directions. In the deconfined phase, on the other hand, finite-volume corrections are suppressed by the fact that the deconfined plasma has a finite screening length. For this reason, some of the $\Delta/T^4$ values computed numerically at a fixed $\Ntau$ and from lattices of different volumes turn out to be fully compatible with each other within their uncertainties: in this case, carrying out a thermodynamic extrapolation by fitting those values to a constant plus a term inversely proportional to the volume reveals that the interpolating straight line is flat, and, in practice, the extrapolated constant is numerically close to the weighted average of the $\Delta/T^4$ values obtained from the simulations at the different volumes. Thirdly, the uncertainties associated with the scale setting---which, beside $T/\Tc$, directly affect the evaluation of $\Delta/T^4$, because the latter is computed according to eq.~\eqref{Delta}, which involves the relation between the lattice spacing and $\beta$---are typically very small: this is mostly due to the good precision of our determination of the non-perturbative scale setting according to eq.~\eqref{scale-setting_fit}, where especially the leading terms $k_0$ and $k_1$ in eqs.~\eqref{scale-setting_fit_results} have very small uncertainties. Next, we also note that the uncertainties associated with the spline interpolations are always very small, too: in this case, this is due to the fact that in our simulations we sampled a very dense set of temperatures in the range of interest, and (except in the proximity of $\Tc$) the behavior of $\Delta/T^4$ as a function of $T/\Tc$ is quite smooth. The fineness of the grid of temperature values that was used in the numerical integration of $p/T^4$ implies that also the uncertainties associated with this quantity are rather small. In practice, it is only close to $\Tc$ (i.e., in the narrow temperature range where the variation of $\Delta/T^4$ is fastest) that the systematics associated with the numerical integration are not completely negligible. For $p/T^4$, it should be noted that the uncertainties tend to accumulate (more precisely: to sum up in quadrature) in the integration process; however, the highly improved numerical scheme that we used, which is defined in eq.~\eqref{integral_discretization}, allowed us to keep them under control. Finally, the extrapolation to the continuum limit (carried out fitting the results extrapolated to the infinite-volume limit for each $\Ntau$ to a constant plus a term proportional to $\Ntau^{-2}$) was found to have a non-negligible impact on the total uncertainty of $\Delta/T^4$, in particular for temperatures corresponding to the maximum of this quantity. Attempting to include also an $\Ntau^{-4}$ correction (at least at the very few temperatures for which we also had results for $\Delta/T^4$ from simulations with $\Ntau=8$ and $\Ntau=9$) did not lead to any improvement in the results, nor on the estimate of the continuum-extrapolation uncertainties.}

\begin{table}[h]
\centering
\phantom{-------}
\begin{tabular}{|l|l|l|l|}
\hline
$T/\Tc$ & notes & $\Delta/T^4$             & uncertainty on $\Delta/T^4$ \\
\hline
\hline
 $0.95$ & confining phase                  & $0.2115$ & $0.0026~\left\{ 
 \begin{array}{l}
 0.002~\mbox{statistics}\\
 0.0005~\mbox{$V \to \infty$ extrap.}\\
 0.0002~\mbox{scale-setting}\\
 <0.0001~\mbox{spline interp.}\\
 0.001~\mbox{$\spac\to 0$ extrap.}
 \end{array}
 \right.$ \\
\hline
 $1.15$ & peak of $\Delta/T^4$             & $2.715$ & $0.015~\left\{ 
 \begin{array}{l}
 0.013~\mbox{statistics}\\
 0.006~\mbox{$V \to \infty$ extrap.}\\
 0.0002~\mbox{scale-setting}\\
 0.0001~\mbox{spline interp.}\\
 0.006~\mbox{$\spac\to 0$ extrap.}
 \end{array}
 \right.$ \\
\hline
 $1.5$  &                                  & $1.867$ & $0.012~\left\{ 
 \begin{array}{l}
 0.01~\mbox{statistics}\\
 0.003~\mbox{$V \to \infty$ extrap.}\\
 0.0002~\mbox{scale-setting}\\
 <0.0001~\mbox{spline interp.}\\
 0.001~\mbox{$\spac\to 0$ extrap.}
 \end{array}
 \right.$ \\
\hline
 $2$    & largest $T$ of our cont. extrap. & $1.094$ & $0.037~\left\{ 
 \begin{array}{l}
 0.028~\mbox{statistics}\\
 0.01~\mbox{$V \to \infty$ extrap.}\\
 0.005~\mbox{scale-setting}\\
 <0.0001~\mbox{spline interp.}\\
 0.006~\mbox{$\spac\to 0$ extrap.}
 \end{array}
 \right.$ \\
\hline            
\end{tabular}
\phantom{-------}
\caption{\revision{Breakdown of the uncertainty budget on the continuum-extrapolated values of $\Delta/T^4$ at a selection of temperatures.}}
\label{tab:Delta_error_budget}
\end{table}

\begin{table}[h]
\centering
\phantom{-------}
\begin{tabular}{|l|l|l|}
\hline
$T/\Tc$ & $p/T^4$ & uncertainty on $p/T^4$ \\
\hline
\hline
 $0.95$ & $0.0139$ & $0.0014~\left\{ 
 \begin{array}{l}
 0.0014~\mbox{uncertainties on $\Delta/T^4$}\\
 <0.0001~\mbox{numerical integ.}\\
 \end{array}
 \right.$ \\
\hline
 $1.15$ & $0.327$  & $0.010~\left\{ 
 \begin{array}{l}
 0.010~\mbox{uncertainties on $\Delta/T^4$}\\
 0.0001~\mbox{numerical integ.}\\
 \end{array}
 \right.$ \\
\hline
 $1.5$  & $0.958$  & $0.017~\left\{ 
 \begin{array}{l}
 0.017~\mbox{uncertainties on $\Delta/T^4$}\\
 <0.0001~\mbox{numerical integ.}\\
 \end{array}
 \right.$ \\
\hline
 $2$    & $1.363$  & $0.027~\left\{ 
 \begin{array}{l}
 0.027~\mbox{uncertainties on $\Delta/T^4$}\\
 <0.0001~\mbox{numerical integ.}\\
 \end{array}
 \right.$ \\
\hline            
\end{tabular}
\phantom{-------}
\caption{\revision{Same as in table~\ref{tab:Delta_error_budget}, but for $p/T^4$, which, as discussed in the text, is obtained by numerical integration of the continuum-extrapolated results for $\Delta/T^4$. Note that the uncertainties on $\Delta/T^4$ have a cumulative effect on $p/T^4$.}}
\label{tab:p_error_budget}
\end{table}

It is interesting to note that, in the temperature range under consideration, the quantities plotted in figure~\ref{fig:extrapolated_equilibrium_quantities} are still far from their Stefan--Boltzmann-limit values (which are denoted by the horizontal bars on the right-hand side of the plot): \revision{this can be clearly seen in the inset plot, displaying $s/T^3$, $\epsilon/T^4$, and $p/T^4$ divided by their Stefan--Boltzmann limit values.} This shows that, for temperatures up to a few times $\Tc$, the deconfined state of matter can hardly be described as an ideal (or nearly ideal) gas of gluon-like particles: instead, the clearly non-vanishing values for the trace of the energy-momentum tensor reveal that interactions among the degrees of freedom present in the deconfined phase are non-negligible. Also, when the temperature is increased, the approach towards the Stefan--Boltzmann limit is expected to be slow: in particular, if one takes the temperature to be the characteristic (hard) momentum scale in the deconfined phase, then eq.~\eqref{beta_function} shows that, at leading order, the coupling depends logarithmically on the temperature, i.e., it decreases very slowly as a function of $T$. These features are common to other non-Abelian gauge theories.

\begin{figure}
    \centering
    \includegraphics[width=0.7\textwidth]{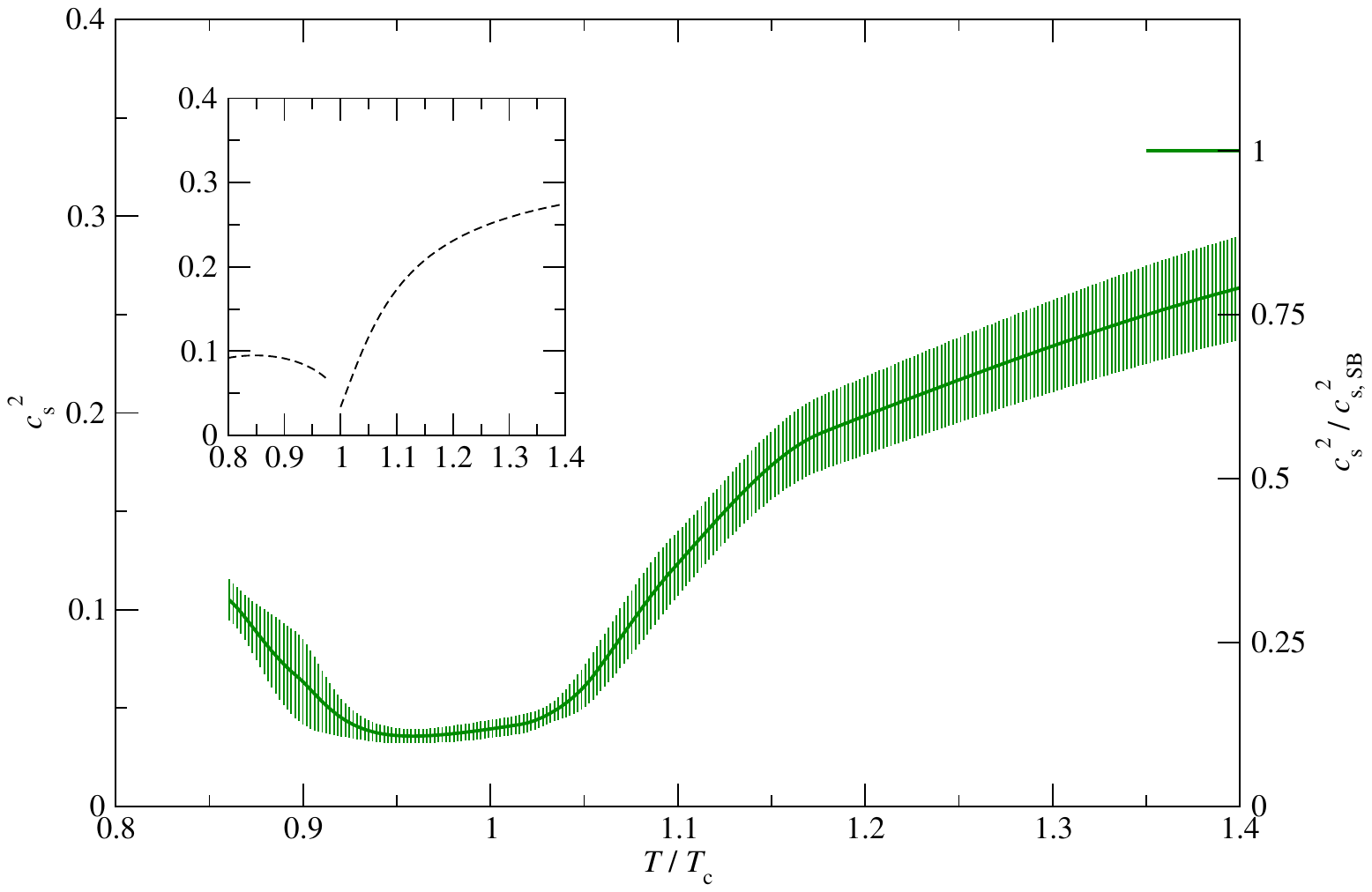}
    \caption{Squared speed of sound in the $\Sp(2)$ theory, plotted against $T/\Tc$; the displayed error band includes both statistical uncertainties and a (crude, and somewhat conservative) estimate of the systematic uncertainties related to the continuum extrapolation of our results. The horizontal bar on the right-hand side of the plot denotes the value of squared speed of sound in the Stefan--Boltzmann limit, which is \revision{$\csSB^2=1/3$, and, in order to help visualize the approach to that limit, the right-hand-side vertical axis shows the values of the $\cs^2/\csSB^2$ ratio.} The inset plot (in which, for the sake of clarity, the uncertainties are not displayed) shows the results \revision{for $\cs^2$} that can be obtained from the lattice calculations of $\SU(3)$ Yang--Mills theory in the deconfined phase~\cite{Borsanyi:2012ve} and in the confining phase~\cite{Giusti:2016iqr}, as discussed in ref.~\cite{Wang_et_al_in_preparation}.}
    \label{fig:cs2}
\end{figure}

A quantity that is of particular interest in the context of the production of gravitational waves in a thermal medium to be discussed in section~\ref{sec:interpretation_as_a_model_for_dark_matter_and_gravitational-wave_production}, and that can be directly derived from the extrapolated equilibrium quantities, is the squared speed of sound, $\cs^2$. Combining its definition according to eq.~\eqref{squared_speed_of_sound} with elementary thermodynamic relations, one obtains
\begin{align}
\label{practical_formula_for_cs2}
\cs^2(T) = \left[\frac{T^4}{s\Tc}\frac{d(s/T^3)}{d(T/\Tc)}+3\right]^{-1},
\end{align}
which can be immediately evaluated from our extrapolated results for the entropy density. The results are shown in the main plot of figure~\ref{fig:cs2}, where the displayed uncertainty band is dominated by the systematic uncertainties affecting these data, which we tentatively estimate to be of the order of $10\%$. For comparison, in the inset plot of figure~\ref{fig:cs2} we also show the analogous results (but without their uncertainties) obtained from lattice simulations of $\SU(3)$ Yang--Mills theory~\cite{Borsanyi:2012ve, Giusti:2016iqr}, and discussed in ref.~\cite{Wang_et_al_in_preparation}. One notes that the behavior of the speed of sound in the two theories is qualitatively, and even quantitatively, very similar.

Another feature of the $\Sp(2)$ equation of state, which is shared with gauge theories based on other non-Abelian groups, is the fact that, at low temperatures in the confining phase, the equilibrium thermodynamic quantities can be accurately modelled in terms of a gas of massive and non-interacting glueballs~\cite{Hagedorn:1965st, Hagedorn:1980kb, Hagedorn:1984hz, Fiore:1984yu, Cleymans:1992zc} (see also ref.~\cite{Trotti:2022knd} for a discussion from a modern perspective). For example, the pressure can be written as the sum of the contributions from the different single-particle glueball states, labelled by $k$, with masses $m_k$:
\begin{align}
\label{glueball_gas_pressure}
p(T) = \frac{T^2}{2 \pi^2} \sum_k d_k m_k^2 \sum_{l=1}^\infty \frac{1}{l^2} K_2\left( \frac{l m_k}{T} \right),
\end{align}
where $K_\nu(z)$ denotes the modified Bessel function of the second kind of argument $z$ and index $\nu$, while $d_k$ is the number of physical states for the particle species $k$. Note that, for the $\Sp(2)$ Yang--Mills theory, the latter reduces to $d_k=2J_k+1$, with $J_k$ the spin of the glueball labelled by $k$.

Using a known asymptotic expansion of Bessel functions~\cite{Abramowitz:1970hom}, at large $m_k/T$ eq.~\eqref{glueball_gas_pressure} can be rewritten as
\begin{align}
\label{glueball_gas_pressure_at_low_temperatures}
\frac{p}{T^4} \simeq \sum_k d_k \left(\frac{m_k}{2 \pi T}\right)^{\frac{3}{2}} \left\{ \Li_{\frac{5}{2}}\left( e^{-\frac{m_k}{T}}\right) + 
\sum_{j=1}^\infty \frac{\Li_{\frac{5}{2}+j}\left( e^{-\frac{m_k}{T}}\right)}{j!} \left(\frac{T}{8m_k}\right)^j \prod_{q=1}^j \left[ 16-(2q-1)^2\right]
\right\}
\end{align}
(where $\Li_\nu(z)$ denotes the polylogarithm of argument $z$ and order $\nu$), that we used in the numerical evaluation of the glueball-gas prediction for the thermodynamics of the $\Sp(2)$ theory.

\begin{figure}
    \centering
    \includegraphics[width=0.7\textwidth]{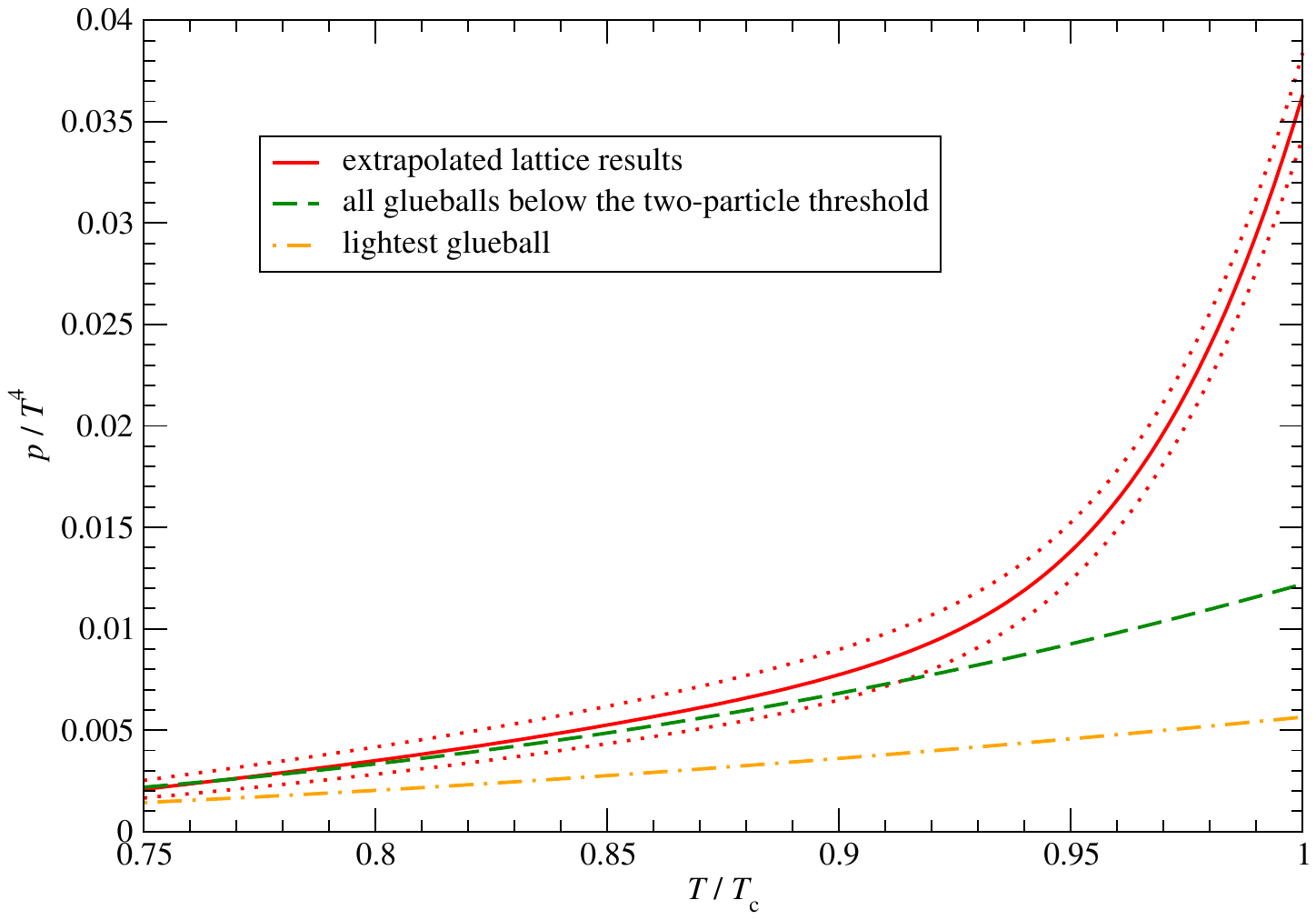}
    \caption{Comparison of our continuum-extrapolated lattice results for the $p/T^4$ ratio in the confining phase (solid red curve) with the predictions from a gas of non-interacting glueballs according to eq.~\eqref{glueball_gas_pressure}, including all glueballs with masses below the two-particle threshold (dashed green curve). The dash-dotted orange curve denotes the contribution from the lightest glueball only.}
    \label{fig:comparison_with_glueball_gas}
\end{figure}

The comparison of our extrapolated simulation results with the prediction from eq.~\eqref{glueball_gas_pressure} is shown in figure~\ref{fig:comparison_with_glueball_gas}, where the dash-dotted line denotes the dominating contribution, from the lightest glueball, to the $p/T^4$ ratio, while the dashed line includes the contributions from all glueballs below the two-particle threshold. As the plot shows, the continuum-extrapolated results obtained from our lattice calculations (denoted by the solid red curve) are consistent with the prediction from the glueball-gas model at low temperatures, but for $T \gtrsim 0.9 \Tc$ the curve obtained from eq.~\eqref{glueball_gas_pressure} underestimates the lattice results. As in $\SU(3)$ Yang--Mills theory~\cite{Meyer:2009tq, Borsanyi:2012ve}, this behavior can be interpreted in terms of a continuous, exponentially growing, Hagedorn-like, density of states. \revision{Note that, while in principle the inclusion of high-precision results obtained from finer lattices ($\Ntau \ge 8$) would be useful to corroborate the continuum extrapolation of the equation of state at low temperatures, in practice the computational costs to reach uncertainties comparable with those of the data evaluated from the other ensembles considered in this work, up to $\Ntau=7$, would quickly become prohibitive. The problem is especially severe at $T\le \Tc$, since in the confining phase the thermodynamic quantities take values that are significantly smaller than in the deconfined phase, and that, furthermore, are exponentially suppressed when the temperature is reduced. For this reason, in figure~\ref{fig:comparison_with_glueball_gas} we did not include, for example, the results of our simulations on lattices with $\Ntau=8$ and $\Ntau=9$ at $T=0.9 \Tc$ and at $T=0.95 \Tc$ that are shown in figure~\ref{fig:Delta_at_finite_Ntau}: they are compatible with zero within their uncertainties, they would not be sufficient to obtain any meaningful interpolating curve for the pressure at either of those $\Ntau$ values (for this purpose, the numerical integration would require a much larger number of data points, at sufficiently close temperatures), and would not add any significant information to the curve displayed in figure~\ref{fig:comparison_with_glueball_gas}.}

\begin{figure}
    \centering
    \includegraphics[width=0.7\textwidth]{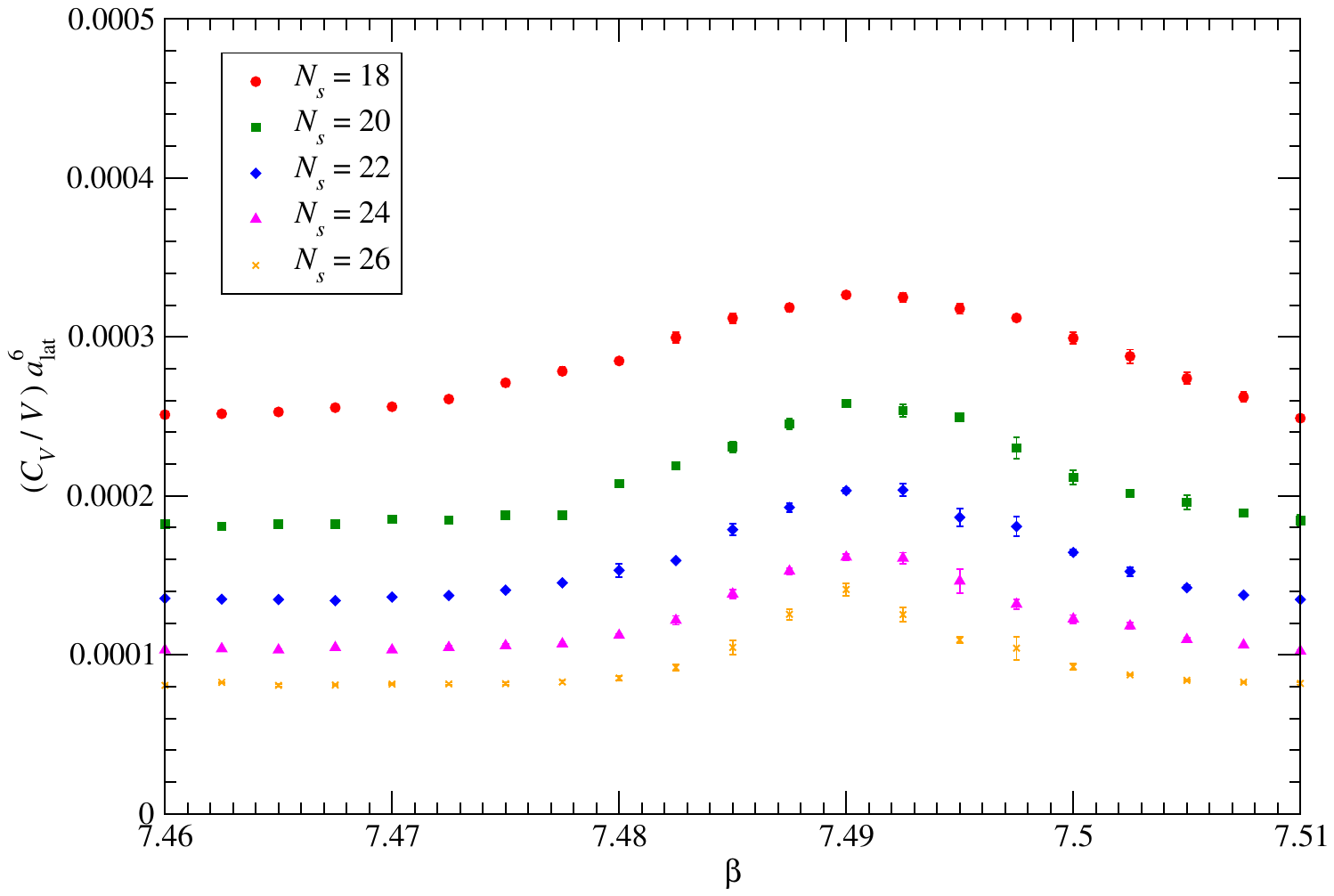}
    \caption{Values of the specific heat at fixed volume, defined according to eq.~\eqref{CV_definition} and in the appropriate units of the lattice spacing, as obtained from simulations on lattices with $\Ntau=5$ and for different values of $\Ns$, denoted by symbols of different shapes and colors.}
    \label{fig:specific_heat_Ntau5}
\end{figure}

We now move on to our numerical results in the study of the latent heat $\Lh$ and of the specific heat at fixed volume $\CV$; as we mentioned in section~\ref{sec:formulation_of_the_model}, these two quantities are directly related with each other via eq.~\eqref{Lh_and_CV}. In figure~\ref{fig:specific_heat_Ntau5} we present an example of our results for the determination of $\CV$: the plot shows $\CV$ as a function of $\beta$, and as obtained from simulations on lattices with $\Ntau=5$ sites in the Euclidean-time direction and for different values of the linear extent of the three ``spatial'' sizes of the lattice (in units of the lattice spacing $\spac$), ranging from $\Ns=18$ to $\Ns=26$, which corresponds to a variation in the lattice volume by a factor larger than three.

\begin{figure}
    \centering
    \includegraphics[width=0.7\textwidth]{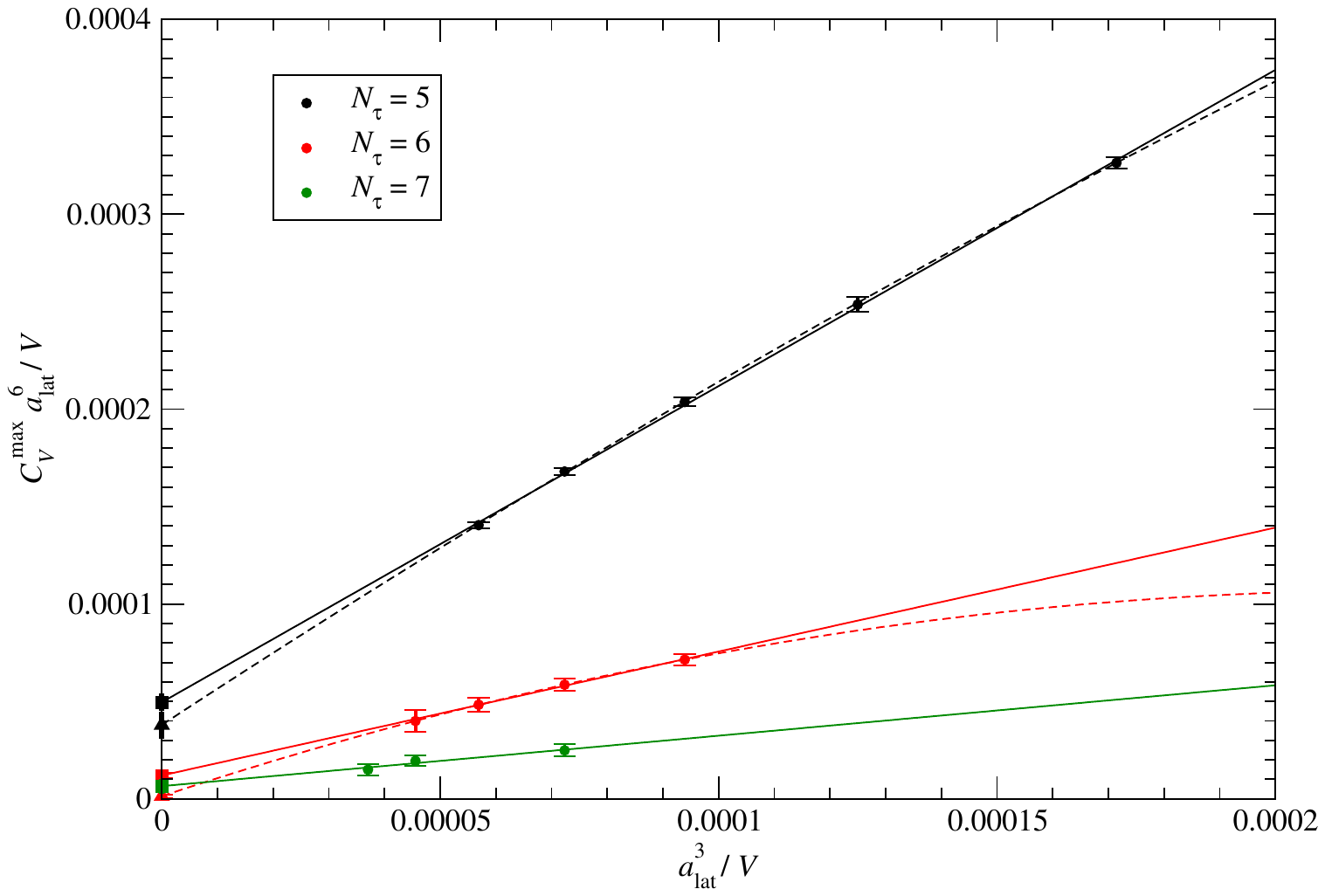}
    \caption{Extrapolation of our lattice results at \revision{$\Ntau=5$, $\Ntau=6$, and $\Ntau=7$} for the maximum value of the specific heat divided by the volume and expressed in the appropriate units of the lattice spacing (circles) to their infinite-volume limit. The dashed \revision{curves (for the $\Ntau=5$ and $\Ntau=6$ ensembles only) are} the result of \revision{parabolic fits} according to eq.~\eqref{parabolic_CV_over_V_fit}, while the \revision{solid straight lines are obtained by linear fits} of the data (i.e., setting $\gamma_2$ to zero). The corresponding extrapolated values, respectively denoted by the \revision{triangles and by the squares, lead to the values for the specific heat reported in the main text, which} are compatible with each other.}
    \label{fig:specific_heat_thermodynamic_extrapolation}
\end{figure}

For each value of $\Ns$, we determined the maximum of the specific heat per volume from the lattice results for $(\CV/V)\spac^6$ at different values of $\beta$, by interpolating the datum corresponding to the largest value of $\CV$, the two data points at its left, and the two at its right, through a quadratic fit in $\beta$, and extracting the maximum value of the fitted parabola. In figure~\ref{fig:specific_heat_thermodynamic_extrapolation} we show our study of the latent heat of the theory, based on eq.~\eqref{Lh_and_CV}: the plot shows the extrapolation of the peak values for the specific heat to the infinite-volume limit according to
\begin{align}
\label{parabolic_CV_over_V_fit}
\frac{\CmaxV \spac^6}{V} = \gamma_0 + \frac{\gamma_1}{V/\spac^3} + \frac{\gamma_2}{(V/\spac^3)^2}.
\end{align}
\revision{For the data extracted from our simulations with $\Ntau=5$,} the fit of the data shown in figure~\ref{fig:specific_heat_thermodynamic_extrapolation} yields
\begin{align}
\label{parabolic_CV_over_V_fit_results}
\gamma_0 = 3.8(7) \cdot 10^{-5}, \qquad \gamma_1 = 1.87(15), \qquad \gamma_2 = -1.1(7) \cdot 10^{3},
\end{align}
with $\redchisq=0.09$. The smallness of the reduced $\chi^2$ and the fact that the fit result for $\gamma_2$ is compatible with zero within approximately one standard deviation and a half indicates that a quadratic fit in $1/(V\spac^3)$ is an overfitting. For comparison, a two-parameter fit with $\gamma_2$ set to zero, leads to
\begin{align}
\label{linear_CV_over_V_fit_results}
\gamma_0 = 5.0(4) \cdot 10^{-5}, \qquad \gamma_1 = 1.62(5),
\end{align}
with $\redchisq=0.67$. Assuming the extrapolation according to this linear fit, and including its discrepancy with respect to the quadratic fit as a systematic uncertainty, the latent heat obtained from our $\Ntau=5$ simulations (and expressed in units of the fourth power of the inverse lattice spacing corresponding to the critical temperature for this lattice) is found to be consistent with the one that can be determined from the discontinuity in the energy density of the theory across the deconfinement temperature, according to the procedure described in refs.~\cite{Lucini:2003zr, Lucini:2005vg}, which in our case leads to $\Lh/\Tc^4=1.76(9)(11)$ for $\Ntau=5$, to $\Lh/\Tc^4=1.5(1)(6)$ for $\Ntau=6$ (where the first error is statistical, whereas the second accounts for systematic uncertainties, including, in particular, those associated with varying the number of data used in the infinite-volume extrapolation, and with performing such extrapolation including or excluding, beside a correction linear in the inverse volume, also a quadratic one). Within their uncertainties (which can be summed in quadrature), these results are consistent with the approximate continuum-extrapolated value
\begin{align}
\label{continuum_estimate_of_Lh_over_Tc4}
\frac{\Lh}{\Tc^4}\simeq 1.69(12).
\end{align}
\revision{As a consistency check, we also carried out a set of computations on finer lattices, with $\Ntau=7$. In this case we run simulations for three volumes only (corresponding to $\Ns=24$, $\Ns=28$, and $\Ns=30$) and, as a consequence, performed the thermodynamic extrapolation of the $\spac^6 \CmaxV/V$ ratio considering only the two-parameter fit with $\gamma_2$ set to zero, obtaining $\gamma_0=6.5(3.7)\cdot 10^{-6}$ and $\gamma_1=0.26(7)$. The corresponding value for the latent heat in units of the fourth power of the deconfinement temperature is $\Lh/\Tc^4=1.8(5)$: even though the uncertainty is large (due to the fact that, for this $\Ntau$ value, the extrapolated $\gamma_0$ value has a relative error of the order of $50\%$), the value is compatible with our continuum estimate reported in eq.~\eqref{continuum_estimate_of_Lh_over_Tc4}. It is important to stress that a reliable estimate of the $\CmaxV/V$ ratio becomes increasingly difficult when $\Ntau$ is increased, because this quantity, when evaluated from a lattice calculation, is expressed as a dimensionless number proportional to $\spac^6$; given that the statistical uncertainties in a Monte~Carlo calculation scale with the inverse of the square root of the number of configurations, this implies that the computational costs to evaluate the $\CmaxV/V$ ratio on the lattice with a fixed relative error are expected to grow at least as $\Ntau^{12}$.

The continuum estimate for the $\Lh/\Tc^4$ in eq.~\eqref{continuum_estimate_of_Lh_over_Tc4}} can be compared with analogous results in $\SU(N)$ Yang--Mills theories. Excluding the $\SU(2)$ theory, which has a second-order deconfinement transition~\cite{Engels:1990vr, Fingberg:1992ju, Engels:1994xj}, and hence a vanishing latent heat, the $\SU(3)$ theory has a relatively weak first-order transition, with $\Lh/\Tc^4 \simeq 1.4$~\cite{Beinlich:1996xg}, while the values of the latent heat observed in $\SU(N)$ gauge theories with more than three color charges scale well with the dimension of the gauge group and are described by \revision{$\Lh/(N^2\Tc^4)=0.360(6)-1.88(17)/N^2$~\cite{Rindlisbacher:2025dqw} (which is consistent with, and significantly more precise than, the earlier result $\Lh/(N^2\Tc^4) \simeq 0.344(72)$~\cite{Lucini:2005vg})}. For the $\Sp(2)$ theory, the dimension of the gauge group is ten, i.e., intermediate between those of $\SU(3)$ and $\SU(4)$, and a discontinuous deconfinement transition is observed; hence, if one would argue that this Yang--Mills theory should behave similarly to those based on unitary gauge groups, and would simply attempt to derive a crude estimate of the $\Lh/\Tc^4$ ratio by rescaling those found for the $\SU(3)$ and $\SU(4)$ theories according to the number of gluon degrees of freedom, one would get a result in the ballpark between approximately $1.5$ and $3$. The estimate~\eqref{continuum_estimate_of_Lh_over_Tc4} obtained from our simulation results lies within this interval.

We conclude this section with a few comments on the implications of our results for the two other reasons of interest for a gauge theory based on a symplectic group (beside its potential viability as a framework for particle dark matter, which is discussed more extensively in section~\ref{sec:interpretation_as_a_model_for_dark_matter_and_gravitational-wave_production}), that we mentioned in section~\ref{sec:introduction}.

One of the reasons why it was interesting for us to study the finite-temperature confining transition in a symplectic Yang--Mills theory consisted in investigating, whether the center of the gauge group, which is widely believed to be directly responsible for the emergence of confinement at large distances in the ground state of non-Abelian gauge theories, via the formation of center vortices~\cite{tHooft:1979rtg, Mack:1980rc, DelDebbio:1996lih, Greensite:2003bk}, also affects the nature of the confinement/deconfinement transition at finite temperature. In particular, general arguments based on the renormalization group indicate that the thermal deconfinement transition for $\Sp(N)$ gauge theories could be of second order, and in the same universality as the three-dimensional Ising model~\cite{Svetitsky:1982gs}. Our results, which are consistent with previous lattice studies of this issue~\cite{Holland:2003kg}, show that this is not the case: the transition is of first order, with a finite latent heat that, when expressed in units of the fourth power of the deconfinement temperature, is of the same order of magnitude as, and slightly larger than, in the $\SU(3)$ Yang--Mills theory. This confirms the conjecture, that the continuous or discontinuous nature of the confinement/deconfinement transition is essentially determined by the imbalance in the number of degrees of freedom in the high- and in the low-temperature phases, i.e., it depends on the size of the gauge group, not directly on the center symmetry~\cite{Holland:2003kg}. We also note that, in fact, the discontinuous nature of the deconfinement transition for the $\Sp(2)$ theory is also supported by analytical arguments, including those based on calculations starting from a supersymmetric version of the theory on a manifold with a compactified spatial direction~\cite{Anber:2014lba}, and those derived within the framework of the functional renormalization group~\cite{Braun:2010cy}.

Another reason of interest to study a non-Abelian gauge theory having a Lie algebra of dimension close to the one of $\SU(3)$, but belonging to a different classical series ($C_n$ in this case) is the set of correspondences that are expected to relate gauge theories based on symplectic, orthogonal, and unitary groups to each other. In particular, it was shown long ago that theories based on these different gauge groups have Wilson loops sharing the same behavior in the large-$N$ limit~\cite{Lovelace:1982hz}; even earlier, hints of this correspondence can be traced back to pioneering work on random matrix theory~\cite{Dyson:1962es}. More recently, the study of such dualities between different gauge theories has been generalized and extended beyond the 't~Hooft limit in various works~\cite{Unsal:2007fb, Mkrtchian:2011oda, Caputa:2013vla, Bond:2019npq}, in particular with the inclusion of a coupling to matter fields, and with the derivation of detailed correspondences between operators in different theories (see also ref.~\cite{Corley:2001zk}). While the numerical simulations that we presented in this section do not allow us to make any claim on the validity or invalidity of these correspondences, we note that our results for the equation of state of the $\Sp(2)$ theory reveal remarkable similarities with those found for $\SU(N)$ gauge theories~\cite{Panero:2009tv}.

\section{Interpretation as a model for dark matter and gravitational-wave production}
\label{sec:interpretation_as_a_model_for_dark_matter_and_gravitational-wave_production}

In this section we discuss some of the implications predicted from our model for the thermal evolution of dark matter in the early universe, and the features of the gravitational-wave power spectrum produced at the confining phase transition. Before doing so, however, an important remark is in order: if the dark sector is not directly coupled with the SM sector, then its temperature can be different from the temperature of the latter. In general, a difference in temperature among the visible and dark sectors could have been generated during reheating after inflation, if inflatons decay mostly into one of the two sectors, or if a sector contains a much larger number of heavy particle species than the other: when a thermalized particle species becomes non-relativistic, its entropy is transferred to the other particle species in the same sector (heating them up), eventually leading to a significant temperature imbalance between the two sectors~\cite{Breitbach:2018ddu}.

If the dark sector is subject to reheating in the early stages of the universe, one expects that it will first be in the deconfined phase at high temperatures, then it will undergo a cosmological phase transition to the confining phase. Thus, in the model considered in this work, the non-Abelian gauge theory describing the dark sector will undergo an evolution qualitatively similar to the one of quantum chromodynamics, from a ``hot'', deconfined phase, to a ``cold'', confining phase.

{\bf Deconfined phase in the early universe.} At sufficiently high temperatures realized in the early universe, the dark sector that we are studying consists of a plasma of $\Sp(2)$ ``dark gluons'', which interact with each other through screened long-range forces. Like in a generic non-Abelian gauge theory (see, for example, refs.~\cite{Laine:2016hma, Ghiglieri:2020dpq} and references therein), at sufficiently high temperatures a hierarchy of scales
\begin{align}
g^2 T \ll gT \ll 2\pi T
\end{align}
develops, so that the ``hard'' scale, parametrically $O(T)$, of the thermal excitations in the plasma becomes well separated from the ``soft'', $O(gT)$, scale, which characterizes the Debye screening of ``chromoelectric'' modes; in turn, the latter gets well separated from the ``ultrasoft'' scale, $O(g^2T)$, typical of the screening of ``chromomagnetic'' modes~\cite{Kajantie:1981hu, Kajantie:1982xx, Toimela:1982ht, Furusawa:1983gb, Arnold:1995bh, Ghisoiu:2015uza}. Note that the screening of ``chromomagnetic'' modes is a phenomenon which retains intrinsically non-perturbative effects~\cite{Linde:1980ts}.

{\bf Confining transition.} As the universe expands and cools down, the equilibrium thermodynamic quantities will decrease as discussed in section~\ref{sec:results}; when its temperature drops below $\Tc$, the dark sector undergoes a change of state, and the dark gluons confine into massive glueballs. This is accompanied by a brisk drop in the equilibrium thermodynamic quantities. In particular the change in the entropy density across $\Tc$ can be interpreted in terms of the sharp reduction in the number of (effective) degrees of freedom of the theory: from $N(2N+1)=10$ ``gluons'', with two polarizations each, in the high-temperature phase, to $O(1)$ massive ``glueballs''. The finiteness of the latent heat implies that this confining transition is a discontinuous one.

{\bf Generation of gravitational waves at the phase transition.} Gravitational-wave astronomy entered the domain of experimental science almost a decade ago~\cite{LIGOScientific:2016aoc}: since then, the Advanced Virgo~\cite{VIRGO:2014yos} and the Advanced Laser Interferometer Gravitational-Wave Observatory~\cite{Harry:2010zz, LIGOScientific:2014pky} have observed a large number of events consistent with gravitational waves produced by transient astrophysical sources: mergers of binary black holes or coalescence of binary neutron stars. A more ambitious goal for the future~\cite{LIGOScientific:2019vic} consists in the detection of gravitational waves produced by cosmological sources~\cite{Maggiore:1999vm, Caprini:2018mtu, Christensen:2018iqi}. In particular, novel ground-based interferometers of third generation, like the Einstein Telescope~\cite{ET:2019dnz} and the Cosmic Explorer~\cite{Reitze:2019iox}, or space-based gravitational wave detectors, like the Laser Interferometer Space Antenna (LISA)~\cite{LISA:2017pwj}, the Big Bang Observer (BBO)~\cite{Harry:2006fi}, Taiji~\cite{Hu:2017mde}, TianQin~\cite{TianQin:2015yph}, and DECIGO~\cite{Kawamura:2020pcg} could detect gravitational waves produced by violent phenomena in the early universe, among which strong cosmological phase transitions~\cite{Weir:2017wfa, Mazumdar:2018dfl, Hindmarsh:2020hop}. The discontinuous nature of a strong first-order phase transition, like the confining phase transition in our $\Sp(2)$ model, implies that the energy-momentum tensor can experience large anisotropic fluctuations, resulting into the generation of a stochastic, ``cosmological'' gravitational-wave background: a phenomenon that was already suggested long ago~\cite{Witten:1984rs, Hogan:1983ixn, Hogan:1986qda, Turner:1990rc}.

\revision{At this point, however, an important remark is in order: the estimates for the spectrum of gravitational waves that will be discussed in this section depend on parameters (including, in particular, the bubble wall velocity and the transition-strength parameter $\alpha$) for which we must necessarily rely on---sometimes relatively crude---estimates, that often involve underlying assumptions. This can be contrasted with the results for the equation of state and for the other quantities discussed in this work, for which we could derive robust \emph{ab~initio} predictions. Ultimately, the reason of this discrepancy lies in the very nature of the phenomena they are related to (and in the way they are studied in the numerical calculation). On the one hand, the pressure, the energy and entropy densities, etc. are equilibrium thermodynamic quantities, which have a natural formulation in terms of Euclidean quantum field theory, and for which a non-perturbative evaluation on a Euclidean lattice is in principle straightforward. On the other hand, quantities related to processes such as the formation, expansion, collision, etc. of bubbles of different phases within a medium encode information about phenomena beyond thermodynamic equilibrium and intrinsically dynamical. In general, the non-perturbative numerical evaluation of such quantities for a quantum field theory regularized on a lattice requires methods that are quite different from those that can be applied to study equilibrium thermodynamics; these include real-time simulations~\cite{Berges:2006xc} or techniques to solve mathematically non-trivial inverse problems~\cite{Rothkopf:2022fyo}, which entail their own peculiar challenges. In cases when the quantum fluctuations of fields are small compared to the thermal ones, one can resort to classical statistical lattice gauge theory simulations, whereby the gauge fields are let evolve in a deterministic way, according to the classical equations of motion, and the process is repeated many times, starting from initial values for the fields sampled from a thermal distribution. This approach allows one to estimate the average value and the variance of various real-time quantities, and has been used in studies of non-equilibrium processes taking places in the early evolution of the system created immediately after a ``small bang'', namely a heavy-ion collision~\cite{Berges:2013fga}.

A particularly important quantity associated with the bubble expansion is the bubble wall velocity, denoted by $\vw$, according to which one can distinguish among deflagrations (when the bubble wall velocity is smaller than the speed of sound, and a ``sound shell'' can form in front of the wall), detonations (when the bubble wall velocity is larger than the speed of sound, and the plasma behind the wall can have non-zero velocity), or hybrid cases. The bubble wall velocity is determined by the competing forces acting on the wall: the latent heat that is released in the phase transition pushes the bubble outwards, while its interaction with the plasma creates friction. When the two forces are balanced, the wall ceases to accelerate. Deriving an accurate estimate for the bubble wall velocity in a certain model, however, is highly non-trivial, as it cannot be determined only from the effective potential associated with a first-order transition, and requires instead knowledge of the friction of the plasma near the bubble wall, which can be determined only if the precise hydrodynamical solution of the plasma profile is accessible. While lattice studies of this problem do exist (see, for instance, ref.~\cite{Krajewski:2023clt} as a recent example), they typically rely on a setup which is entirely different from the one of the lattice calculations reported in the present work---which, instead, are based on the ``microscopic'' formulation of the theory in terms of fundamental fields. Extending our work to include this kind of lattice calculations lies beyond the scope of the present study and would entail significant computational costs, hence we leave it for the future. \revision{It is nevertheless possible to provide at least a rough, order-of-magnitude estimate of $\vw$, based on the following argument~\cite{Witten:1984rs, Asadi:2021pwo, Gouttenoire:2023roe}: let $x$ denote the fraction of the dark sector in the confining phase. By the definition of the latent heat, when this fraction increases by an amount $dx$, this is accompanied by the release (the confining transition being an exothermic process) of an amount of energy per volume $\Lh dx$, which corresponds to a temperature variation
\begin{align}
dT=\frac{\Lh}{\frac{d \epsilon}{d T}} dx = \frac{15}{2\pi^2 \gstareff} \frac{\Lh }{T^3} dx,
\end{align}
where $\gstareff$ denotes the effective number of relativistic degrees of freedom in the thermal bath: for the dark sector, it can be expressed as the product of the number of gluon polarizations ($2$) times the multiplicity factor associated with the dimension of the Lie algebra of the gauge group (which is $10$ for the $\Sp(2)$ theory, or $N(2N+1)$ for a generic $\Sp(N)$ theory), times a coefficient of order $1$---encoding the fact that, as our results for the equation of state show, at temperatures of the order of $\Tc$ the energy density in the deconfined phase is smaller than, albeit comparable with, the energy density of a free gluon plasma, due to the fact that gluon-gluon interactions are not negligible---times a $(1-x)$ factor that accounts for the fraction that is in the deconfined phase. The evolution of the temperature of the dark sector in the primordial universe is thus determined by the competition between the heating due to the latent heat conversion into thermal energy and the adiabatic cooling induced by the Hubble expansion:
\begin{align}
\label{temporal_evolution_of_T}
\frac{d T}{d t} \simeq \frac{3 R^2 \Lh T}{8 \epsilon R_0^3} \vw -HT,
\end{align}
where $R$ denotes the radius of the bubble, $R_0$ is its value at bubble percolation, which can be defined by the condition $x|_{R=R_0}=1/2$, and we used the fact that in general $x$ is proportional to the volume of the bubble (which we take to be spherical), so that $x=\frac{R^3}{2R_0^3}$, the bubble wall velocity is defined as $\vw=\frac{d R}{d t}$, and, following the literature, we approximated $\frac{d \epsilon}{d T}$ as $\frac{4\epsilon}{T}$, under the assumption that the $\frac{\epsilon}{T^4}$ ratio has only a relatively mild dependence on the temperature. Then, the condition that the temperature remains constant as a function of time leads to
\begin{align}
\label{vw_estimate}
\vw \simeq \frac{8 \epsilon R_0^3 H}{3 R^2 \Lh}.
\end{align}
}

Another parameter on which the predictions for the gravitational-wave spectrum depend is the transition-strength parameter $\alpha$: it can be defined as the ratio of the trace-anomaly difference between the two phases to the thermal energy density, evaluated at the nucleation temperature (that is, at the temperature realized when the volume-averaged bubble nucleation rate reaches its maximum)~\cite{Correia:2025qif}. It should be noted, however, that the nucleation of bubbles is a dynamical process, and the determination of the associated rate is highly non-trivial (for a detailed discussion see, for instance, refs.~\cite{Langer:1967ax, Langer:1969bc, Coleman:1977py, Callan:1977pt, Affleck:1980ac, Linde:1980tt, Arnold:1996dy, Moore:2000jw}, which describe the general aspects of this problem as well as some specific details relevant for a possible electroweak phase transition). Also in this case, our estimate for $\alpha$, that is discussed below, cannot be regarded as a fully first-principle prediction; as a consequence, the estimate for the spectrum of gravitational waves should be considered as a qualitative (or, at best, semi-quantitative) estimate, rather than a hard quantitative prediction.

Finally, a third parameter entering the determination of the gravitational-wave spectrum is the interface tension $\sigmacd$ between the confining and deconfined phases. In this case, the status of the estimate for this quantity is somehow better, given that a dedicated lattice determination of $\sigmacd/\Tc^3$ for the $\Sp(2)$ theory was recently published in ref.~\cite{Bennett:2024bhy}. Even though the final result reported therein was obtained from simulations at fixed $\Ntau=4$ (and thus may be significantly different from the continuum-extrapolated result), extending the calculation that in ref.~\cite{Bennett:2024bhy} was carried out for a single, relatively coarse lattice spacing---which deserved a publication on its own---to several finer lattices would require an amount of computational resources beyond the scope of the present work.

In view of these caveats, the results presented in this section should be considered as benchmark examples, rather than as genuinely first-principle predictions.}

The quantity of interest to characterize the spectrum of gravitational waves is the power spectrum, defined as the fraction of the total energy density per interval of the logarithm of the frequency and denoted as $h^2 \Omega_{\rm GW}$. Note that, for gravitational waves of cosmological origin, the minimal frequency is simply given by the Hubble rate $H$ at the time of their production. Due to the expansion of the universe, the frequency $f$ of a gravitational wave produced at time $t$ gets redshifted by the ratio of the Robertson--Walker scale factor (denoted by $a$) at time $t$ to the scale factor at the present time $t_0$, so that the observed frequency reads $f a(t)/a(t_0)$. From this, one can derive an approximate estimate of the frequencies of interest for the visible sector: in the case of a first-order phase transition at the electroweak scale, of the order of $10^2$~GeV, taking place at a time $t\simeq 10^{-11}$~s after the Big~Bang, the characteristic frequencies of the gravitational waves would be in the ballpark of $10^{-4}$~Hz. For a first-order phase transition at the scale characteristic of the strong interaction, about $10^2$~MeV, and taking place at the end of the so-called ``quark epoch'', i.e., at $t\simeq 10^{-6}$~s, the resulting observable frequencies today would be even lower, in the range of $10^{-7}$~Hz.

The generation of gravitational waves can be described by the (linearized) Einstein equation for tensor perturbations in a Friedmann background; following the notations and conventions of ref.~\cite{Caprini:2010xv}, and writing the metric as $ds^2=a^2(t)\left(-dt^2+d{\mathbf x}^2\right)$, one has
\begin{align}
\ddot{h}_\pm + 2\frac{\dot{a}}{a} \dot{h}_\pm +k^2 h_\pm = 8\pi G a^2\rho\Pi_\pm ,
\end{align}
where dots denote derivatives with respect to the conformal time, $G$ is the gravitational constant, $\rho$ is the background cosmological energy density, $\Pi_\pm$ are the tensor helicity modes of the (dimensionless) anisotropic stress sourcing the gravitational waves, and $h_\pm$ represent the helicity modes of the gravitational waves.

The production of gravitational waves can take place through different mechanisms (see, for example, ref.~\cite[section 8]{Caprini:2018mtu} for a review): the simplest of these is the collision of true-vacuum bubbles~\cite{Kosowsky:1991ua, Kosowsky:1992vn, Kamionkowski:1993fg, Caprini:2007xq, Huber:2008hg, Gould:2021dpm}. Bubbles expanding in a thermal medium, however, are subject to friction: the latent heat released in the phase transition pushes the bubbles to expand, but their interaction with the plasma constituents opposes this expansion~\cite{Weir:2017wfa}. When the two forces are balanced, the bubble walls slow down to an approximately constant speed, and the energy transferred to the plasma generates sound waves, which can be another source of gravitational waves~\cite{Hogan:1986qda, Espinosa:2010hh, Hindmarsh:2013xza, Hindmarsh:2015qta, Hindmarsh:2016lnk, Hindmarsh:2017gnf, Hindmarsh:2019phv}. In addition, the bubble-merging process can also induce hydrodynamic turbulence, which is another potential source of gravitational waves~\cite{Kosowsky:2001xp, Dolgov:2002ra, Caprini:2006jb, Gogoberidze:2007an, Kahniashvili:2008pe, Kahniashvili:2009mf, Caprini:2009yp, Auclair:2022jod}.

After production, the propagation of gravitational waves is essentially free, with the phenomenologically very interesting implication that their experimental detection could provide a unique probe of the early universe, given that primordial gravitational waves decouple much earlier than photons or neutrinos. More precisely, the frequency of gravitational waves produced at the phase transition decreases as the inverse of the Robertson--Walker scale factor, while their fraction of the critical energy density decreases as $a^{-4}$. Following the discussion in ref.~\cite{Schwaller:2015tja}, one can then derive that entropy conservation implies that 
\begin{align}
\frac{a_\star}{a_0} = \left(\frac{g_{0,s}}{g_{\star,s}}\right)^\frac{1}{3} \frac{\TCMB}{T_\star},
\end{align}
where $\TCMB = 2.34864(49) \cdot 10^{-13}$~GeV denotes the temperature of the cosmic microwave background, while $g_s$ is the effective number of relativistic degrees of freedom, the $\star$ subscript refers to the time of production, while the $0$ subscript refers to the time of observation.

{\bf Gravitational waves from bubble collisions.} Gravitational waves produced by bubble collisions are often modelled in the thin-wall and envelope
approximations~\cite{Kosowsky:1991ua, Kosowsky:1992rz, Kosowsky:1992vn, Kamionkowski:1993fg} (see also ref.~\cite{Jinno:2016vai}); the power spectrum can be expressed as~\cite{Huber:2008hg, Caprini:2009fx, Caprini:2010xv}
\begin{align}
\label{bubble-collision_spectrum}
\OmegaGW^{\mathrm{(bc)}} h^2 \simeq \frac{2}{3\pi} h^2 \Omega_{\mbox{\tiny{r}}0}\Theta^{\mathrm{(bc)}} {\cal H}_\star^2 \tpt^2 \Omega_{S\star}^2 v^3 \frac{(k\tpt)^3}{1+(k\tpt)^4},
\end{align}
where $h=0.6766(42)$~\cite{Planck:2018vyg} denotes the dimensionless Hubble parameter today (related to the dimensionful Hubble parameter today via the formula $H_0=h\cdot 100$~km$\cdot$s$^{-1}\cdot$Mpc$^{-1}$), $k=2\pi a f$ and ${\cal H}= aH$ respectively denote the conformal wave number and the conformal Hubble parameter, $\Omega_{\mbox{\tiny{r}}}$ is the radiation energy density, and we also include a phenomenological effective attenuation factor $\Theta^{\mathrm{(bc)}}$ (for which we make the conservative \emph{Ansatz} $\Theta^{\mathrm{(bc)}}\simeq 10^{-2}$), in order to put the contribution to the power spectrum from bubble collisions on a similar footing as the sound-wave power spectrum in eq.~\eqref{soundwave_power_spectrum}, for which attenuation effects due to reheating and those due to the finiteness of the sound-wave lifetime are taken into account. $\tpt$ denotes the duration of the phase transition, $\Omega_{S\star} =\rho_{S\star}/\rho_{\star,\rm crit}$ is the relative energy density in the source, and $v$ is the velocity of bubble walls in their expansion. Note that the determination of the latter in a first-order phase transition is non-trivial~\cite{Laurent:2022jrs}; recently, however, ref.~\cite{Ai:2024btx} proposed to constrain it using bounds based on the hypothesis of local thermal equilibrium and on the ballistic limit (for an example of application in the context of the strong nuclear interaction, see also ref.~\cite{Cline:2025bwe}). The duration of the phase transition can be defined in terms of the bubble nucleation rate (per unit time and unit volume) $\Gamma$ via~\cite{Gould:2022ran}
\begin{align}
\label{tpt_and_bubble_nucleation_rate}
\frac{1}{\tpt} = \frac{d}{dt} \left( \frac{\Gamma}{T^4} \right),
\end{align}
and the spectrum of gravitational waves produced by bubble collisions depends on the temperature of the phase transition through the dependence of ${\cal H}_\star$ on $T_\star$. It should also be noted that the bubble nucleation rate can be affected by the presence of topological defects, which act as local impurities triggering the phase transition: a detailed discussion about this aspect and the related phenomenological implications for gravitational waves can be found in ref.~\cite{Blasi:2022woz}.

\begin{figure}
    \centering
    \includegraphics[width=0.7\textwidth]{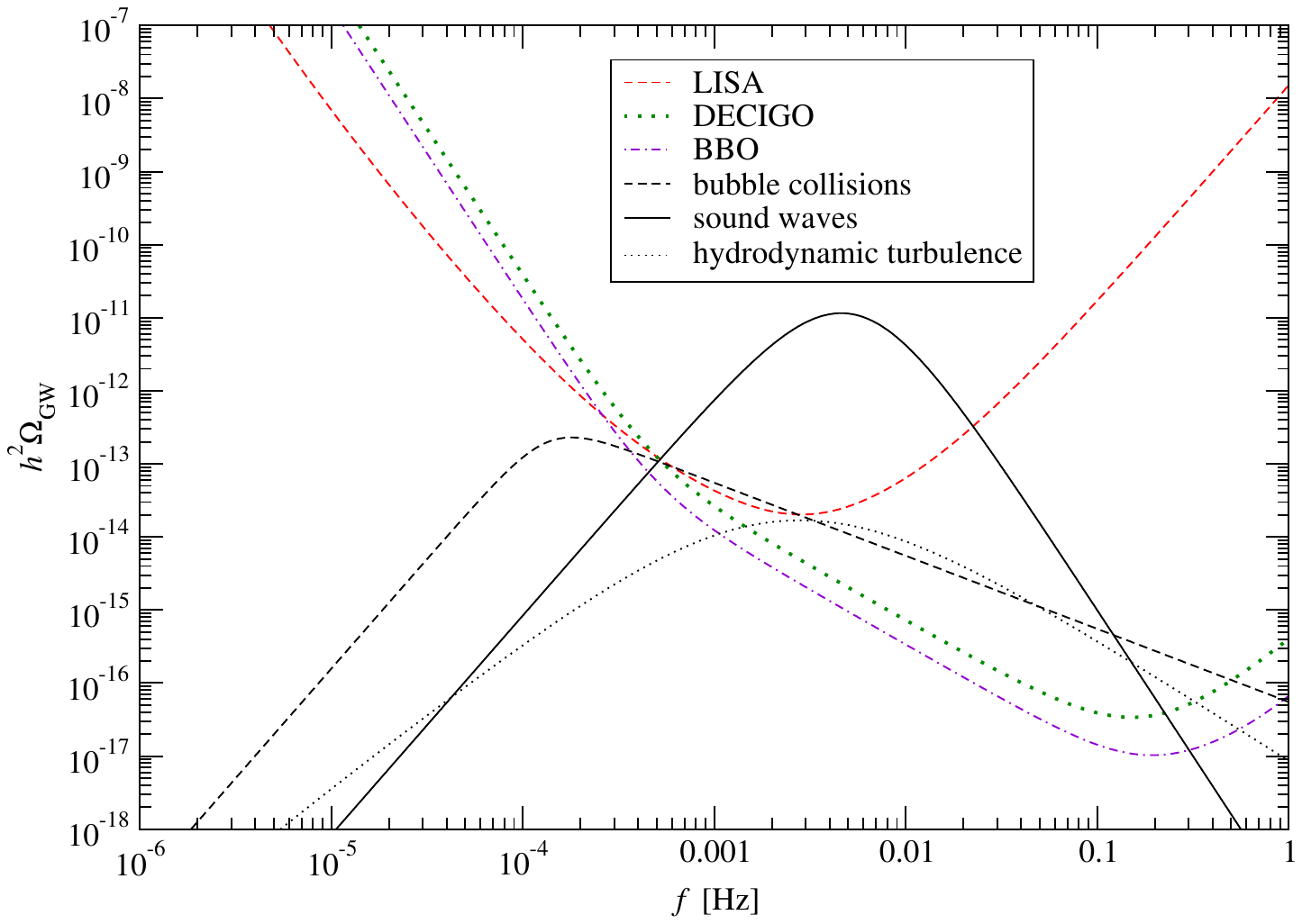}
    \caption{Power spectrum of gravitational waves produced at the first-order confining phase transition of the $\Sp(2)$ model. The dashed, solid, and dotted concave black curves respectively represent the contributions from gravitational waves produced through bubble collisions, sound waves, and hydrodynamic turbulence, as discussed in the text, \revision{and using the estimate $\alpha \simeq 0.62(4)$ for the transition-strength parameter}. In turn, the dashed red, dotted green, and dash-dotted purple curves denote the expected \revision{peak-integrated} LISA, DECIGO, and BBO sensitivities~\cite{Schmitz:2020syl}.}
    \label{fig:spectrum}
\end{figure}

Following ref.~\cite{Schwaller:2015tja}, the peak frequency of the power spectrum associated with gravitational waves produced through bubble collisions in our model (for a confining transition at a temperature of the order of $500$~GeV) can be estimated to be around $1.8\cdot 10^{-4}$~Hz. Using the value of the radiation density $h^2 \Omega_{\mbox{\tiny{r}}0}=2.47 \cdot 10^{-5}$~\cite{Lahav:2024npe}, one obtains the contribution to the gravitational-wave power spectrum displayed in fig.~\ref{fig:spectrum}, alongside those from sound waves and from hydrodynamic turbulence, that we address next.

{\bf Gravitational waves from sound waves.} To discuss the spectrum of gravitational waves produced from sound waves in our $\Sp(2)$ theory, one can follow the analysis presented in ref.~\cite{Morgante:2022zvc} for a dark $\SU(3)$ Yang--Mills model, which, in turn, relies on the formalism summarized in ref.~\cite{Caprini:2015zlo} (see also refs.~\cite{Hindmarsh:2013xza, Giblin:2013kea, Giblin:2014qia}).

The bubble generation and growth process involves two characteristic temperatures: the nucleation temperature and the percolation temperature. The former represents the temperature at which, on average, one bubble per Hubble time forms within a Hubble volume, while the latter can be defined as the temperature at which the probability\footnote{Fixing the precise numerical value of this probability is somewhat arbitrary, and different choices are possible~\cite{Enqvist:1991xw, Ellis:2018mja, Croon:2020cgk}, but this ambiguity does not have a large impact on the estimate of the percolation temperature.} of the system being in the confining vacuum (corresponding to the disappearance of supercooled regions of deconfined vacuum, that is, to the end of the phase transition) becomes comparable to one. As will be discussed below, it turns out that both the nucleation temperature and the percolation temperature are close to $\Tc$; a heuristic argument suggesting this is based on the observation that the nucleation rate is exponentially suppressed when the temperature is decreased~\cite{Linde:1981zj}.

Other important parameters relevant for the gravitational-wave spectrum include the inverse of the product of the phase-transition duration times the Hubble parameter at the time when the gravitational waves are produced, $1/(\tpt H_\star)$, as well as the ratio of the vacuum energy density released in the confining transition to the energy density of the deconfined plasma, denoted by $\alpha$. Note that the $1/(\tpt H_\star)$ ratio can be estimated evaluating $-\frac{d \ln \Gamma}{d \ln T}$ at the percolation temperature, where $\Gamma$ is related to the Euclidean action associated with a spherically symmetric bubble configuration at finite temperature; the larger the value of $\tpt H_\star$, the stronger the phase transition and the gravitational-wave signal. The $\alpha$ parameter, on the other hand, encodes the fraction of energy released during the phase transition that can be converted into fluid motion in the plasma, and can also be interpreted as a parameter describing the transition strength. It can be defined as
\begin{align}
\label{alpha_definition}
\alpha = \frac{4}{3} \cdot \frac{\theta_+ - \theta_-}{w_+},
\end{align}
where $\theta=\Delta/4$, while $w=\epsilon+p$ denotes the enthalpy density, and the $+$ and $-$ subscripts respectively indicate that the quantity is evaluated just in front or just behind the bubble wall. In our setup, and under the assumption of small supercooling, one can take the values in the neighborhood of the transition temperature, in the deconfined (high-temperature) or in the confining (low-temperature) phase as proxies for these quantities. Note that, using the continuity of the pressure, eq.~\eqref{alpha_definition} can be rewritten as
\begin{align}
\label{alpha_definition_bis}
\alpha = \frac{\frac{\Lh}{\Tc^4}}{3\left(\frac{s}{T^3}\right)_+}.
\end{align}
The results of our lattice simulations indicate that $\alpha \simeq 0.62(4)$ for the $\Sp(2)$ Yang--Mills theory. Even though the error budget of this quantity involves systematic uncertainties that are non-trivial to quantify, taken at face value, this value would correspond to a transition that, in the terminology of ref.~\cite{Cutting:2019zws}, could be described as ``intermediate'' to ``strong''. For comparison, the value of the same parameter for the $\SU(3)$ gauge theory is approximately $0.343$~\cite{Morgante:2022zvc}.

Finally, another parameter of relevance for the study of gravitational waves produced through sound waves generated after bubble collisions is the speed of propagation of the bubble wall, evaluated in the rest frame of the fluid, at a sufficiently large distance from the bubble~\cite{Steinhardt:1981ct}, that we denoted as $\vw$. For bubbles expanding in a thermal medium, such speed is expected to be limited by frictional forces and to be associated with non-trivial effects~\cite{Liu:1992tn, Moore:1995ua, Moore:1995si, Baker:2019ndr, Hoche:2020ysm, Azatov:2020ufh, Baldes:2020kam, Janik:2022wsx, DeCurtis:2022hlx, Laurent:2022jrs, Krajewski:2023clt, Baldes:2023cih, Branchina:2025jou, Heikinheimo_et_al_in_preparation}.

The power spectrum of gravitational waves produced through sound waves can be written as 
\begin{align}
\label{soundwave_power_spectrum}
h^2\Omega^{\mathrm{(sw)}}_{\mathrm{gw}}=2.061 h^2 \Fgwzero  K^{2}\left( \Hn R_{\star} \right) \tilde{\Omega}_{\mathrm{gw}} C(f) \Upsilon(y) \Sigma ,
\end{align}
where $\Fgwzero$ is the attenuation factor 
\begin{align}
\Fgwzero = 3.57(5) \cdot 10^{-5} \cdot \left( \frac{100}{g_{\star,s}} \right)^{\frac{1}{3}},
\end{align}
where $K$ is the kinetic-energy fraction of the plasma around the expanding bubbles, which can be defined from the ratio of the mean value of $w \gammaf^2 \vf^2$ (with $\vf$ the spatial velocity of the fluid, and $\gammaf$ the associated relativistic factor) over the mean energy density of the fluid, and can be estimated from the hydrodynamics of expanding spherical bubbles~\cite{Kamionkowski:1993fg, Espinosa:2010hh}. $K$ is computed as a function of the $\Hn R_{\star}$ product~\cite{Giese:2020znk}, where $\Hn$ represents the Hubble parameter at the bubble nucleation time, while $R_\star$ is the mean bubble separation, for which we use $R_\star = 2 \sqrt[3]{\pi}\vw \tpt$~\cite{Caprini:2019egz}. For the $\tilde{\Omega}_{\mathrm{gw}}$ constant, we use the numerical estimate $\tilde{\Omega}_{\mathrm{gw}}=10^{-2}$ from ref.~\cite{Hindmarsh:2017gnf}, from which we also take the form of the  spectral shape function
\begin{align}
C(f) = \left(\frac{f}{\fpzero}\right)^3\left[\frac{7}{4+3\left(\frac{f}{\fpzero}\right)^2}\right]^{\frac{7}{2}},
\end{align}
with the peak frequency
\begin{align}
\label{peak_frequency}
\fpzero \simeq 26.2 \left( \frac{1}{\Hn R_\star} \right) \left( \frac{\Tn}{100~\text{GeV}} \right) \left( \frac{g_{\star,s}}{100} \right)^{\frac{1}{6}} 10^{-6}~\text{Hz} . 
\end{align}
The $\Upsilon(y)$ term in eq.~\eqref{soundwave_power_spectrum} is related to the finiteness of the sound-wave lifetime~\cite{Guo:2020grp} and is defined as
\begin{align}
\Upsilon(y)=1-\frac{1}{\sqrt{1+2y}};
\end{align}
the argument of this function is the sound-wave lifetime in units of the Hubble time, $y=\Hn R_\star/\sqrt{K}$; we take this quantity to be of the order of $10^{-2}$. The last factor on the right-hand side of eq.~\eqref{soundwave_power_spectrum}, $\Sigma$, accounts for attenuation effects caused by reheating~\cite{Cutting:2019zws}: it depends on the wall velocity and $\alpha$ and on $\vw$, and, in the most \revision{unfavorable} cases, can take values as small as $10^{-3}$. For the relatively large value of $\alpha$ for our model, assuming intermediate values of $\vw \simeq 0.5$, we take $\Sigma \simeq 0.3$ as a crude estimate of this parameter.

{\bf Gravitational waves from turbulence.} Finally, the contribution to the spectrum of gravitational waves due to hydrodynamic turbulence can be estimated as~\cite{Caprini:2009yp, Schwaller:2015tja}
\begin{align}
\label{turbulence_spectrum}
h^2 \Omega^{\mathrm{(ht)}}_{\rm GW} & \simeq \frac{8}{\pi^6} h^2 \Omega_{r0} \Theta^{\mathrm{(ht)}} {\cal H}_\star \tpt \Omega_{S\star}^{3/2} v^4 \, \frac{ (k\tpt)^3}{\left(1+\frac{4 k}{{\cal H}_\star}\right)\left(1+\frac{vk\tpt}{\pi^2} \right)^{11/3}},
\end{align}
where, as in eq.~\eqref{bubble-collision_spectrum}, we explicitly included an effective suppression factor, set to $\Theta^{\mathrm{(ht)}}=10^{-2}$.

The gravitational-wave power spectra associated with bubble collisions, with sound waves, and with hydrodynamic turbulence are shown in fig.~\ref{fig:spectrum}, where they are compared with the expected \revision{peak-integrated} sensitivities of the LISA, DECIGO, and BBO observatories, as estimated in ref.~\cite{Schmitz:2020syl}. The plot shows that the dominant contribution is expected to be the one resulting from sound waves: for the model under consideration (with the chosen parameters), this contribution to the gravitational-wave background peaks in the mHz frequency, and is predicted to be experimentally detectable. Conversely, the contributions from bubble collisions and from turbulence are expected to be subleading: they may evade detection by LISA, but could be within reach of DECIGO and BBO.

Before continuing the discussion of the confining phase of our model, it is appropriate to make some remarks about the parameters involved in the tentative estimate of the gravitational-wave power spectra produced at the confining transition of the $\Sp(2)$ model. In general, the non-perturbative determination of quantities related to time-dependent phenomena and/or beyond thermodynamic equilibrium from lattice simulations is possible, but it requires techniques beyond those used for the equation of state. While Monte~Carlo simulations of real-time dynamics would be hampered by a numerical sign problem, one can resort to an alternative method, focused on the study of a low-energy effective field theory describing the long-wavelength modes only\footnote{A similar approach was proposed in ref.~\cite{Caron-Huot:2008zna} and later implemented in refs.~\cite{Panero:2013pla, Moore:2019lgw, Moore:2021jwe} to study the transverse-momentum broadening associated with the jet quenching phenomenon in the QCD plasma.} through Langevin equations~\cite{Bodeker:1998hm}: this approach was used in refs.~\cite{Moore:2000jw, Gould:2022ran} to study the bubble nucleation rate (and the duration of the phase transition, which can be obtained from the latter via eq.~\eqref{tpt_and_bubble_nucleation_rate}) at the electroweak phase transition. As discussed in ref.~\cite{Schwaller:2015tja}, the value of the gravitational-wave frequency corresponding to the maximum in signal intensity observable today depends on $\tpt$, as does the value of the signal-intensity maximum itself.

Cosmological first-order phase transitions and the related bubble-nucleation processes have been studied extensively since the 1980's~\cite{Olive:1980dy, Suhonen:1982ee, Hogan:1983ixn, Gyulassy:1983rq, DeGrand:1984uq, Bonometto:1984hh, Witten:1984rs, Kurki-Suonio:1984zeb, Applegate:1985qt, Kajantie:1986hq}. To obtain an approximate quantitative description of bubble nucleation at the phase transition, one can follow the same arguments as in ref.~\cite{GarciaGarcia:2015fol} (that we reproduce here, for the readers' convenience), applying them to the $\Sp(2)$ theory. For the $\SU(3)$ theory, the latent heat and the interface tension $\sigmacd$ between the deconfined phase and the confining phase were estimated in ref.~\cite{Lucini:2005vg}, where it was found that $\Lh/(\sigmacd\Tc) \simeq 70$; one can assume that this ratio takes a similar value in the $\Sp(2)$ theory, too. This order-of-magnitude assumption is corroborated by the observation that, combining our result for the latent heat with the result for the $\sigmacd/\Tc^3$ ratio in the $\Sp(2)$ theory, as obtained from simulations at a finite lattice spacing corresponding to $\Ntau=4$, that was reported in ref.~\cite{Bennett:2024bhy}, namely $(\sigmacd/\Tc^3)_{\Ntau=4}=0.01718(32)$, one obtains a roughly comparable estimate for the $\Lh/(\sigmacd\Tc)$ ratio. Even though the result for the $\sigmacd/\Tc^3$ reported in ref.~\cite{Bennett:2024bhy} is not extrapolated to the continuum limit, and corrections due to the finiteness of the lattice spacing in simulations with $\Ntau=4$ may be non-negligible, we take this as an indication of the validity of our assumption about the value of the $\Lh/(\sigmacd\Tc)$ ratio in the $\Sp(2)$ theory.

Assume that, while the temperature decreases, the deconfined phase supercools down to a dark-sector temperature $T=(1-\delta)\Tc$, with $\delta$ a small, positive real number.\footnote{A small value of $\delta$ was indeed suggested, for example, by early studies modelling the cosmological change of phase in the strong-interaction sector of the Standard Model as a first-order transition~\cite{Ignatius:1993qn, Ignatius:1994fr}.} Then, if the confining phase sets in within a small region of space (a bubble), this region can start growing, as long as the force due to the difference in pressure overcomes the force due to the interface tension between the two phases. For a spherically symmetric bubble, this occurs when its radius $r$ exceeds a critical value $\rc=2\sigmacd/(\Delta p)$, where $\Delta p$ denotes the difference between the pressure $\pc(T)$ of the confining phase at temperature $T$ and the pressure of the deconfined phase $\pd(T)$, at the same temperature. Note that, since $\delta$ is small, $\Delta p$ can be approximated by working out the Taylor expansions of $\pc(T)$ and $\pd(T)$ around $\Tc$ at the leading non-trivial order:
\begin{align}
\Delta p &= \pc(T)-\pd(T) =\pc(\Tc-\delta \Tc)-\pd(\Tc-\delta \Tc) \nonumber \\
&=\pc(\Tc)-\delta \Tc\left.\frac{\partial \pc}{\partial T}\right|_{T=\Tc} - \pd(\Tc)+\delta \Tc\left.\frac{\partial \pd}{\partial T}\right|_{T=\Tc} + O\left(\delta^2\right)\nonumber \\
&=\delta \Tc \left[ \entropydensityd(\Tc) - \entropydensityc(\Tc) \right] + O\left(\delta^2\right) = \delta \Lh + O\left(\delta^2\right),
\end{align}
in which $\entropydensityc$ and $\entropydensityd$ respectively denote the entropy densities in the confining and in the deconfined phases, and where we used the thermodynamic definition of the entropy $S$ in the canonical ensemble as $S=-\frac{\partial F}{\partial T}$, the identity~\eqref{pressure_free-energy_density}, which holds in the thermodynamic limit, the continuity of the pressure at the critical temperature, and the fact that the latent heat per unit volume equals the product of $\Tc$ times the discontinuity in entropy density at criticality. Accordingly, the free-energy cost of a bubble of radius $r=\rc$ of confining phase within a supercooled deconfined phase can be approximated as
\begin{align}
\label{free-energy_cost_of_bubble}
\Delta F &= 4 \pi \rc^2 \sigmacd - \frac{4}{3}\pi \rc^3 \Delta p = \frac{4}{3}\pi \rc^2 \sigmacd = \frac{16}{3}\pi \frac{\sigmacd^3}{(\Delta p)^2} \simeq \frac{16}{3}\pi \frac{\Lh}{\Tc^3 \xi^3 \delta^2},
\end{align}
where we denoted the $\Lh/(\sigmacd \Tc)$ ratio for the $\Sp(2)$ theory as $\xi$. Note that in the denominator of the last fraction in eq.~\eqref{free-energy_cost_of_bubble} there is a potential competition (or at least partial compensation) between the third power of the large term $\xi$ and the square of the small parameter $\delta$. Using our estimate for $\Lh/\Tc^4$ given in eq.~\eqref{continuum_estimate_of_Lh_over_Tc4}, and working under the aforementioned assumption $\xi \simeq 70$,
eq.~\eqref{free-energy_cost_of_bubble} yields 
\begin{align}
\label{numerical_free-energy_cost_of_bubble_over_T}
\frac{\Delta F}{\Tc} \simeq 8.3 \cdot 10^{-5} \cdot \frac{1}{\delta^2}.
\end{align}
Note that the dimensionless numerical prefactor on the right-hand side of eq.~\eqref{numerical_free-energy_cost_of_bubble_over_T} is larger than, albeit of the same order of, the estimate of the same quantity reported in ref.~\cite{Schwarz:2003du} for the case of purely gluonic $\SU(3)$ Yang--Mills theory.

Recalling that the bubble nucleation rate per unit time and unit volume can be related to the free-energy cost of a bubble of critical radius via~\cite{Csernai:1992tj}
\begin{align}
\Gamma \simeq \Tc^4 \exp\left( -\frac{\Delta F}{\Tc}\right)
\end{align}
(for a detailed discussion, see also ref.~\cite[section~9.2]{Laine:2016hma}), one finds that $\Gamma \simeq H^4$ (corresponding, on average, to one bubble nucleation per Hubble time taking place within one Hubble volume) when
\begin{align}
\label{DeltaF_over_Tc_vs_4ln}
\frac{\Delta F}{\Tc} \simeq 4 \ln\left[ \frac{\Tc}{H(\Tc)} \right].
\end{align}
It is interesting to consider the implications of eq.~\eqref{DeltaF_over_Tc_vs_4ln} for the (unphysical) case of first-order transitions in the Standard Model, through a back-of-the-envelope calculation based on the values for the Hubble radius reported in ref.~\cite{Schwarz:2003du}. If the electroweak crossover were a first-order transition, then a crude estimate for the logarithm on the right-hand side of eq.~\eqref{DeltaF_over_Tc_vs_4ln} would be of the order of $37$; similarly, if the QCD crossover were a first-order transition, then the logarithm on the right-hand side of eq.~\eqref{DeltaF_over_Tc_vs_4ln} would be of the order of $44$. Thus, in both cases the $\frac{\Delta F}{\Tc}$ ratio would be rather large, corresponding to small values of $\delta$, of the order of $O(10^{-4})$, implying that even for modest supercooling the heat released by the formation and expansion of bubbles would increase again the temperature above the critical one, unless the transition could be completed within a very small fraction of a Hubble time\footnote{For a first-order transition in the strong-interaction sector of the Standard Model, this could happen if the heat capacity of the quark-gluon plasma was sufficiently large, which would allow the supercooled deconfined medium to absorb the heat released by a bubble in its expansion.}~\cite{GarciaGarcia:2015fol}.

This reasoning, however, cannot be directly extended to our dark $\Sp(2)$ glueball model, given that \emph{a priori} there is no obvious reason to assume that its temperature coincides with the one of the visible sector, and thus, in particular, there is no compelling argument allowing one to derive an estimate for the right-hand side of eq.~\eqref{DeltaF_over_Tc_vs_4ln}. As a consequence, if supercooling down to temperatures not too close to $\Tc$ could take place in our model, then a non-negligible amount of entropy could be produced during the bubble-growth process, in which the system is out of thermodynamic equilibrium. This could potentially lead to a detectable effect on relic densities.\footnote{The non-equilibrium nature of bubble nucleation and expansion can be related with a wide variety of interesting phenomenological implications~\cite{Schramm:1984bt}. For example, in refs.~\cite{Azatov:2020ufh, Azatov:2021ifm, Baldes:2021vyz} it was pointed out that, when the bubble wall velocity is sufficiently large, a non-thermal dark-matter production mechanism can set in, which could lead to a large relic abundance of particles with masses much larger than the scale of the transition, or to a non-trivial baryon-asymmetry generation.} Note that in this respect our model is different from the ``twin $\SU(3)$'' model that was discussed in ref.~\cite{GarciaGarcia:2015fol}.

{\bf Glueballs in the confining phase as dark matter.} At temperatures lower than $\Tc$, the thermodynamics of the $\Sp(2)$ gauge theory can be modelled in terms of a gas of massive glueballs: after confinement, a freeze-out process takes place, leading to a gas dominated by the contribution of the lightest state in the spectrum, i.e., the scalar glueball. If the dynamics of the latter can be described in terms of an effective Lagrangian of the form
\begin{align}
\label{effective_Lagrangian_for_lightest_glueball}
\mathcal{L}=\frac{1}{2}(\partial_\mu \phi)(\partial^\mu \phi) - \frac{1}{2}m^2 \phi^2 - \sum_{n=3}^\infty \frac{1}{n!} \lambda_n \phi^n ,
\end{align}
which accounts for number-changing glueball self-interaction processes, like $3 \to 2$ scatterings, then the number density of the glueball is expected to decrease as a function of time. While the $\lambda_n$ coefficients appearing in eq.~\eqref{effective_Lagrangian_for_lightest_glueball} are \emph{a priori} unknown, one can extract information about them by simple arguments. In particular, from na\"{\i}ve dimensional analysis follows that the energy dimension of the generic $\lambda_n$ coefficient is $4-n$. A slightly less trivial observation is that, since $\phi$ represents a glueball state in a Yang-Mills theory in which the ``color'' multiplicity of the dark gluons is $10$, one could invoke large-$N$ arguments~\cite{tHooft:1973alw} (see also ref.~\cite[section~2.2]{Lucini:2012gg}) to argue that $\lambda_n=O(N^{2-n})$, so that the dominant glueball self-interaction processes are encoded in the cubic term appearing in eq.~\eqref{effective_Lagrangian_for_lightest_glueball}. As long as the self-interaction rate is larger than the expansion rate of the universe, the comoving entropy density of the gas of dark glueballs is conserved. Noting that the results for the glueball masses in table~\ref{tab:glueball_masses} and the value of the $\frac{\Tc}{\sqrt{\sigma}}$ ratio reported in ref.~\cite{Holland:2003kg} indicate that in the confining phase one can use a non-relativistic approximation for the equation of state, and including only the leading contribution for the lightest glueball (denoting its mass as $m$), the entropy density reduces to
\begin{align}
\label{approximate_entropy_density}
s \simeq m^2 \sqrt{\frac{mT}{8\pi^3}} \exp\left(-\frac{m}{T}\right).
\end{align}
As a consequence, entropy conservation in a comoving volume $a^3$ implies that the temperature of dark matter decreases only logarithmically with $a$:
\begin{align}
T \simeq \frac{m}{3\ln(a/a_0)},
\end{align}
where $a_0$ is a constant~\cite{Carlson:1992fn}. Note that this can lead to a difference in temperature between the dark and the visible sectors.

Kinetic equilibrium implies that, for a generic glueball of mass $m$ at temperature $T$ and for chemical potential $\mu$, the particle density is expected to be
\begin{align}
\label{particle_density}
n = \frac{Tm^2}{2 \pi^2} \sum_{l=1}^\infty \frac{1}{l} \exp\left(\frac{l \mu}{T} \right) K_2\left( \frac{l m}{T} \right).
\end{align}
In the non-relativistic limit eq.~\eqref{particle_density} reduces to
\begin{align}
\label{non-relativistic_particle_density}
n \simeq \left( \frac{mT}{2 \pi} \right)^{3/2} \exp\left(-\frac{m-\mu}{T} \right).
\end{align}

Depending on phase space and kinematics, at this stage heavier dark glueballs may decay to lighter ones, until a freeze-out condition is reached. In addition, even in the absence of direct couplings to SM fields (a particularly intriguing scenario, in which the characteristic mass scale of dark matter could evade current lower bounds from collider experiments~\cite{Pospelov:2008zw, Bjorken:2009mm}), the dark sector can be connected to the visible sector via non-renormalizable operators~\cite{Juknevich:2009ji, Juknevich:2009gg}, which can transfer energy and possibly allow dark glueballs (at least those in some channels) to decay. The existence of such ``connector operators'' is generic, as some degree of mixing between the visible and dark sectors is expected to be induced at least by quantum-gravity effects, but in this case the mixing between sectors would be suppressed by some power of the ratio of the dark confinement scale to the Planck scale~\cite{Giddings:1988cx}. In particular, dark glueballs are expected to decay to gravitons, with a decay to two gravitons being the dominant channel, but the corresponding decay rate would scale proportionally to the inverse of the fourth power of the Planck scale and like the fifth power of the glueball mass. Following the arguments of ref.~\cite{Soni:2016gzf}, this would imply a glueball lifetime exceeding the age of the universe for glueball masses below $10^7~\GeV$. By contrast, a stronger mixing could appear in the presence of additional fields, directly coupled to both sectors. Depending on the strength of the couplings of the connector operators, dark glueballs decaying to the visible sector could then yield potentially observable contributions to the electromagnetic radiation emitted by astrophysical sources~\cite{Cohen:2016uyg}, to the cosmic microwave background~\cite{Chen:2003gz}, and to Big-Bang nucleosynthesis~\cite{Kawasaki:2004qu, Jedamzik:2006xz, Iocco:2008va}.

If the mixing between the visible and dark sectors due to connector operators is negligible, then the two sectors can evolve with different temperatures. In particular, the relic dark matter abundance may have been determined by asymmetric reheating, whereby the same mechanism that sets the SM abundances in the early universe is at play for the dark sector, too, but at a different temperature~\cite{Adshead:2016xxj}. Generalizing the discussion for a dark $\SU(N)$ Yang--Mills theory presented in ref.~\cite{Forestell:2017wov} to the $\Sp(2)$ model, the ratio $R$ of the entropy density of the dark sector to the entropy density of the SM (whose value is determined by the details of the reheating process) is then expected to remain constant during the evolution immediately after reheating; its value is related to the temperatures of the two sectors via
\begin{align}
\label{R_at_high_temperatures}
R = \frac{20}{g_{\star S}} \left(\frac{\Tdark}{\TSM}\right)^3,
\end{align}
where $\Tdark$ and $\TSM$ respectively denote the temperature of the dark sector and the temperature of the visible sector, $g_{\star S}$ is the (effective) number of SM degrees of freedom, while $20=2 \times 10$ arises from the product of the number of ``color'' degrees of freedom for the dark $\Sp(2)$ gluons times two transverse polarizations.

Equation~\eqref{R_at_high_temperatures} is expected to hold at temperatures $\Tdark$, that are sufficiently high, and well above the temperature of the confinement/deconfinement transition of the dark sector. However, as we discussed, when the temperature of the dark sector is reduced to $\Tc$, the entropy density has a discontinuity, $s|_{T=\Tc^-}=s|_{T=\Tc^+}-\Lh/\Tc$, and the system enters the confining phase, in which its equilibrium-thermodynamics can be described in terms of a gas of glueballs, according to eq.~\eqref{glueball_gas_pressure}.

As the system evolves further, the temperature continues to decrease, with dark glueball interactions taking place; these interactions, in particular, include the annihilation of heavier glueballs, which occurs primarily through $2 \to 2$ decays~\cite{Pappadopulo:2016pkp, Farina:2016llk}, as well as the self-annihilation of the lightest glueball, e.g., via $3 \to 2$ processes. Such interactions continue until freeze-out, which takes place when the rate of $3 \to 2$ self-annihilation processes becomes too small compared to the Hubble expansion rate. From that point, the densities of the various dark glueball species remain fixed and simply dilute with the expansion of the universe, and $2 \to 2$ elastic scatterings (which, from general arguments based on effective field theory, are expected to be parametrically favored with respect to $3 \to 2$ decays) maintain kinetic equilibrium, so that stable relic densities are obtained. 

For a dark $\SU(N)$ Yang--Mills theory (with $N>2$), in ref.~\cite{Forestell:2016qhc} it was shown that the time evolution of the densities of the lightest glueball species in the spectrum, corresponding to lowest-mass states in the $J^{PC}=0^{++}$ and $J^{PC}=1^{+-}$ channels, can be approximated well by two coupled equations, relating them to each other via the thermally averaged cross sections for glueball interactions. For our $\Sp(2)$ model, however, the discussion has to be modified, as the theory features no $C=-1$ states. Restricting, for simplicity, the analysis to the lightest glueball state only, the associated particle density $n_{0^+}$ is expected to evolve in time according to
\begin{align}
\label{scalar-glueball_density_evolution}
\frac{dn_{0^+}}{dt} +3Hn_{0^+} = -\langle\sigma_{3 \to 2}v^2\rangle n_{0^+}^2(n_{0^+}-\bar{n}_{0^+}),
\end{align}
where $\bar{n}_{0^+}$ denotes the average value of the number density for vanishing glueball chemical potential, and an order-of-magnitude estimate for the thermally averaged cross section of $3 \to 2$ self-annihilation processes $\langle\sigma_{3 \to 2}v^2\rangle$ can be obtained from the effective description in terms of eq.~\eqref{effective_Lagrangian_for_lightest_glueball} and from large-$N$ arguments:
\begin{align}
\label{EFT_estimate_for_sigma32v2}
\langle\sigma_{3 \to 2}v^2\rangle \simeq \left(\frac{4\pi}{N^2}\right)^3 \frac{1}{m^5},
\end{align}
where we neglected coefficients that are expected to be of order one, and where the $N^2$ term can be traded for the ``color'' multiplicity of the dark gluons, which is $10$ in the $\Sp(2)$ theory.

The temperature at which the rate for $3 \to 2$ decays becomes less than the fractional rate of change of $n_{0^+} a^3$ defines the freeze-out temperature for the lightest glueball~\cite{Carlson:1992fn}:
\begin{align}
\label{freeze-out_temperature}
\Tfo \simeq \frac{m_{0^+}}{3H} \langle\sigma_{3 \to 2}v^2\rangle \bar{n}_{0^+}^2.
\end{align}
In turn, eq.~\eqref{freeze-out_temperature} can be related to the temperature of the visible sector, by combining it with the $R$ ratio introduced in eq.~\eqref{R_at_high_temperatures}. In particular, if one assumes that at the time of dark-matter freeze-out the largest contribution to the total energy density comes from visible radiation, and uses the expression of $\bar{n}_{0^+}$ in its non-relativistic limit from eq.~\eqref{non-relativistic_particle_density} as well as the ratio of entropy densities in the dark and visible sectors denoted as $R$, one can follow the same reasoning that in ref.~\cite{Forestell:2016qhc} was applied to a dark $\SU(3)$ glueball model to compute the mass-weighted relic yield. Note that the latter is a monotonically increasing function of $R$ and of the dark-glueball mass, both of which can in principle vary over several orders of magnitude; however, their combination (or, equivalently, the $R\Tfo$ product) can be constrained using the updated experimental value reported in ref.~\cite{Planck:2018vyg} for the dark-matter density parameter:
\begin{align}
\label{OmegaDM_and_RTfo}
\OmegaDM h^2 = 0.1200(12) \cdot \left( 2.314 \cdot 10^9 \right) \frac{R\Tfo}{\GeV}.
\end{align}

In passing, it is worth noting that refs.~\cite{Frey:2024jqy, Villa:2024jbf} discussed the spectrum of gravitational waves that one can obtain in a theory with a Hagedorn-like spectrum. This may be of relevance for our model, too, since our results for the equation of state in the confining phase do provide evidence for an exponential growth of the density of states at temperatures close to $\Tc$.

{\bf Comments on the \revision{parameter space of the model}.} Finally, we point out some comments on \revision{the parameter space of the theory, including, in particular, on the  possible values of} the temperature of the dark sector, which is directly relevant for the existence of a phase transition, for the details of the spectrum of gravitational waves that the latter can produce, as well as for the freeze-out yields of candidate dark-matter states. The issue was discussed in ref.~\cite{Kolesova:2023yfp}, which pointed out various important aspects. Firstly, the very definition of a ``temperature'' for the dark sector implicitly assumes that thermodynamic equilibrium can be reached; for a non-Abelian gauge theory, the equilibration rate is expected to be parametrically of the form $g^2 T$, and this quantity should be larger than the Hubble rate. In turn, the latter is determined by the square of the largest, between the temperature of the visible sector and the temperature of the dark sector, divided by the Planck mass, so that the dark sector can reach thermal equilibrium for temperatures up to $g^2 \mPlanck$, with $\mPlanck$ being the Planck mass. Secondly, it was shown that more constraining bounds on the allowed temperatures for the dark sector can be obtained, by assuming a ``natural inflation'' scenario~\cite{Freese:1990rb}, whereby an axion-like inflaton (with a simple cosine potential with parameters determined from experimental constraints~\cite{Planck:2018jri}) directly couples to the dark sector: the reasoning starts from the observation that the smallness of equilibrium thermodynamic functions in the confining phase implies that even a modest energy transfer from the inflaton field can lead to a significant increase of the dark-sector temperature. By solving the equations determining the time evolution of the plasma temperature and of the average value of the inflaton, the authors of ref.~\cite{Kolesova:2023yfp} found that if the ratio of the confining scale of the dark sector to $\mPlanck$ is $O(10^{-8})$ or larger, then the deconfinement temperature may not be accessible to the dark sector. Thirdly, they found that, for sufficiently high temperatures, the thermal, monotonically increasing intermediate-frequency contribution to the gravitational-wave background discussed in ref.~\cite{Klose:2022rxh} may be close to the detectability range of forthcoming experiments. Finally, they also discussed the gravitational waves produced at the confinement/deconfinement phase transition, finding that the peak frequencies predicted by their calculations could be observable by the Einstein Telescope and/or DECIGO.\footnote{A further, detailed discussion about this subject was later presented in ref.~\cite{Kolesova:2023mno}.} While the discussion of ref.~\cite{Kolesova:2023yfp} was carried out for a dark $\SU(3)$ Yang--Mills theory, in view of the many similarities with the $\Sp(2)$ theory, one may argue that their results can be expected to be qualitatively (and probably also semi-quantitatively) relevant for our model, too.

Further implications of a dark $\SU(N)$ theory coupled to an axion-like inflaton, resulting in the excitation of the Yang--Mills fields already in the slow-roll inflationary phase (the ``minimal warm inflation'' scenario~\cite{Berghaus:2019whh, Kamali:2023lzq}), were later analyzed in ref.~\cite{Biondini:2024cpf}. In particular, this work studied in detail the cosmological evolution, from inflation to SM nucleosynthesis, of a dark sector that gets populated first by inflaton decays, and then, in turn, leads to production of particles of the Standard Model through portal interactions~\cite{Berlin:2016vnh, Tenkanen:2016jic, Biondini:2020xcj}. For an $\SU(N\ge 3)$ theory, the authors of ref.~\cite{Biondini:2024cpf} demonstrated that the lightest glueball state with charge-conjugation quantum number $C=-1$ could be a viable dark-matter candidate, with appropriate relic abundance, and with a correct reheating temperature. A key insight leading to these findings is that, while the lightest $C$-even state can couple to the Standard Model Higgs field through a dimension-six operator, the lightest $C$-odd state can only decay via operators of dimension at least equal to eight (for related work, see also refs.~\cite{Redi:2020ffc, Gross:2020zam, McKeen:2024trt}). This implies that the decay width of the axial-vector $C=-1$ glueball to a dark $J^{PC}=0^{++}$ glueball and to electroweak bosons is strongly suppressed, thus the lightest $C$-odd state can have a lifetime longer than the age of the universe. This mechanism, whereby the stability of the dark-matter candidate on cosmological time scales is protected by charge conjugation, is very appealing, and, as discussed in ref.~\cite{Biondini:2024cpf}, leads to very interesting phenomenology, reconciling the requirements for dark matter with those from Standard-Model nucleosynthesis. It should be noted that, unfortunately, this mechanism cannot be at work in our model, which, being based on an $\Sp(N)$ group, cannot accommodate any $C$-odd state.

\revision{A different situation can be realized under the hypothesis that the dark sector is not reheated by the decay of the inflaton, and is instead populated only by gravitational interactions with the visible sector at very high temperatures, deep in the deconfined phase~\cite{Garny:2015sjg, Tang:2016vch}. In particular, it was shown in ref.~\cite[section~3]{Redi:2020ffc} that in this scenario the problem of dark-matter overproduction can be evaded.

In general, understanding whether the parameter space of a certain theory is consistent with observational constraints (in particular those from Big-Bang nucleosynthesis and from the cosmic microwave background) is a central question of any particle dark matter model. For the gauge theory that we considered, we can distinguish the relevant parameters into two groups:
\begin{itemize}
\item parameters related to general, ``static'' or ``equilibrium'' properties of the theory: these include, in particular, the masses of physical states in the confining phase, the deconfinement temperature, the latent heat, etc.;
\item parameters related to ``dynamical'' phenomena: examples of these include, for instance, glueball decay and/or annihilation rates.
\end{itemize}
While the lattice calculations that we presented in this work allow one to determine unambiguously, in a gauge-invariant, fully non-perturbative and systematically improvable way all of the parameters within the first group (except one, which has to be used to set the scale), they do not give us direct access to those in the second group, for which we have to rely on general arguments or on crude tentative estimates.

As we remarked above, dark matter contributes to the total energy density of the universe, and thus may affect, in particular, the primordial deuterium and helium-$4$ yields; at the same time, it may also affect the cosmic microwave background, since it can modify the tail of temperature-polarization spectra. It should be noted that relativistic particle species dominate the early‑Universe expansion, as their contribution to the energy density is proportional to the fourth power of the temperature. This case is relevant in the high-temperature regime of the $\Sp(2)$ model that we discussed here, in which our lattice results for the equation of state show that the energy density scales approximately like the fourth power of the temperature (up to a further, mild dependence on the temperature, which accounts for the fact that, for temperatures not much higher than the deconfinement temperature, the ``gluon-like'' quasiparticles in the deconfined phase are not completely free; rather, the coupling describing their mutual interactions is expected to scale with the inverse of the logarithm of the temperature). Conversely, the contribution to the energy density from massive species is exponentially suppressed: when the temperature drops below the particle mass, the particle becomes non-relativistic, and the kinetic energy can be neglected with respect to the rest-mass energy: this implies that the particle's contribution to the energy density becomes proportional to the exponential of the ratio of the particle mass to the temperature, times a power of the temperature---see also the expression for the number density in eq.~\eqref{non-relativistic_particle_density}. As we discussed at the beginning of section~\ref{sec:results}, for the $\Sp(2)$ gauge theory the masses of all physical states in the spectrum are significantly larger than the critical temperature at which the theory enters its confining phase and develops a mass gap: this implies that, as soon as the temperature of the theory drops below $\Tc$, the mass-to-temperature ratio is quite large (larger than $5$ for the lightest glueball species), and thus the energy-density contribution from the dark sector becomes strongly suppressed.

A concrete example of a confining dark sector, in which the problem of dark-matter overproduction can be evaded, was presented in ref.~\cite{Redi:2020ffc}: while the focus of that work was on a dark $\SU(3)$ Yang--Mills theory, we note that most of their arguments apply in the case of a dark $\Sp(2)$ gauge theory, too, up to small quantitative changes: for example, the central charge of the theory in the plasma phase is simply rescaled by a factor equal to the ratio of the number of gluon degrees of freedom in the two theories, i.e., $10/8=1.25$, and similar changes can be applied for the quantities relevant for the glueball phase, like the value of the mass of the lightest glueball and the latent heat in the appropriate units of the deconfinement temperature. In particular, it is interesting to discuss two different situations, namely the case in which the number density of glueballs produced at the confining transition is not significantly modified by subsequent number-changing processes, and the case in which, instead, $3 \to 2$ processes alter it: since in this case dark matter particles gain energy ``by eating each other'', this latter scenario is often referred to as ``cannibalism''~\cite{Carlson:1992fn}.

If dark glueball interactions are sufficiently weak and one can neglect processes that change the glueball number density, the latter can be estimated from energy conservation at the phase transition (under the assumption used above, that the nucleation temperature is very close to the critical temperature). Note that, as we mentioned above, the assumption of negligible glueball interactions in a gauge theory without elementary matter fields in the fundamental representation is at least qualitatively supported by large-$N$ arguments~\cite{tHooft:1973alw}, and the most recent lattice results suggest that even the purely geometric cross-section of glueballs may be small~\cite{Abbott:2025irb}. Comparison with experimental results from the strong-interaction sector of the Standard Model, on the other hand, is of little guidance, since in QCD glueballs are affected by significant mixing with flavor-singlet mesons, and, in fact, even the proper identification of glueball states in QCD remains somewhat elusive~\cite{Mathieu:2008me, Crede:2008vw, Ochs:2013gi, Vadacchino:2023vnc}. In any case, if one neglects glueball interactions, then, following the discussion in ref.~\cite[section~3]{Redi:2020ffc} and noting how the parameters of the $\Sp(2)$ theory are related to those of the $\SU(3)$ theory, one obtains a tentative estimate for the yield:
\begin{align}
\label{yield_estimate}
Y \simeq 2 \cdot 10^{-3} \left( \frac{\Treh}{\mPlanck} \right)^{9/4},
\end{align}
where $\Treh$ denotes the reheating temperature, and a corresponding crude estimate for the dark matter abundance:
\begin{align}
\label{dark_matter_abudance_for_free_glueballs}
\frac{\Omega h^2}{0.12} \simeq \frac{m_{0^+}}{10~\mbox{GeV}} \left( \frac{\Treh}{10^{15}~\mbox{GeV}}\right)^{9/4}.
\end{align}
It should be stressed that, in view of the inherent uncertainties on the assumption of negligible glueball interactions, eq.~\eqref{yield_estimate} and eq.~\eqref{dark_matter_abudance_for_free_glueballs} should be considered as order-of-magnitude estimates only.

On the other hand, if the glueball interactions are to be taken into account, one can estimate the yield by considering the effect of $3 \to 2$ interactions, which are expected to be those responsible for the leading glueball number density changing processes. Note that, parametrically, the cross-section associated with these interactions is inversely proportional to the fifth power of the glueball mass, hence it is strongly suppressed if the dynamically generated scale of the dark sector is large. Nevertheless, $3 \to 2$ processes can be sufficiently frequent (as compared to the expansion rate of the universe) if the coefficient encoding the quantitative magnitude of these interactions and the ratio of the energy density of the dark sector to the one of the visible sector are not too small, and the glueball mass is not too large. In this case, cannibalism allows glueballs to remain in thermal equilibrium, and the temperature of the dark sector varies slowly, with the logarithm of the cosmological scale factor~\cite{Carlson:1992fn, Forestell:2017wov}, and, as discussed in detail in ref.~\cite{Redi:2020ffc}, the dark matter equilibrium yield is then linear in the temperature of the dark sector. As the temperature continues to decrease, the rate of $3 \to 2$ processes eventually drops below the Hubble rate, at which point cannibalism stops, leaving a dark matter yield directly related to the ratio of the entropy densities of the dark and visible sectors. In order to make precise quantitative predictions about this type of dynamics, it would be crucial to know exactly the value of the cross-section of $3 \to 2$ interactions. Unfortunately, however, studying this type of processes on the lattice is particularly challenging~\cite{Briceno:2017tce, Briceno:2017max, Hansen:2019nir}. At best, we can only make semiqualitative arguments, and/or rely on the expectation that estimates obtained in the case of a dark $\SU(3)$ theory could approximately hold also for the $\Sp(2)$ model. In particular, for the $\SU(3)$ theory, the matrix elements associated to the decay of the lightest glueball were studied in refs.~\cite{Chen:2005mg, Meyer:2008tr}; the relative uncertainties affecting this type of computations are in the ballpark of $20\%$, i.e., significantly larger than those of the glueball spectrum. We expect that the systematic uncertainty in the assumption that the decay matrix elements of the $\Sp(2)$ theory take values similar to those of the $\SU(3)$ does not exceed such uncertainties.

In short, the similarities between the model that we considered herein and a dark $\SU(3)$ Yang--Mills theory suggest that the conclusions reached in ref.~\cite{Redi:2020ffc} are relevant for our model, too, and that, for example, a plot of the viable values for the confining temperature of gravitationally produced $\Sp(2)$ glueball dark matter as a function of the reheating temperature of the visible sector would be quantitatively very similar to the one shown in ref.~\cite[figure~1]{Redi:2020ffc}. 

A different, and more involved, discussion would be necessary, instead, if one were to explicitly consider also the variety of scenarios by which dark glueballs can eventually decay to the Standard Model via connector operators. For a dark $\SU(3)$ Yang--Mills theory, these possibilities were discussed in detail in ref.~\cite{Forestell:2017wov}, considering, in particular, operators of different dimensions (and with broken or unbroken discrete charge-conjugation symmetry, which is not present in our model). In particular, if the lifetimes of dark glueballs are sufficiently short, their decays can induce changes in the primordial yields of light elements produced through Big Bang nucleosynthesis~\cite{Jedamzik:2006xz, Kawasaki:2017bqm}, 
they can modify the cosmic microwave background~\cite{Chen:2003gz, Slatyer:2016qyl}, and alter the spectrum of cosmic rays~\cite{Cohen:2016uyg}. In the absence of an accurate, quantitative estimate for the parameters relevant for such decays in the $\Sp(2)$ model, however, we do not make any comments on this possibility, except that, once again, we expect the overall picture to be roughly similar to the $\SU(3)$ case, except for the absence of states with negative charge-conjugation quantum number.

To summarize, while in the present work we addressed the explicit lattice calculation of a number of physical quantities for the $\Sp(2)$ theory, and could therefore derive robust predictions for the equation of state and for (at least certain aspects related to) the gravitational-wave production at the confining transition of this dark Yang--Mills model, for observables that depend on physical quantities that were not addressed in the present lattice calculation, we necessarily have to rely on qualitative or semi-quantitative estimates only, based on theoretical arguments (the large-$N$ limit and effective field theory) or from some empirical numerical evidence (including, in particular, the similar values of quantities computed from lattice simulations of the $\Sp(2)$ theory and of the $\SU(3)$ theory). While this implies that, as we discussed above, some of the expectations derived from the $\Sp(2)$ model at present remain at the level of educated guesses, and/or are meant to be more qualitative than quantitative, in principle the analysis could still be refined through a sequence of dedicated studies in a longer-term plan. Examples of quantities for which this would be in principle possible include, for instance, the matrix elements associated with glueball decays, for which one could repeat a study similar to those that were carried out for the pure-glue sector of QCD in refs.~\cite{Chen:2005mg, Meyer:2008tr}. Even this small step, however, would require significant computational time and resources, and we postpone it to the future.

}

\section{Conclusions}
\label{sec:conclusions}

The possibility that dark matter consists of bound states of a confining non-Abelian gauge theory is theoretically appealing. Beside the fact that in the Standard Model this is the same mechanism that explains the existence of the largest fraction of visible matter, in addition to the other aspects that we discussed in section~\ref{sec:introduction}, we would like to stress that a ``purely gluonic'' dark Yang--Mills theory, like the one that we considered in the present paper, can be regarded as a particularly ``economic'' model, depending on a single dimensionful parameter. As such, it has potentially strong predictive power. 

Note that an interpretation of dark matter in terms of bound states of a strongly coupled gauge theory is complementary with respect to models, in which dark matter is explained by extensions of the Standard Model through a minimal set of additional fundamental multiplets, coupled to SM particles only through gauge interactions, and with spin, isospin and hypercharge quantum numbers allowing the stability and neutrality of the lightest multiplet component, and the consistency with experimental bounds; for a systematic discussion of such models, we recommend ref.~\cite{Cirelli:2005uq}.

While in the past several works have already considered the study of a ``dark $\SU(N)$ theory'', and discussed its potential implications for the evolution of the early universe, much less attention has been devoted to dark-matter models based on confining theories constructed from other non-Abelian gauge groups. This motivated us to consider, in this work, the thermal evolution of a dark-matter model based on the $\Sp(2)$ theory: the latter shares many qualitative (and, in part, quantitative) features with those based on a special unitary group, but also differs from them under some aspects. For the purposes of this work, the most interesting differences between $\Sp(2)$ and $\SU(N)$ (for $N \ge 3$) Yang--Mills theories are probably the absence of states with a negative quantum number under charge conjugation, and the fact that the center of the group is $\Z_2$. In particular, the non-existence of $C=-1$ glueball states precludes the viability of the scenario discussed in ref.~\cite{Biondini:2024cpf}, which can simultaneously account for correct values of dark matter abundance and for a sufficiently high reheating temperature, and which has the attractive feature of ``protecting'' the existence of the lightest $C=-1$ state by a discrete symmetry. This mechanism allows the dark matter candidate to survive and be cosmologically stable, even if all other dark glueball states, including the $C=1$ glueballs, decay before primordial nucleosynthesis.

The study of the equation of state of the dark $\Sp(2)$ theory that we presented in section~\ref{sec:results} is entirely non-perturbative, and was carried out in a completely gauge-invariant approach based on first principles, in the lattice regularization~\cite{Wilson:1974sk}. We stress that this approach provides a mathematically rigorous definition of the theory, which, in contrast to phenomenological models, is free from uncontrolled approximations and is systematically improvable; in the thermodynamic and continuum limits, all discretization artifacts (including, in particular, those related to the explicit breaking of Lorentz--Poincar\'e invariance by the regularization on a grid) vanish. The motivation for carrying out the present study of the $\Sp(2)$ equation of state on the lattice is that, at the temperatures close to $\Tc$ that we considered, the physics of non-Abelian gauge theories involves non-trivial non-perturbative aspects, which cannot be captured by weak-coupling expansions. While perturbative predictions in high-temperature gauge theories can be improved by combining them with effective field theories~\cite{Braaten:1995jr, Farakos:1994xh, Braaten:1995cm, Kajantie:1995dw} constructed via dimensional reduction~\cite{Ginsparg:1980ef, Appelquist:1981vg}, and this approach is indeed actively pursued in the study of cosmological phase transitions~\cite{Croon:2020cgk}, our present study does not rely on any weak- (nor strong-) coupling approximation.

The results for the equilibrium thermodynamic quantities of the $\Sp(2)$ theory that we presented in section~\ref{sec:results} show remarkable similarities with those obtained in $\SU(N)$ gauge theories~\cite{Panero:2009tv}, and especially with those from the $\SU(3)$ theory~\cite{Boyd:1996bx, Borsanyi:2012ve}. In both cases, the confinement/deconfinement transition is of first order, with a finite, but not excessively large latent heat: in the symplectic theory, the dimensionless ratio $\Lh/\Tc^4$ takes a value that is larger than, but of the same order of, the one it has in the $\SU(3)$ theory. These similarities between the theories may be interpreted invoking large-$N$ arguments, and/or the quite similar dimensions of the underlying Lie algebras ($10$ \emph{versus} $8$). 

In section~\ref{sec:interpretation_as_a_model_for_dark_matter_and_gravitational-wave_production} we then discussed some topics relevant for the interpretation of our model as a framework for a dark-matter candidate, focusing on its evolution in the early universe. In particular, we concentrated on aspects related to phenomena beyond thermodynamic equilibrium, including some that may be of relevance for the production of gravitational waves. It is important to remark that gravitational waves generated by cosmological processes in the early universe will appear today as randomly distributed in all directions, with many unresolved sources. As a matter of fact, the number of Hubble patches at a given redshift $z$ that fit the current universe scales as $(1+z)^3$: this is a huge number for gravitational waves generated at large $z$, so the superposition of all incoming unresolved primordial gravitational waves results into a stochastic gravitational-wave background~\cite{Domenech:2021ztg}, and, depending on the temperature of the universe at the time of the gravitational wave production, the corresponding peak frequencies can be in the range accessible to different detectors. As we discussed, gravitational waves can be sourced by bubble collisions, by sound waves, or by hydrodynamic fluctuations of a plasma of particles at high temperature. The latter mechanism has been studied in detail and extended, in an approach directly based on the ``microscopic'' formulation of the Standard Model, in ref.~\cite{Ghiglieri:2015nfa}, where it was found that the energy density carried by thermally produced gravitational radiation is maximal for gravitational waves with frequencies of the order of $10^{11}$~Hz. Such frequencies are very far from the optimal operational range of present experiments, and, while this may change with future detectors~\cite{Cruise:2006zt, Tong:2008rz, Robbins:2018thb, Aggarwal:2020olq}, reaching the necessary sensitivity would likely remain a challenge. As was pointed out in ref.~\cite{Muia:2023wru}, however, the situation may be quite different in the presence of a new dark sector, like the one that we considered in the present work.

Albeit a more systematic lattice study of several physical quantities mentioned in section~\ref{sec:interpretation_as_a_model_for_dark_matter_and_gravitational-wave_production} for the $\Sp(2)$ theory remains beyond the scope of the present work (as it would require simulations based on non-trivial generalizations of those that we performed for the equation of state, and demanding a substantial amount of CPU time on supercomputers), \revision{so, as we already mentioned, the results presented in section~\ref{sec:interpretation_as_a_model_for_dark_matter_and_gravitational-wave_production} should be interpreted as benchmark examples,} we \revision{remark that our approach followed the main} arguments that had already been \revision{put forward} for dark $\SU(N)$ theories, trying to highlight analogies and differences relevant for the $\Sp(2)$ model. Our results can also be compared with those obtained using different techniques to study strongly coupled gauge theories, like the gauge/gravity duality~\cite{Maldacena:1997re, Gubser:1998bc, Witten:1998qj}: examples of such studies, with a particular focus on the spectrum of gravitational waves, include those reported in refs.~\cite{Bigazzi:2020avc, Ares:2020lbt}. All in all, our findings can be summarized by stating that the $\Sp(2)$ theory considered in this work is a viable and potentially interesting dark-matter model, and that, depending on the value of the energy scale characterizing the model, the gravitational-wave spectrum produced at its first-order deconfinement/confinement phase transition could have a potentially observable signature, which could reveal important information on the early stages of the universe evolution.

While we tried to make our numerical study as complete as was feasible, and the discussion in this paper as exhaustive as possible, there are several directions in which the present work could be generalized. Our lattice determination of equilibrium thermodynamic quantities could be refined with simulations on larger lattices (to improve the extrapolations to the thermodynamic limit) and at finer spacings (to improve the extrapolations to the continuum limit); in addition, the temperature range that was probed could be expanded. These extensions of the numerical calculations presented herein would reduce the systematic and statistical uncertainties of our results for the equation of state of the $\Sp(2)$ theory, but would require a non-negligible amount of computational resources. Even though the lattice study of systems featuring a first-order transition involves its own challenges, the technology to tackle them is now available~\cite{Bennett:2024bhy}. A different type of generalization of this work would consist in trying to carry out a non-perturbative computation of quantities beyond equilibrium thermodynamics, like the bubble nucleation rate~\cite{Moore:2000jw} or the interface tension between the confining and deconfined phases~\cite{Moore:1996bn}. While such calculations have already been performed for $\SU(N)$ gauge theories (for recent work, see refs.~\cite{Seppa:2025lud, Salami:2025iqq, Hallfors:2025key}), their extension to a symplectic Yang--Mills theory would be new. Then, it would also be interesting to repeat calculations such as those that we discussed in section~\ref{sec:interpretation_as_a_model_for_dark_matter_and_gravitational-wave_production} using the parameters that can be extracted (or, at least, estimated) from dedicated lattice simulations of the $\Sp(2)$ theory. Finally, other possible generalizations of this work could be to investigate other $\Sp(N>2)$ theories, or theories based on different gauge groups, and/or coupling the theory with dynamical matter fields. We leave all of these interesting possibilities for future research.

\acknowledgments
We thank Simone Blasi, Pierluca Carenza, Nicolao Fornengo, Mark Hindmarsh, Helena Kole\v{s}ov\'{a}, Kari Rummukainen, Gonzalo Villa, and Zhi-Wei Wang for helpful discussions. The numerical simulations were run on machines of the Consorzio Interuniversitario per il Calcolo Automatico dell'Italia Nord Orientale (CINECA) and of the European Organization for Nuclear Research (CERN). We acknowledge support from the SFT Scientific Initiative of INFN. 
At the beginning of this project, M.B. was supported by the Italian program for young researchers ``Rita Levi Montalcini''. 
The work of N.F. is supported by the UKRI Science and Technology Facilities Council (STFC) grant ST/X508834/1. 
The work of M.P. has been partially supported by the Italian PRIN ``Progetti di Ricerca di Rilevante Interesse Nazionale -- Bando 2022'', prot. 2022TJFCYB, CUP D53D23002970006. 
M.B. and M.P. have been partially supported by the Italian Research Centre in High-Performance Computing, Big Data and Quantum Computing (ICSC), funded by the European Union -- NextGenerationEU. 
The work of A.S. was supported by the STFC consolidated grant
No. ST/X000648/1.\\

\noindent {\bf Research Data Access Statement.} The data generated for this work and the analysis code can be downloaded from refs.~\cite{sp2eos_data_release} and~\cite{sp2eos_analysis_code}, respectively.\\

\noindent {\bf Open Access Statement.} For the purpose of open access, the authors have applied a Creative Commons Attribution (CC BY) licence to any Author Accepted Manuscript version arising.

\appendix

\section{Generalities about the compact symplectic group}
\label{app:generalities_about_the_compact_symplectic_group}
\renewcommand{\theequation}{A.\arabic{equation}}
\setcounter{equation}{0}

Given the $2N \times 2N$ skew-symmetric matrix
\begin{align}
\Omega=\left(
\begin{array}{cc}
\zeromat_N & \ide_N \\
-\ide_N & \zeromat_N
\end{array}
\right) = i\sigma_2 \otimes \ide_N
\end{align}
(where $\zeromat_N$ and $\ide_N$ denote the $N \times N$ null matrix and the $N \times N$ identity matrix, respectively, and $\sigma_2$ is the purely imaginary Pauli matrix), the symplectic group over a generic field $\F$, denoted as $\Sp(2N,\F)$, can be defined\footnote{Strictly speaking, the group is defined by abstraction from the set of matrices defined here.} as the set of linear transformations of the $2N$-dimensional vector space over $\F$ preserving $\Omega$, with matrix multiplication as the group product:
\begin{align}
\Sp(2N,\F)=\left\{ M \in \Mat(2N,\F): \transp{M} \Omega M = \Omega \right\}.
\end{align}
Note that the definition implies that every matrix of $\Sp(2N,\F)$ has unit determinant, and that $\Sp(2N,\F)$ is non-compact. Typically, the field $\F$ is taken to be the field of real or the field of complex numbers; accordingly, $\Sp(2N,\F)$ is a Lie group of real or complex dimension $N(2N+1)$, respectively, and its center is $\left\{\ide_{2N}, -\ide_{2N}\right\} \cong \Z_2$, so that $\Sp(2N,\F)$ is a simple Lie group.

The compact symplectic group $\Sp(N)$ can be defined as
\begin{align}
\Sp(N)=\Sp(2N,\C) \cap \U(2N).
\end{align}
Note that $\Omega$ itself is both an element of $\Sp(2N,\C)$ and a unitary matrix, and hence an element of $\Sp(N)$. Since $\Omega^{-1}=-\Omega=\transp{\Omega}=\Omega^\dagger$, it is easy to show that a generic $U\in \Sp(N)$ satisfies the condition
\begin{align}
\label{charge_conjugation_is_global_gauge_transform}
U^\star=\Omega U \Omega^\dagger,
\end{align}
which means that in an $\Sp(N)$ gauge theory charge conjugation is nothing but a global gauge transform.

Writing a generic element $U\in \Sp(N)$ in terms of $N \times N$ blocks
\begin{align}
\label{spn_matrix_in_blocks}
U=\left(
\begin{array}{cc}
A & B \\
C & D
\end{array}
\right),
\end{align}
from the requirement $U \in \Sp(2N,\C)$ follows that $\transp{A}C$ and $\transp{B}D$ must be symmetric matrices, and that $\transp{A}D-\transp{C}B=\ide_N$. On the other hand, the unitarity condition implies $AA^\dagger+BB^\dagger=CC^\dagger+DD^\dagger=\ide_N$ and $AC^\dagger=-BD^\dagger$. Combining these constraints, eq.~\eqref{spn_matrix_in_blocks} can be rewritten as
\begin{align}
\label{spn_matrix_in_blocks_bis}
U=\left(
\begin{array}{cc}
A & B \\
-B^\star & A^\star
\end{array}
\right),
\end{align}
with $AA^\dagger+BB^\dagger=\ide_N$ and $A\transp{B}$ symmetric. Note that $\Sp(1)=\SU(2)$, and indeed for $N=1$ eq.~\eqref{spn_matrix_in_blocks_bis} reduces to a familiar parameterization for $\SU(2)$ matrices in terms of a three-sphere.

For every $N$, the compact symplectic group $\Sp(N)$ is a simply connected Lie group (and, hence, is its own universal covering group). Locally, the $\Sp(N)$ group manifold can be written as the product of spheres:
\begin{align}
\Sp(N)= \prod_{k=1}^N S^{4k-1},
\end{align}
so that, in particular, $\Sp(1)=S^3=\SU(2)$, whereas the local structure of the $\Sp(2)$ manifold can be described as the $S^3 \times S^7$ product. Other interesting properties include the fact that the center of the $\Sp(N)$ group is always $\Z_2$; the fundamental group and the second homotopy group are trivial, $\pi_1(\Sp(N))=\pi_2(\Sp(N))=\Z_1$, while the third homotopy group is the group of integers: $\pi_3(\Sp(N))=\Z$~\cite{Bott:1959tsh, Mimura:1963hgo}.

Writing a generic element of the $\Sp(N)$ group as $U=\exp(iX)$, with $X$ an Hermitian and traceless $2N \times 2N$ matrix (so that $U \in \SU(2N)$), the condition~\eqref{charge_conjugation_is_global_gauge_transform} implies the additional requirement $X^\star=-\Omega X \transp{\Omega}$. Equivalently, the Lie algebra $\mathrm{sp}(N)$ of the compact symplectic group can be thought of as the set of quaternion-valued $N \times N$ matrices satisfying $X=-X^\dagger$ (where the dagger denotes the transpose of the quaternion-conjugate matrix).

The $\mathrm{sp}(N)$ algebra has dimension $N(2N+1)$ and rank $N$; it can be encoded in the Dynkin diagram
\begin{displaymath}
\begin{array}{c}
\begin{picture}(130,10)
\put(15,5){\line(1,0){20}}
\put(95,5){\line(1,0){20}}
\put(123,8){\line(1,0){24}}
\put(123,2){\line(1,0){24}}
\put(45,5){\line(1,0){2}}
\put(49,5){\line(1,0){2}}
\put(53,5){\line(1,0){2}}
\put(57,5){\line(1,0){2}}
\put(61,5){\line(1,0){2}}
\put(65,5){\line(1,0){2}}
\put(69,5){\line(1,0){2}}
\put(73,5){\line(1,0){2}}
\put(77,5){\line(1,0){2}}
\put(81,5){\line(1,0){2}}
\put(10,5){\circle*{10}}
\put(40,5){\circle*{10}}
\put(90,5){\circle*{10}}
\put(120,5){\circle*{10}}
\put(151,5){\circle{10}}
\end{picture}
\end{array}
\hspace{9mm} .
\end{displaymath}
For every $\mathrm{sp}(N)$, each irreducible representation $R$ is either real (when the sum of the components $a_i$ of the Dynkin label of $R$ with an odd index $i$ is even) or pseudo-real (when the sum of the odd-index components of the Dynkin label is odd). The dimension of a generic  representation $R$ is given by the Weyl formula
\begin{align}
\label{dim_R}
\dim R = \left[ \prod_{k=1}^N \prod_{i=1}^k \left( 1 + \frac{\sum_{j=i}^k a_j}{k-i+1} \right) \right] \cdot \prod_{k=0}^{N-2} \prod_{i=k+1}^{N-1} \left[ 1+ \frac{2 \left(\sum_{j=0}^k a_{N-j} \right) + \left(\sum_{j=k+1}^i a_{N-j} \right)}{k+i+2} \right]
\end{align}
(which, for $\mathrm{sp}(1)=\su(2)$, reduces to the well-known relation $\dim R = a_1 + 1 = 2J + 1$ for a spin-$J$ irreducible representation associated with a Young diagram of $a_1$ boxes). 

Focusing on the $N=2$ case, which is the one of relevance for our present work, one notes that the $\mathrm{sp}(2)$ Lie algebra is isomorphic to the $\so(5)$ algebra. In the conventions of ref.~\cite{Keski-Vakkuri:2022hhl}, the Cartan matrix of the $\mathrm{sp}(2)$ algebra is
\begin{align}
\label{Cartan_matrix}
A=\left(
\begin{array}{cc}
2 & -1 \\
-2 & 2
\end{array}
\right)
\end{align}
and the roots are $(\pm1,0)$, $(0,\pm1)$, $\pm\left(\frac{1}{2},\frac{1}{2}\right)$, $\pm\left(\frac{1}{2},-\frac{1}{2}\right)$; the simple roots are $\alpha^1=\left(\frac{1}{2},-\frac{1}{2}\right)$ and $\alpha^2=(0,1)$, and their coroots are $\alpha^{1\vee}=(2,-2)$ and $\alpha^{2\vee}=(0,2)$. Correspondingly, the highest-weight vectors of the two fundamental representations, with Dynkin labels $(1\,0)$ and $(0\,1)$, are $\mu^1=\left(\frac{1}{2},0\right)$ and $\mu^2=\left(\frac{1}{2},\frac{1}{2}\right)$, respectively. Denoting reflections with respect to the straight line orthogonal to the simple root $\alpha^i$ and going through the origin as $\sigma_i$, and noting that $r_+=\sigma_1 \sigma_2$ is a rotation by $\pi/2$ (with $r_+^{-1}=r_-=\sigma_2 \sigma_1$), it is easy to see that the Weyl group $W$ is the dihedral group $D_4$.

Recalling Kostant's formula~\cite{Kostant:1958aff}
\begin{align}
m_\mu(\nu) = \sum_{\sigma \in W} \sgn(\sigma) \mathcal{P} \left[ \sigma\left( \rho + \mu \right) - \left( \rho + \nu\right) \right],
\end{align}
which expresses the multiplicity of a weight $\nu$ in an irreducible representation with highest-weight vector $\mu$ in terms of a sum over the elements of the Weyl group, with $\sgn(\sigma)$ denoting the signature of $\sigma$ (which is $1$ when $\sigma$ can be written as the product of an even number of $\sigma_1$ and $\sigma_2$ factors, while it is $-1$ when $\sigma$ is given by the product of an odd number of $\sigma_1$ and $\sigma_2$ factors), $\rho$ denoting half of the sum of the positive roots, and $\mathcal{P}(\zeta)$ the Kostant partition function, defined as the number of ways to express $\zeta$ as a linear combination of the positive roots with natural coefficients, it is straightforward to derive the weight multiplicities of any irreducible representation.

The smallest irreducible representations of the $\mathrm{sp}(2)$ algebra are listed in table~\ref{tab:irreducible_representations}, where $\mathcal{C}_2(R)$ denotes the eigenvalue of the quadratic Casimir operator (namely, the sum of the squares of the generators in a generic representation $R$ is equal to $\mathcal{C}_2(R)\ide$) and $\lambda_R$ is the Dynkin index (defined as the trace of the square of each generator in the representation $R$---with the products of two different generators being traceless). For a more extensive list of irreducible representations of this algebra (up to dimension $94605$), see ref.~\cite{Yamatsu:2015npn}.

\begin{table}[h]
\centering
\phantom{-------}
\begin{tabular}{|c|c|c|c|c|}  
\hline
Dynkin label & dimension & $\mathcal{C}_2(R)$ & $\lambda_R$ & notes \\
\hline
\hline
 $(0\,0)$ & $1$                  & $0$            & $0$             & trivial, real \\
 $(1\,0)$ & $4$                  & $\frac{5}{4}$  & $\frac{1}{2}$   & defining, fundamental, pseudo-real \\ 
 $(0\,1)$ & $5$                  & $2$            & $1$             & fundamental, real \\
 $(2\,0)$ & $10$                 & $3$            & $3$             & adjoint, real \\
 $(0\,2)$ & $14$                 & $5$            & $7$             & real \\
 $(1\,1)$ & $16$                 & $\frac{15}{4}$ & $6$             & pseudo-real \\
 $(3\,0)$ & $20$                 & $\frac{21}{4}$ & $\frac{21}{2}$  & pseudo-real \\
 $(0\,3)$ & $30$                 & $9$            & $27$            & real \\
 $(2\,1)$ & $35$                 & $6$            & $21$            & real \\
 $(4\,0)$ & $35^\prime$          & $8$            & $28$            & real \\
 $(1\,2)$ & $40$                 & $\frac{29}{4}$ & $29$            & pseudo-real \\
 $(0\,4)$ & $55$                 & $14$           & $77$            & real \\
 $(5\,0)$ & $56$                 & $\frac{45}{4}$ & $63$            & pseudo-real \\
 $(3\,1)$ & $64$                 & $\frac{35}{4}$ & $56$            & pseudo-real \\
 $(1\,3)$ & $80$                 & $\frac{47}{4}$ & $94$            & pseudo-real \\
 $(2\,2)$ & $81$                 & $10$           & $81$            & real \\
 $(6\,0)$ & $84$                 & $15$           & $126$           & real \\
 $(0\,5)$ & $91$                 & $20$           & $182$           & real \\
 $(4\,1)$ & $105$                & $12$           & $126$           & real \\
 $(7\,0)$ & $120$                & $\frac{77}{4}$ & $231$           & pseudo-real \\
 $(3\,2)$ & $140$                & $\frac{53}{4}$ & $\frac{371}{2}$ & pseudo-real \\
 $(1\,4)$ & $140^\prime$         & $\frac{69}{4}$ & $\frac{483}{2}$ & pseudo-real \\
 $(0\,6)$ & $140^{\prime\prime}$ & $27$           & $378$           & real \\
 $(2\,3)$ & $154$                & $15$           & $231$           & real \\
 $(5\,1)$ & $160$                & $\frac{63}{4}$ & $252$           & pseudo-real \\
\hline                 
\end{tabular}
\phantom{-------}
\caption{The smallest irreducible representations of the $\mathrm{sp}(2)$ algebra and their main properties; $\mathcal{C}_2(R)$ denotes the eigenvalue of the quadratic Casimir operator, while $\lambda_R$ denotes the Dynkin index (with the conventional normalization to $\frac{1}{2}$ for the defining representation).}
\label{tab:irreducible_representations}
\end{table}

Some examples of decomposition of tensor products of irreducible representations of the $\mathrm{sp}(2)$ algebra (denoted by their Dynkin labels) include:
\begin{align}
(1\,0)\otimes(1\,0)&=(2\,0)\oplus(0\,1)\oplus(0\,0), \label{two_defining_give_singlet} \\
(0\,1)\otimes(0\,1)&=(0\,2)\oplus(2\,0)\oplus(0\,0), \label{also_two_fives_give_singlet} \\
(1\,0)\otimes(0\,1)&=(1\,1)\oplus(1\,0), \label{no_singlet_from_two_different_fundamentals} \\
(2\,0)\otimes(2\,0)&=(4\,0)\oplus(2\,1)\oplus(0\,2)\oplus(2\,0)\oplus(1\,0)\oplus 2(0\,0), \label{two_adjoint_give_singlet} 
\end{align}
showing that the singlet can be obtained from the product of any even number of copies of the defining representation, or of the fundamental representation of dimension $5$; it can also be obtained by multiplying the latter with any even, positive number of copies of the defining representation, but not from its product with just one fundamental representation of dimension $4$. Similarly, the singlet can also be obtained from the product of at least two copies of the adjoint representation. For an alternative method to work out the tensor products of irreducible representations, see also refs.~\cite{Antoine:1962zz, Antoine:1964zz, Vlasii:2015swq}.

\section{Details of the simulation updates} 
\label{app:details_of_the_simulation_updates}

Our Markov-chain Monte~Carlo algorithm is based on local updates of the $U_\mu(x)$ link matrices; more precisely, the matrices are updated through a combination of one heat-bath~\cite{Creutz:1980zw, Kennedy:1985nu} followed by several overrelaxation steps~\cite{Adler:1981sn, Brown:1987rra} on all of the link variables of the lattice: this defines one ``sweep''. In turn, the heat-bath and overrelaxation updates are implemented by applying them to a set of $\SU(2)$ subgroups of the $\Sp(N)$ matrices, according to method first proposed in ref.~\cite{Cabibbo:1982zn} for $\SU(N)$ lattice gauge theories. To ensure ergodicity and to reduce autocorrelation times, these updates are applied to a sufficiently large number of $\SU(2)$ subgroups for each $\Sp(N)$ matrix. In practice, the procedure involves multiplying the original $U_\mu(x)$ matrices by $\SU(2)$ elements embedded in $\Sp(N)$ matrices. Given a generic $\SU(2)$ element
\begin{align}
u=\left(
\begin{array}{cc}
\alpha & \beta \\
-\beta^\star & \alpha^\star
\end{array}
\right) , \qquad \alpha , \beta \in \C, \quad |\alpha|^2+|\beta|^2=1,
\end{align}
there exist different, unitarily inequivalent ways to embed it into an $\Sp(N)$ matrix, which are determined by the two indices $(i,j)$ specifying the rows and columns where the $\SU(2)$ element appears:
\begin{enumerate}
\item if the $\SU(2)$ element appears in the top-left block (and, as a consequence, its complex conjugate appears in the same location, in the bottom-right block), i.e., if $1 \le i <j \le N$, one obtains $\Sp(N)$ matrices $\mathcal{M}$ with elements:
\begin{align}
\mathcal{M}_{k,l}=\left\{
\begin{array}{ll}
\alpha & \mbox{when $k=l=i$ or $k=l=j+N$} \\
\alpha^\star & \mbox{when $k=l=j$ or $k=l=i+N$} \\
\beta & \mbox{when $k=i$ and $l=j$} \\
-\beta^\star & \mbox{when $k=j$ and $l=i$} \\
\beta^\star & \mbox{when $k=i+N$ and $l=j+N$} \\
-\beta & \mbox{when $k=j+N$ and $l=i+N$} \\
1 & \mbox{when $k=l$, with $k$ different from $i$, $j$, $i+N$ and $j+N$} \\
0 & \mbox{otherwise}
\end{array}
\right. ;
\end{align}
there exist $N(N-1)/2$ families of matrices of this type, which is the number of ways to choose the strictly ordered pair of indices between $1$ and $N$;
\item if $1\le i \le N$, while $N+1 \le j \le 2N$, with $j-i\neq N$, one obtains $\Sp(N)$ matrices $\mathcal{N}$ with elements:
\begin{align}
\mathcal{N}_{k,l}=\left\{
\begin{array}{ll}
\alpha & \mbox{when $k=l=i$, or when $k=l=j-N$} \\
\alpha^\star & \mbox{when $k=l=j$, or when $k=l=i+N$} \\
\beta & \mbox{when $k=i$ and $l=j$, or when $k=j-N$ and $l=i+N$} \\
-\beta^\star & \mbox{when $k=j$ and $l=i$, or when $k=i+N$ and $l=j-N$} \\
1 & \mbox{when $k=l$, with $k$ different from $i$, $j-N$, $i+N$ and $j$} \\
0 & \mbox{otherwise}
\end{array}
\right. ;
\end{align}
again, there exist $N(N-1)/2$ families of matrices of this type: this results from the product of the number of ways to choose the index $i$ between $1$ and $N$, times the number of ways to choose the index $j$ between $N+1$ and $2N$, with the constraint $j-i\neq N$, divided by two, to account for the fact that the resulting matrix is invariant under the interchange of $i$ with $j-N$;
\item finally, for $1\le i \le N$ and $j=i+N$, one obtains $\Sp(N)$ matrices $\mathcal{R}$ with elements:
\begin{align}
\mathcal{R}_{k,l}=\left\{
\begin{array}{ll}
\alpha & \mbox{when $k=l=i$} \\
\beta & \mbox{when $k=i$ and $l=i+N$} \\
\alpha^\star & \mbox{when $k=l=i+N$} \\
-\beta^\star & \mbox{when $k=i+N$ and $l=i$} \\
1 & \mbox{when $k=l$, with $k$ different from $i$ and $i+N$} \\
0 & \mbox{otherwise}
\end{array}
\right. ;
\end{align}
there exist $N$ families of matrices of this type (the number of ways to choose the index $i$ between $1$ and $N$).
\end{enumerate}
Specifically, our simulation code combines updates through matrices of type $\mathcal{M}$ and $\mathcal{R}$.

\bibliography{paper}

\end{document}